\documentclass[12pt, DIV10,a4paper]{article}
\usepackage{color}
\usepackage{float}
\usepackage[bf,small]{caption2}
\usepackage{comment}
\usepackage{amsmath}
\usepackage{bigints}
\usepackage{rotating, booktabs}
\usepackage{amsopn}
\usepackage{amsfonts}
\usepackage{amsthm}
\usepackage{amssymb}
\usepackage{amsbsy}
\usepackage{gensymb}
\usepackage{multirow}
\usepackage{titling}
\usepackage{natbib}
\usepackage{tocbibind}
\usepackage[a4paper,colorlinks,breaklinks,bookmarksopen,bookmarksnumbered]{hyperref}
\usepackage{epsfig,psfrag}
\usepackage{graphicx}
\usepackage{amsmath}
\usepackage{epstopdf}
\usepackage{tabularx}
\usepackage{subfigure}
\usepackage[mathscr]{euscript}
\usepackage[margin=1in]{geometry}
\usepackage{amsfonts} % Mathesymbole
\usepackage{calc}
\usepackage{titlesec}

\newcommand{\ueq}[1][]{%
  \if\relax\detokenize{#1}\relax
    \sbox0{$\underbrace{=}_{}$}%
    \mathrel{\mathmakebox[\wd0]{=}}
  \else
    \mathrel{\underbrace{=}_{\mathclap{#1}}}
  \fi}

\newcommand{\bzero}{\boldsymbol{0}}

\newcommand {\ctn}{\cite}

\usepackage{url}
\newcommand{\btheta}{\boldsymbol{\theta}}

\newcommand{\bbeta}{\boldsymbol{\beta}}

\newcommand{\bgamma}{\boldsymbol{\gamma}}
\newcommand{\bSigma}{\boldsymbol{\Sigma}}

\newcommand{\bmu}{\boldsymbol{\mu}}
\newcommand{\bnu}{\boldsymbol{\nu}}

\newcommand{\bB}{\boldsymbol{B}}

\newcommand{\bA}{\boldsymbol{A}}
\newcommand{\bS}{\boldsymbol{S}}

\newcommand{\bU}{\boldsymbol{U}}
\newcommand{\bx}{\boldsymbol{x}}
\newcommand{\bX}{\boldsymbol{X}}

\newcommand{\bY}{\boldsymbol{Y}}

\newcommand{\topline}{\hrule height 1pt width \textwidth \vspace*{2pt}}
\newcommand{\botline}{\vspace*{2pt}\hrule height 1pt width \textwidth \vspace*{4pt}}
\newtheorem{algo}{Algorithm} 

\begin{document}

\title{\vspace{-0.8in}
\textbf{IID Sampling from Intractable Distributions}}
\author{Sourabh Bhattacharya}
%\thanks{
%Sourabh Bhattacharya is a Professor in Interdisciplinary Statistical Research Unit, Indian Statistical
%Institute, 203, B. T. Road, Kolkata 700108.
%Corresponding e-mail: sourabh@isical.ac.in.}}
\date{\vspace{-0.5in}}
\maketitle%
\begin{center}
Indian Statistical Institute\\
Corresponding email:  \href{mailto: bhsourabh@gmail.com}{bhsourabh@gmail.com}
\footnote{This article is intended for Dr. C. R. Rao Special Issue.}
%\href{mailto: kshldey@gmail.com}{kshldey@gmail.com}
%,  \href{mailto: bhsourabh@gmail.com}{bhsourabh@gmail.com}\\
\end{center}
	
\begin{abstract}
In this article, we propose a novel methodology for drawing $iid$ realizations from any target distribution on the $d$-dimensional Euclidean space, for any $d \geq 1$. 
No assumption of compact support is required for the validity of our theory or method. The key idea is to construct an infinite sequence of concentric closed ellipsoids, 
motivated by the insight that the central ellipsoid tends to capture the modal region, while the regions between successive ellipsoids (ellipsoidal annuli) increasingly 
represent the tail regions of the distribution.

Representing the target distribution as an infinite mixture of component distributions defined on the central ellipsoid and the annuli, we propose a simulation strategy 
in which a component is first selected with its mixing probability, and then sampled exactly using perfect simulation. The perfect sampling scheme is built upon a 
minorization inequality for the general Metropolis--Hastings algorithm driven by uniform proposal distributions on the compact ellipsoid and annuli.

Unlike most existing work on perfect sampling, our method is not only theoretically valid but also practically applicable to any target distribution on $\mathbb{R}^d$, 
and is readily parallelizable. We validate the practicality of our approach by generating 10,000 $iid$ realizations from standard distributions such as 
the normal, Student’s $t$ with 5 degrees of freedom, and Cauchy, for dimensions $d = 1, 5, 10, 50, 100$, as well as from a 50-dimensional normal mixture distribution. 
In all cases, implementation times are reasonable, often less than a minute in our parallel setup, and the results are highly accurate.

We further demonstrate the method by generating 10,000 $iid$ realizations from posterior distributions associated with the well-known Challenger data, a Salmonella dataset, 
and a 160-dimensional spatial example of radionuclide count data on Rongelap Island. In each case, the results are encouraging and computation times remain very reasonable.
\\[2mm]
AMS (2000) subject classification Primary: 65C05; 62F15; Secondary: 60J22; 65C40; 62H20; 68W10.
\\[2mm]
\noindent \textbf{Keywords:} Ellipsoid; Minorization; Parallel computing; Perfect sampling; Residual distribution; Transformation-based Markov Chain Monte Carlo.
\end{abstract}
	
%\tableofcontents
\pagebreak

\section{Introduction}
\label{sec:introduction}

That Markov Chain Monte Carlo (MCMC) has revolutionized Bayesian computation is an understatement. Yet, despite its impact, MCMC 
has faced renewed scrutiny due to persistent convergence assessment challenges.
%seems to be gradually losing its appeal due to the persistent challenges of assessing convergence in practice. 
Even the most experienced MCMC theorists and practitioners often resort to deterministic 
or ad-hoc approximations to posterior distributions, considering the difficulties of diagnosing convergence insurmountable.

The concept of perfect simulation, introduced by \ctn{Propp96}, initially appeared promising to the MCMC community as a solution to the convergence problem. However, 
the apparent infeasibility of direct practical implementation in general Bayesian problems led to skepticism and, in many cases, outright disbelief. For instance, when 
the article by \ctn{Sabya12} on perfect sampling in mixture contexts was submitted to a reputed journal, reviewers rejected it, deeming the idea overly ambitious. 
To our knowledge, little promising research on perfect sampling has appeared in recent years.

In this article, we address the task of producing $iid$ realizations from any target distribution on $\mathbb{R}^d$, with $d \geq 1$, where $\mathbb R$ is the real line. 
Our key idea is to construct an infinite sequence of closed ellipsoids and represent the target distribution as an infinite mixture on the central ellipsoid and the annuli 
between successive ellipsoids. This construction reflects our insight that the central ellipsoid essentially captures the modal region, while the annuli represent the tail 
regions of the target distribution.

Our proposal for generating $iid$ realizations involves selecting a mixture component with its mixing probability and then simulating exactly from the chosen component 
using a perfect sampling scheme. To develop this scheme, we first consider the general Metropolis--Hastings algorithm on compact ellipsoids and annuli, driven by uniform 
proposal distributions. Exploiting compactness, we derive a suitable minorization inequality for each set in the sequence. Representing the Metropolis--Hastings chain 
as a mixture of two components, we then construct a perfect sampler for the density on its relevant support. Moreover, by adopting and rectifying an idea proposed in 
Murdoch and Green (1998), we avoid the computationally expensive coupling from the past (CFTP) algorithm of \ctn{Propp96}. Bypassing CFTP makes our approach computationally 
feasible. Integrating these ideas, we develop a generic parallel algorithm for simulating $iid$ realizations from general target distributions.

To illustrate the feasibility of our methodology, we apply the parallel algorithm to several standard distributions: normal, Student’s $t$ with 5 degrees of freedom, 
and Cauchy, for $d = 1, 5, 10, 50, 100$. We also consider a 50-dimensional mixture of normal distributions. In all cases, we simulate 10,000 $iid$ realizations, and the 
results are highly accurate. Furthermore, computation times are reasonable, with several experiments completing in less than a minute.

We also apply our method to posterior distributions based on real data. Specifically, we generate 10,000 $iid$ realizations from the posteriors corresponding to the 
Challenger data and the Salmonella data, with two and three parameters respectively. Finally, we address a particularly challenging case: generating 10,000 $iid$ 
realizations from the 160-dimensional posterior distribution for radionuclide count data on Rongelap Island, a problem widely regarded as difficult for MCMC convergence. 
Remarkably, we obtain $iid$ samples within 30 minutes, and the results are highly reliable.

The remainder of the article is organized as follows. Section~\ref{sec:idea} introduces our key idea. Section~\ref{subsec:ptmcmc} develops the set-wise perfect sampling theory 
and method. Section~\ref{sec:complete_algo} presents our complete parallel algorithm for $iid$ sampling from general target distributions. 
Section~\ref{sec:simstudy} reports simulation experiments with standard univariate and multivariate distributions, while Section~\ref{sec:realdata} details 
applications to posterior distributions based on real data. Finally, Section~\ref{sec:conclusion} provides concluding remarks.

\section{The $iid$ sampling idea}
\label{sec:idea}

Let $\pi(\btheta)$ be the target distribution supported on $\mathbb R^d$, where $d\geq 1$. 
Here $\btheta=(\theta_1,\ldots,\theta_d)^T$. Note that the distribution can be represented as
\begin{equation}
	\pi(\btheta)=\sum_{i=1}^{\infty}\pi(\bA_i)\pi_i(\btheta),
	\label{eq:p1}
\end{equation}
where $\bA_i$ are disjoint compact subsets of $\mathbb R^d$ such that $\cup_{i=1}^{\infty}\bA_i=\mathbb R^d$, and
\begin{equation}
	\pi_i(\btheta)=\frac{\pi(\btheta)}{\pi(\bA_i)}I_{\bA_i}(\btheta), 
	\label{eq:p2}
\end{equation}
is the distribution of $\btheta$ restricted on $\bA_i$; $I_{\bA_i}$ being the indicator function of $\bA_i$.
In (\ref{eq:p1}), $\pi(\bA_i)=\int_{\bA_i}\pi(d\btheta)\geq 0$. Clearly, $\sum_{i=1}^{\infty}\pi(\bA_i)=1$.

The key idea of generating $iid$ realizations from $\pi(\btheta)$ is to randomly select $\pi_i$ with probability $\pi(\bA_i)$ and then to perfectly simulate
from $\pi_i$. 

\subsection{Choice of the sets $\bA_i$}
\label{subsec:choice_sets}
For some appropriate $d$-dimensional vector $\bmu$ and $d\times d$ positive definite scale matrix $\bSigma$,
we shall set $\bA_i=\{\btheta:c_{i-1}\leq (\btheta-\bmu)^T\bSigma^{-1}(\btheta-\bmu)\leq c_i\}$ for $i=1,2,\ldots$, where $0=c_0<c_1<c_2<\cdots$. 
Note that $\bA_1=\{\btheta:(\btheta-\bmu)^T\bSigma^{-1}(\btheta-\bmu)\leq c_1\}$, and for $i\geq 2$, 
$\bA_i=\{\btheta:(\btheta-\bmu)^T\bSigma^{-1}(\btheta-\bmu)\leq c_i\}\setminus\cup_{j=1}^{i-1}\bA_j$.
Observe that ideally one should set 
$\bA_i=\{\btheta:c_{i-1}<(\btheta-\bmu)^T\bSigma^{-1}(\btheta-\bmu)\leq c_i\}$, but since
$\btheta$ has continuous distribution in our setup, we shall continue with $\bA_i=\{\btheta:c_{i-1}\leq (\btheta-\bmu)^T\bSigma^{-1}(\btheta-\bmu)\leq c_i\}$.

Thus, the first member of the sequence of sets $\bA_i$; $i\geq 1$, is a closed ellipsoid, while the others are closed annuli, the regions between two 
successive concentric closed ellipsoids.
The compact ellipsoid $\bA_1$ tends to support the modal region of the target distribution $\pi$ and for increasing $i\geq 2$, the compact 
annuli $\bA_i$ tend to support the tail regions
of the target distribution. Thus in practice, it is useful to choose $\sqrt{c_1}$ associated with $\bA_1$ to be relatively large compared
to $\sqrt{c_i}$ for $i\geq 2$. As will be seen in Section \ref{subsubsec:uniform_A}, 
for $i\geq 1$, $\sqrt{c_i}$ are the radii of $d$-dimensional balls of the form $\left\{\bY:\bY^T\bY\leq c_i\right\}$ which has linear correspondence
with $d$-dimensional ellipsoids of the form $\{\btheta:(\btheta-\bmu)^T\bSigma^{-1}(\btheta-\bmu)\leq c_i\}$. 
Hence, we shall refer to $\sqrt{c_i}$; $i\geq 1$, as radii.

\subsection{Choice of $\bmu$ and $\bSigma$}
\label{subsec:choice_mu_Sigma}
Theoretically, any choice of $\bmu\in\mathbb R^d$ and any $d\times d$ positive definite scale matrix $\bSigma$ is valid for our method. However,
judicious choices of $\bmu$ and $\bSigma$ play important roles in the efficiency of our $iid$ algorithm. In this regard,
in practice, $\bmu$ and $\bSigma$ will be the estimates of the mean (if it exists, or co-ordinate-wise median otherwise) 
and covariance structure of $\pi$ (it it exists, or some appropriate scale matrix otherwise). These estimates will be assumed to be obtained from some 
reliably implemented efficient MCMC algorithm. In this regard, we recommend the transformation based Markov Chain Monte Carlo (TMCMC)
methodology introduced by \ctn{Dutta14} that simultaneously updates all the variables using suitable deterministic transformations of some low-dimensional (often
one-dimensional) random variable defined on some relevant support. Since the method is driven by the low-dimensional random variable, this leads to
drastic dimension reduction in effect, with great improvement in acceptance rates and computational efficiency, while ensuring adequate convergence
properties. Our experiences also reveal that addition of a further deterministic step to TMCMC, in a similar vein as in \ctn{Liu00} often 
dramatically enhances the convergence properties of TMCMC.

\subsection{Dealing with the mixing probabilities $\pi(\bA_i)$}
\label{subsec:mixing_probs}
Recall that the key idea of $iid$ sampling from $\pi(\btheta)$ is to randomly select $\pi_i$ with probability $\pi(\bA_i)$ and then to exactly simulate
from $\pi_i$. However, the mixing probabilities $\pi(\bA_i)$ are not available prior to simulation. Since there are infinite number of such probabilities, 
even estimation of all these probabilities using only a finite TMCMC sample from $\pi$ is out of the question.  
But estimation of $\pi(\bA_i)$ using Monte Carlo samples drawn uniformly from $\bA_i$ makes sense, and the strategy applies to all $i\geq 1$. For the time being assume
that we have an infinite number of parallel processors, and the $i$-th processor is used to estimate $\pi(\bA_i)$ using Monte Carlo sampling up to a constant.
%let the resultant estimate be denoted by $\widehat{\pi(\bA_i)}$. 
To elaborate, let $\pi(\btheta)=C\tilde\pi(\btheta)$, where $C>0$ is the unknown normalizing constant. Then for any Borel set $\bA$ in the Borel $\sigma$-field
of $\mathbb R^d$, letting $\mathcal L(\bA)$ denote the
Lebesgue measure of $\bA$, observe that
\begin{equation}
	\pi(\bA_i)=C\mathcal L(\bA_i)\int \tilde\pi(\btheta)\frac{1}{\mathcal L(\bA_i)}I_{\bA_i}(\btheta)d\btheta = C\mathcal L(\bA_i)E\left[\tilde\pi(\btheta)\right],
	\label{eq:mc1}
\end{equation}
the right hand side being $C\mathcal L(\bA_i)$ times the expectation of $\tilde\pi(\btheta)$ with respect to the uniform distribution on $\bA_i$. This expectation
can be estimated by generating realizations of $\btheta$ from the uniform distribution on $\bA_i$, evaluating $\tilde\pi(\btheta)$ for the realizations and taking their
average. Let $\widehat{\tilde\pi(\bA_i)}$ denote the Monte Carlo average times $\mathcal L(\bA_i)$, so that 
$\widehat{\tilde\pi(\bA_i)}$ is an estimate of $\tilde\pi(\bA_i)$.
The required uniform sample generation from $\bA_i$ and computation of $\mathcal L(\bA_i)$ can be achieved in straightforward ways, which we elucidate below.

\subsubsection{Uniform sampling from $\bA_i$}
\label{subsubsec:uniform_A}
First consider uniform generation from the $d$-dimensional ball $\{\bY:\bY^T\bY\leq c_1\}$, which consists of generation of 
$X_k\stackrel{iid}{\sim}N(0,1)$, for $k=1,\ldots,d$, $U\sim U(0,1)$, the uniform distribution on $[0,1]$,
and setting $\bY=\sqrt{c_1}U^{1/d}\bX/\|\bX\|$, where $\bX=(X_1,\ldots,X_d)^T$ and $\|\cdot\|$ is the Euclidean norm. 
Then, letting $\bSigma=\bB\bB^T$, where $\bB$ is the lower triangular
Cholesky factor of $\bSigma$, $\btheta=\bmu+\bB\bY$ is a uniform realization from $\bA_1$.
To generate uniformly from $\bA_i=\{\btheta:c_{i-1}\leq (\btheta-\bmu)^T\bSigma^{-1}(\btheta-\bmu)\leq c_i\}
=\{\btheta:(\btheta-\bmu)^T\bSigma^{-1}(\btheta-\bmu)\leq c_i\}\setminus \{\btheta:(\btheta-\bmu)^T\bSigma^{-1}(\btheta-\bmu)\leq c_{i-1}\}$ for $i\geq 2$, 
we shall continue uniform simulation of $\tilde\btheta$ from $\{\btheta:(\btheta-\bmu)^T\bSigma^{-1}(\btheta-\bmu)\leq c_i\}$ until $\tilde\btheta$ satisfies 
$(\tilde\btheta-\bmu)^T\bSigma^{-1}(\tilde\btheta-\bmu)> c_{i-1}$. 

Note that this basic rejection sampling scheme will be inefficient with high rejection probabilities for $i\geq 2$ when $\sqrt{c_i}-\sqrt{c_{i-1}}$ are very small. 
In such cases, it is important to replace this algorithm with methods that avoid the rejection mechanism.
We indeed devise such a method for situations with small $\sqrt{c_i}-\sqrt{c_{i-1}}$ for $i\geq 2$ 
that completely bypasses rejection sampling; see Section \ref{sec:realdata}. 

\subsubsection{Lebesgue measure of $\bA_i$}
\label{subsubsec:Leb_A}
%Note that the right hand side of (\ref{eq:mc1}) requires calculation of $\mathcal L(\bA_i)$, the Lebesgue measure of $\bA_i$. 
It is well-known 
(see, for example, \ctn{Giraud15}) that the volume of a $d$-dimensional ball of radius $c$ is given by $\frac{\pi^{d/2}}{\Gamma(d/2+1)}c^d$, where
$\Gamma(x)=\int_0^{\infty}t^{x-1}\exp(-t)dt$, for $x>0$, is the Gamma function. Hence, in our case, letting $|\bB|$ denote the determinant of $\bB$,
%and with $\|\btheta\|=\sqrt{\btheta^T\btheta}$, 
$$\mathcal L(\bA_1)=|\bB|\times\mathcal L(\{\btheta:\|\btheta\|\leq \sqrt{c_1}\})=\frac{|\bB|\pi^{d/2}}{\Gamma(d/2+1)}c^{d/2}_1,$$ and for $i\geq 2$,
%$$\bA_i=\{\btheta:c_{i-1}\leq(\btheta-\bmu)^T\bSigma^{-1}(\btheta-\bmu)\leq c_i\}
%=\{\btheta:(\btheta-\bmu)^T\bSigma^{-1}(\btheta-\bmu)\leq c_i\}\setminus \{\btheta:(\btheta-\bmu)^T\bSigma^{-1}(\btheta-\bmu)\leq c_{i-1}\}$$ 
$\bA_i$ has Lebesgue measure
\begin{align}
	\mathcal L(\bA_i)&=|\bB|\times\mathcal L(\{\btheta:\|\btheta\|\leq \sqrt{c_i}\})-|\bB|\times\mathcal L(\{\btheta:\|\btheta\|\leq \sqrt{c_{i-1}}\})\notag\\
	&=|\bB|\times\frac{\pi^{d/2}}{\Gamma(d/2+1)}\left(c^{d/2}_i-c^{d/2}_{i-1}\right).\notag
\end{align}

The aforementioned ideas show that all the estimates $\widehat{\tilde\pi(\bA_i)}$, for $i\geq 1$, are obtainable simultaneously by parallel processing.
Also assume that the Monte Carlo sample size $N_i$ is large enough for each $i$ such that for any $\epsilon>0$, 
\begin{equation}
	1-\epsilon\leq\frac{\tilde\pi(\bA_i)}{\widehat{\tilde\pi(\bA_i)}}\leq 1+\epsilon.
	\label{eq:p3}
\end{equation}

We further assume that 
%$\widehat{\tilde\pi(\bA_i)}$ satisfy
%\begin{equation}
%	C\sum_{i=1}^{\infty}\widehat{\tilde\pi(\bA_i)}=1,
%	\label{eq:p4}
%\end{equation}
depending on $\epsilon$, there exists $M_0\geq 1$ such that for all $M\geq M_0$,
\begin{equation}
	1-\epsilon\leq C\sum_{i=1}^M\widehat{\tilde\pi(\bA_i)}\leq 1+\epsilon.
	\label{eq:p4}
\end{equation}
Letting
\begin{equation}
	\hat\pi(\btheta)=C\sum_{i=1}^{\infty}\widehat{\tilde\pi(\bA_i)}\pi_i(\btheta),
	\label{eq:p6}
\end{equation}
which is a well-defined density due to (\ref{eq:p4}),
we next provide a rejection sampling based argument to establish that it is sufficient to perfectly simulate from $\hat\pi(\btheta)$ in order to sample perfectly from
$\pi(\btheta)$.

\subsection{A rejection sampling argument for validation of $\hat\pi(\btheta)$} 
\label{subsec:rejection_sampling}
Due to (\ref{eq:p3}), for all $\btheta\in\mathbb R^d$, 
\begin{equation*}
	\pi(\btheta)=C\sum_{i=1}^{\infty}\widehat{\tilde\pi(\bA_i)}\times\frac{\tilde\pi(\bA_i)}{\widehat{\tilde\pi(\bA_i)}}\times\pi_i(\btheta)
	\leq C(1+\epsilon)\sum_{i=1}^{\infty}\widehat{\tilde\pi(\bA_i)}\pi_i(\btheta)=(1+\epsilon)\hat\pi(\btheta).
\end{equation*}
That is, for all $\btheta\in\mathbb R^d$,
\begin{equation}
	%\frac{\pi(\btheta)}{\sum_{i=1}^{\infty}\widehat{\pi(\bA_i)}\pi_i(\btheta)}\leq 1+\epsilon.
	\frac{\pi(\btheta)}{\hat\pi(\btheta)}\leq 1+\epsilon,
	\label{eq:p5}
\end{equation}
showing that rejection sampling is applicable
for generating samples from $\pi(\btheta)$ by sampling $\btheta$ from $\hat\pi(\btheta)$ and generating $U\sim U(0,1)$ until the following is satisfied:
\begin{equation}
	U<\frac{\pi(\btheta)}{(1+\epsilon)\hat\pi(\btheta)}.
	\label{eq:p7}
\end{equation}
In (\ref{eq:p7}), $\btheta\in\bA_i$ for some $i\geq 1$. Hence, due to (\ref{eq:p3}), 
\begin{equation}
\frac{\pi(\btheta)}{(1+\epsilon)\hat\pi(\btheta)}=\frac{1}{(1+\epsilon)}\frac{\tilde\pi(\bA_i)}{\widehat{\tilde\pi(\bA_i)}}>\frac{1-\epsilon}{1+\epsilon}.
	\label{eq:p8}
\end{equation}
The right most side of (\ref{eq:p8}) can be made arbitrarily close to $1$ for sufficiently small $\epsilon>0$. Now, since $U\leq 1$
with probability $1$, this entails that (\ref{eq:p7}) is satisfied with probability almost equal to one. 
Thus, for sufficiently small $\epsilon>0$, we can safely assume that (\ref{eq:p7}) holds for all practical applications, so that any simulation of $\btheta$ from
$\hat\pi(\btheta)$ would be accepted in the rejection sampling step. 
%Indeed, it can be checked that for $\epsilon\leq 10^{-8}$, the quantity   
In other words, to generate samples from the target density 
$\pi(\btheta)$ it is sufficient to simply generate from $\hat\pi(\btheta)$, for sufficiently small $\epsilon$.

\section{Perfect sampling from $\hat\pi(\btheta)$}
\label{subsec:ptmcmc}
For perfect sampling from $\hat\pi(\btheta)$ we first need to select $\pi_i$ with probability $C\widehat{\tilde\pi(\bA_i)}$ for some $i\geq 1$, 
and then need to sample from $\pi_i$ in the perfect sense.
Now $\pi_i$ would be selected if 
\begin{equation}
	\frac{\sum_{j=1}^{i-1}\widehat{\tilde\pi(\bA_j)}}{\sum_{j=1}^{\infty}\widehat{\tilde\pi(\bA_j)}}<U<
	\frac{\sum_{j=1}^{i}\widehat{\tilde\pi(\bA_j)}}{\sum_{j=1}^{\infty}\widehat{\tilde\pi(\bA_j)}}, 
	\label{eq:pp1}
\end{equation}
where $U\sim U(0,1)$ and $\widehat{\tilde\pi(\bA_0)}=0$. 
But the denominator of the left and right hand sides of (\ref{eq:pp1}) involves an infinite sum, which can not be computed. However, note that due to (\ref{eq:p4}),
\begin{equation}
	1-\epsilon\leq \frac{\sum_{j=1}^M\widehat{\tilde\pi(\bA_j)}}{\sum_{j=1}^{\infty}\widehat{\tilde\pi(\bA_j)}}\leq 1+\epsilon.
	\label{eq:pp2}
\end{equation}
The following hold due to (\ref{eq:pp2}):
\begin{align}
	&(1-\epsilon)\frac{\sum_{j=1}^{i-1}\widehat{\tilde\pi(\bA_j)}}{\sum_{j=1}^{M}\widehat{\tilde\pi(\bA_j)}}
	\leq\frac{\sum_{j=1}^{i-1}\widehat{\tilde\pi(\bA_j)}}{\sum_{j=1}^{\infty}\widehat{\tilde\pi(\bA_j)}}
	\leq (1+\epsilon)\frac{\sum_{j=1}^{i-1}\widehat{\tilde\pi(\bA_j)}}{\sum_{j=1}^{M}\widehat{\tilde\pi(\bA_j)}};\label{eq:pp3}\\
	&(1-\epsilon)\frac{\sum_{j=1}^{i}\widehat{\tilde\pi(\bA_j)}}{\sum_{j=1}^{M}\widehat{\tilde\pi(\bA_j)}}
	\leq\frac{\sum_{j=1}^{i}\widehat{\tilde\pi(\bA_j)}}{\sum_{j=1}^{\infty}\widehat{\tilde\pi(\bA_j)}}
	\leq (1+\epsilon)\frac{\sum_{j=1}^{i}\widehat{\tilde\pi(\bA_j)}}{\sum_{j=1}^{M}\widehat{\tilde\pi(\bA_j)}}.\label{eq:pp4}
\end{align}
Due to (\ref{eq:pp3}) and (\ref{eq:pp4}), for sufficiently small $\epsilon>0$, in all practical implementations (\ref{eq:pp1}) holds if and only if
\begin{equation}
	\frac{\sum_{j=1}^{i-1}\widehat{\tilde\pi(\bA_j)}}{\sum_{j=1}^{M}\widehat{\tilde\pi(\bA_j)}}<U<
	\frac{\sum_{j=1}^{i}\widehat{\tilde\pi(\bA_j)}}{\sum_{j=1}^{M}\widehat{\tilde\pi(\bA_j)}}
	\label{eq:pp5}
\end{equation}
holds. If, in any case, $i=M$, then we shall compute $\widehat{\tilde\pi(\bA_j)}$; $j=M+1,\ldots,2M$, and re-check (\ref{eq:pp5}) for the same set of uniform random numbers
$U\sim U(0,1)$ as before by replacing $M$ with $2M$, and continue the procedure until all the indices $i$ are less than $M$. 

%In the traditional way, the probailities $\widehat{\pi(\bA_j)}$; $j\geq 1$, are to be computed first and then
%$U$ is generated, using which (\ref{eq:pp1}) is checked. But this requires computation of infinite number of probabilities, which is infeasible. To avoid this problem,
%we shall first draw $U\sim U(0,1)$ and then compute $\widehat{\pi(\bA_j)}$ until (\ref{eq:pp1}) is satisfied for some $i\geq 1$, 
%Indeed, assuming that $M$ parallel processors are available to us, where $P$ is considerably large, we shall compute $\widehat{\pi(\bA_j)}$; $j=1,\ldots,P$ in parallel
%in the first place, and check (\ref{eq:pp1}). If $i>M$, then we shall compute $\widehat{\pi(\bA_j)}$; $j=M+1,\ldots,i$ and ahain check (\ref{eq:pp1}). This way
%we avoid computation of infinite number of probabilities.

Once $\pi_i$ is selected by the above strategy, we can proceed to sample perfectly from $\pi_i$. We progress towards building an effective perfect sampling strategy
by first establishing a minorization condition for the Markov transition kernel.

\subsection{Minorization for $\pi_i$}
\label{subsec:minorization}

For $\pi_i$, we consider the Metropolis-Hastings algorithm in which all the co-ordinates $\theta_k$; $k=1,\ldots,d$, are updated in a single block 
using an independence uniform proposal density on $\bA_i$ given by
\begin{equation}
	q_i(\btheta')=\frac{1}{\mathcal L(\bA_i)}I_{\bA_i}(\btheta').
	\label{eq:proposal}
\end{equation}
Although independence proposal distributions are usually not recommended, particularly in high dimensions, here this makes good sense. Indeed,
since $\bA_i$ are so constructed that the densities $\pi_i$ are expected to be relatively flat on these sets, 
the uniform proposal is expected to provide good approximation
to the target $\pi_i$. Furthermore, as we shall show, the uniform proposal will be responsible for obtaining significantly large lower bound for our
minorization, which, in turn, will play the defining role in our construction of an efficient perfect sampler for $\pi_i$.

%fixed at their current values. Note that this is equivalent to symmetric random walk MCMC where the co-ordinates are updated successively, one at a time.
%For $\btheta\in\bA_i$, let $P_i(\btheta,\bB\cap\bA_i)$ denote the corresponding symmetric random walk Markov transition probability for $\pi_i$, 
%where $\bB\subseteq\mathbb R^d$ is any Borel set. 
%For $k=1,\ldots,d$, let $\theta'_k=\theta_k+\epsilon_k$, where $\epsilon_k\stackrel{iid}{\sim}U(-a_i,a_i)$, for some $a_i>0$.
%Let $g_k(\theta'_k|\theta_k)$ denote the proposal density of $\theta'_k$ given $\theta_k$, induced by the distribution of $\epsilon_k$. Clearly,
%\begin{equation}
%	g_k(\theta'_k|\theta_k)=\frac{1}{2a}I_{[-a_i+\theta_k,a_i+\theta_k]}(\theta'_k).
%	\label{eq:minor1}
%\end{equation}
For $\btheta\in\bA_i$, for any Borel set $\mathbb B$ in the Borel $\sigma$-field of $\mathbb R^d$, 
let $P_i(\btheta,\mathbb B\cap\bA_i)$ denote the corresponding Metropolis-Hastings transition probability for $\pi_i$. 
Let $s_i=\underset{\btheta\in\bA_i}{\inf}~\tilde\pi(\btheta)$ and $S_i=\underset{\btheta\in\bA_i}{\sup}~\tilde\pi(\btheta)$.
Then, with (\ref{eq:proposal}) as the proposal density we have 
\begin{align}
	P_i(\btheta,\mathbb B\cap\bA_i)&\geq\int_{\mathbb B\cap\bA_i}
	\min\left\{1,\frac{\tilde\pi(\btheta')}{\tilde\pi(\btheta)}\right\}q_i(\btheta')d\btheta'\notag\\
	&\geq\left(\frac{s_i}{S_i}\right)\times\frac{\mathcal L(\mathbb B\cap\bA_i)}{\mathcal L(\bA_i)}\notag\\
	&=p_i~Q_i(\mathbb B \cap\bA_i),
	\label{eq:minor1}
\end{align}
where $p_i=s_i/S_i$ and
%For any desired $p_i\in (0,1)$, let $a_i$ be so chosen that
%\begin{align}
%	&\left(\frac{s_i}{2a_iS_i}\right)^d\mathcal L(\bA_i)=p_i\notag\\
%	&\Leftrightarrow a_i=\left(\frac{\mathcal L(\bA_i)}{p_i}\right)^{1/d}\times\left(\frac{s_i}{2S_i}\right).
%	\label{eq:minor3}
%\end{align}
%Combining (\ref{eq:minor2}) and (\ref{eq:minor3}) we obtain
%\begin{equation}
%	P_i(\btheta,\bB\cap\bA_i)\geq p_i~ Q_i(\bB\cap\bA_i),
%	\label{eq:minor4}
%\end{equation}
\begin{equation*}
	Q_i(\mathbb B\cap\bA_i)=\frac{\mathcal L(\mathbb B\cap\bA_i)}{\mathcal L(\bA_i)}
	%\label{eq:minor2}
\end{equation*}
is the uniform probability measure corresponding to (\ref{eq:proposal}).
%Thus, $Q_i$ is the uniform distribution on $\bA_i$.
Since (\ref{eq:minor1}) holds for all $\btheta\in\bA_i$, the entire set $\bA_i$ is a small set.

Here are some important remarks regarding justification of the choice of the uniform proposal density. First, since the uniform proposal is the independence
sampler, independent of the previous state of the underlying Markov chain, the right hand side of (\ref{eq:minor1}) is rendered independent of $\btheta$, which
is a requisite condition for minorization. For proposal distributions of the form $q_i(\btheta'|\btheta)$, infimum of $q_i(\btheta'|\btheta)$ for $\btheta,\btheta'\in\bA_i$
would be necessary, which would lead to much smaller lower bound for $P_i(\btheta,\mathbb B\cap\bA_i)$ compared to the right hand side of (\ref{eq:minor1}).  
This is precisely the reason why we do not consider TMCMC for $iid$ sampling, whose move types depend upon the previous state. 
Any independence proposal density other than the uniform will not get cancelled in the Metropolis-Hastings acceptance ratio and the corresponding $s_i/S_i$ may be 
vanishingly small if $\tilde\pi$ is not well-approximated by the independence proposal on $\bA_i$. This would make perfect sampling infeasible. 
Furthermore, as will be seen in Section \ref{subsec:coupling}, for perfect sampling it is required to simulate from the proposal density. 
But unlike the case of uniform proposal, it may be difficult to simulate from the proposal on $\bA_i$, 
which has closed ellipsoidal or closed annulus structure.

Observe that $s_i$ and $S_i$ associated with (\ref{eq:minor1}) 
will usually not be available in closed forms. In this regard, let $\hat s_i$ and $\hat S_i$ denote the minimum and maximum of
$\tilde\pi(\cdot)$ over the Monte Carlo samples drawn uniformly from $\bA_i$ in course of estimating $\tilde\pi(\bA_i)$ by $\widehat{\tilde\pi(\bA_i)}$.
Then $\frac{s_i}{S_i}\leq\frac{\hat s_i}{\hat S_i}$. Hence, there exists $\eta_i>0$ such that $1\geq\frac{s_i}{S_i}\geq\frac{\hat s_i}{\hat S_i}-\eta_i>0$.
Let $\hat p_i=\frac{\hat s_i}{\hat S_i}-\eta_i$. Then it follows from (\ref{eq:minor1}) that
\begin{equation}
	P_i(\btheta,\mathbb B\cap\bA_i\geq \hat p_i~Q_i(\mathbb B\cap\bA_i),
	\label{eq:minor1_hat}
\end{equation}
which we shall consider for our purpose from now on.
In practice, $\eta_i$ is expected to be very close to zero, since the Monte Carlo sample size would be sufficiently large. 
Thus, $\hat p_i$ is expected to be very close to $p_i$.

%Note that choosing $a_i$ as per (\ref{eq:minor3}) requires calculation of $\mathcal L(\bA_i)$, the Lebesgue measure of $\bA_i$. It is well-known 
%(see, for example, \ctn{Giraud15}) that the volume of a $d$-dimensional ball of radius $c$ is given by $\frac{\pi^{d/2}}{\Gamma(d/2+1)}c^d$, where
%$\Gamma(x)=\int_0^{\infty}t^{x-1}\exp(-t)dt$, for $x>0$, is the Gamma function. Hence, in our case, 
%$\mathcal L(\bA_1)=\mathcal L(\{\btheta:\|\btheta\|\leq c_1\})=\frac{\pi^{d/2}}{\Gamma(d/2+1)}c^d_1$, and for $i\geq 2$,
%$\bA_i=\{\btheta:c_{i-1}\leq\|\btheta\|\leq c_i\}=\{\btheta:\|\btheta\|\leq c_i\}\setminus \{\btheta:\|\btheta\|\leq c_{i-1}\}$ has Lebesgue measure
%$\mathcal L(\bA_i)=\mathcal L(\{\btheta:\|\btheta\|\leq c_i\})-\mathcal L(\{\btheta:\|\btheta\|\leq c_{i-1}\})=
%\frac{\pi^{d/2}}{\Gamma(d/2+1)}\left(c^d_i-c^d_{i-1}\right)$.

\subsection{Split chain}
\label{subsec:coupling}
Due to the minorization (\ref{eq:minor1_hat}), $P_i(\btheta,\mathbb B\cap\bA_i)$ admits the following decomposition for all $\btheta\in\bA_i$:
\begin{equation}
	P_i(\btheta,\mathbb B\cap\bA_i)=\hat p_i~ Q_i(\mathbb B\cap\bA_i)+(1-\hat p_i)~R_i(\btheta,\mathbb B\cap\bA_i),
	\label{eq:split1}
\end{equation}
where
\begin{equation}
	R_i(\btheta,\mathbb B\cap\bA_i)=\frac{P_i(\btheta,\mathbb B\cap\bA_i)-\hat p_i~ Q_i(\mathbb B\cap\bA_i)}{1-\hat p_i}
	\label{eq:split2}
\end{equation}
is the residual distribution.

Therefore, to implement the Markov chain $P_i(\btheta,\mathbb B\cap\bA_i)$, rather than proceeding directly with the uniform proposal based 
Metropolis-Hastings algorithm, we can
use the split (\ref{eq:split1}) to generate realizations from $P_i(\btheta,\mathbb B\cap\bA_i)$. That is, given $\btheta$, we can 
simulate from $Q_i$ with probability $\hat p_i$, and with the remaining
probability, can generate from $R_i(\btheta,\cdot)$. 
%Generating samples from $Q_i$, that is, uniform simulation on $\bA_i$ is simple.
%Indeed, note that uniform generation from $\bA_1=\{\btheta:\|\btheta\|\leq c_1\}$ consists of generation of $X_k\stackrel{iid}{\sim}N(0,1)$, for $k=1,\ldots,d$, $U\sim U(0,1)$
%and setting $\btheta=c_1U^{1/d}\bX/\|\bX\|$, where $\bX=(X_1,\ldots,X_d)^T$.
%To generate uniformly from $\bA_i=\{\btheta:c_{i-1}\leq\|\btheta\|\leq c_i\}=\{\btheta:\|\btheta\|\leq c_i\}\setminus \{\btheta:\|\btheta\|\leq c_{i-1}\}$ for $i\geq 2$, 
%we shall continue uniform simulation from $\{\btheta:\|\btheta\|\leq c_i\}$ until $\btheta$ satisfies $\|\btheta\|\geq c_{i-1}$.

To simulate from the residual density $R_i(\btheta,\cdot)$ we devise the following rejection sampling scheme. 
Let $\tilde R_i(\btheta,\btheta')$ and $\tilde P_i(\btheta,\btheta')$ %and $\tilde Q_i(\btheta')$ 
denote the densities of $\btheta'$ with respect to
$R_i(\btheta,\cdot)$ and $P_i(\btheta,\cdot)$, %and $Q_i(\cdot)$, 
respectively. Then it follows from (\ref{eq:split1}) and (\ref{eq:split2}) that for all $\btheta\in\bA_i$,
\begin{align*}
	&\tilde R_i(\btheta,\btheta')=\frac{\tilde P_i(\btheta,\btheta')-\hat p_i~ q_i(\btheta')}{1-\hat p_i}\leq \frac{\tilde P_i(\btheta,\btheta')}{1-\hat p_i}\notag\\
	&\Leftrightarrow \frac{\tilde R_i(\btheta,\btheta')}{\tilde P_i(\btheta,\btheta')}\leq \frac{1}{1-\hat p_i},~\mbox{for all}~\btheta'\in\bA_i.
	%\label{eq:rs1}
\end{align*}
Hence, given $\btheta$ we can continue to simulate $\btheta'\sim \tilde P_i(\btheta,\cdot)$ using the uniform proposal distribution (\ref{eq:proposal}) 
and generate $U\sim U(0,1)$ until 
\begin{equation}
U<\frac{(1-\hat p_i)\tilde R_i(\btheta,\btheta')}{\tilde P_i(\btheta,\btheta')}
	\label{eq:rejection_sampling2}
\end{equation}
is satisfied, at which point we accept $\btheta'$ as a realization from $\tilde R_i(\btheta,\cdot)$.

Now 
\begin{align}
	\tilde P_i(\btheta,\btheta')&=q_i(\btheta')\min\left\{1,\frac{\pi(\btheta')}{\pi(\btheta)}\right\}+r_i(\btheta)I_{\btheta}(\btheta')\notag\\
	&=\frac{1}{\mathcal L(\bA_i)}\min\left\{1,\frac{\pi(\btheta')}{\pi(\btheta)}\right\}+r_i(\btheta)I_{\btheta}(\btheta'),\notag
	%\label{eq:mc_kernel1}
\end{align}
where
\begin{align}
	r_i(\btheta)&=1-\int_{\bA_i}\min\left\{1,\frac{\pi(\tilde\btheta)}{\pi(\btheta)}\right\}q_i(\tilde\btheta)d\tilde\btheta\notag\\
	&= 1-\int_{\bA_i}\min\left\{1,\frac{\pi(\tilde\btheta)}{\pi(\btheta)}\right\}\frac{1}{\mathcal L(\bA_i)}d\tilde\btheta.
	\label{eq:mc_kernel2}
\end{align}
Let $\hat r_i(\btheta)$ denote the Monte Carlo estimate of $r_i(\btheta)$ obtained by simulating $\tilde\btheta$ from the uniform distribution on $\bA_i$ and
taking the average of $\min\left\{1,\frac{\pi(\tilde\btheta)}{\pi(\btheta)}\right\}$ in (\ref{eq:mc_kernel2}).
Let
\begin{align}
	\hat P_i(\btheta,\btheta') &=\frac{1}{\mathcal L(\bA_i)}\min\left\{1,\frac{\pi(\btheta')}{\pi(\btheta)}\right\}+\hat r_i(\btheta)I_{\btheta}(\btheta'),~\mbox{and}\notag\\
	%\label{eq:mc_kernel3}\\
	\hat R_i(\btheta,\btheta')&=\frac{\hat P_i(\btheta,\btheta')-\hat p_i~ q_i(\btheta')}{1-\hat p_i}.\notag
	%\label{eq:mc_kernel4}
\end{align}
Given $\btheta$ and any $\epsilon>0$, let the Monte Carlo sample size be large enough such that 
\begin{align}
	&	1-\epsilon\leq \frac{\tilde P_i(\btheta,\btheta')}{\hat P_i(\btheta,\btheta')}\leq 1+\epsilon,~\mbox{and}\notag\\
	&1-\epsilon\leq \frac{\tilde R_i(\btheta,\btheta')}{\hat R_i(\btheta,\btheta')}\leq 1+\epsilon.\notag
\end{align}
Then
\begin{equation}
	\left(\frac{1-\epsilon}{1+\epsilon}\right)\frac{\tilde R_i(\btheta,\btheta')}{\tilde P_i(\btheta,\btheta')}
	\leq\frac{\hat R_i(\btheta,\btheta')}{\hat P_i(\btheta,\btheta')}\leq \left(\frac{1+\epsilon}{1-\epsilon}\right)
	\frac{\tilde R_i(\btheta,\btheta')}{\tilde P_i(\btheta,\btheta')}.
	\label{eq:bound1}
\end{equation}
Due to (\ref{eq:bound1}), in all practical implementations, for sufficiently small $\epsilon>0$, (\ref{eq:rejection_sampling2}) holds if and only if 
\begin{equation}
U<\frac{(1-\hat p_i)\hat R_i(\btheta,\btheta')}{\hat P_i(\btheta,\btheta')}
	\label{eq:rejection_sampling3}
\end{equation}
holds. Hence, we shall carry out our implementations with (\ref{eq:rejection_sampling3}).

\subsection{Perfect simulation from $\pi_i$}
\label{subsec:perfect}
For perfect simulation we exploit the following idea first presented in \ctn{Murdoch98}.
From (\ref{eq:split1}) note that at any given positive time $T_i=t$, $\btheta'$ will be drawn
from $Q_i(\cdot)$ with probability $\hat p_i$. Also observe that $T_i$ can not take the value $0$, since only initialization is done at the zero-th time point. 
Hence, $T_i$ follows a geometric distribution given by
\begin{equation}
P(T_i=t)=\hat p_i (1-\hat p_i)^{t-1};~t=1,2,\ldots.
	\label{eq:geo}
\end{equation}
This is in contrast with the form of the geometric distribution 
provided in \ctn{Murdoch98}, which gives positive probability to $T_i=0$ (in our context, 
$P(T_i=t)=\hat p_i (1-\hat p_i)^{t};~t=0,1,2,\ldots$, with respect to their specification). Indeed, all our implementations %, detailed below,
yielded correct $iid$ simulations only for the form (\ref{eq:geo}) and failed to provide correct answers for that of \ctn{Murdoch98}.

Because of (\ref{eq:geo}), it is possible to simulate
$T_i$ from the geometric distribution and then draw $\btheta^{(-T_i)}\sim Q_i(\cdot)$. Then the chain only
needs to be carried forward in time till $t=0$, using $\btheta^{(t+1)}=\psi_i(\btheta^{(t)},\bU^{(t+1)}_i)$,
where $\psi_i(\btheta^{(t)},\bU^{(t+1)}_i)$ is the deterministic function corresponding to the simulation
of $\btheta^{(t+1)}$ from $\tilde R_i(\btheta^{(t)},\cdot)$ using the set of appropriate random numbers $\bU^{(t+1)}_i$;
the sequence $\{\bU^{(t)}_i;t=0,-1,-2,\ldots\}$ being assumed to be available before beginning the perfect
sampling implementation. The resulting draw $\btheta^{(0)}$ sampled at time $t=0$ is a perfect sample
from $\pi_i(\cdot)$. 

In practice, storing the uniform random numbers $\{\bU^{(t)}_i;t=0,-1,-2,\ldots\}$ or explicitly considering the deterministic relationship 
$\btheta^{(t+1)}=\psi_i(\btheta^{(t)},\bU^{(t+1)}_i)$,
are not required. These would be required only if we had taken the search approach, namely, iteratively starting the Markov chain at all initial values at negative times
and carrying the sample paths to zero.

\section{The complete algorithm for $iid$ sample generation from $\pi$}
\label{sec:complete_algo}
We present the complete algorithm for generating $iid$ realizations from the target distribution $\pi$ as Algorithm \ref{algo:perfect}.
\begin{algo}\label{algo:perfect}\topline
IID sampling from the target distribution $\pi$ \botline \normalfont \ttfamily
\begin{itemize}
	\item[(1)] Using TMCMC based estimates, obtain $\bmu$ and $\bSigma$ required for the sets $\bA_i$; $i\geq 1$.
	\item[(2)] Fix $M$ to be sufficiently large. 
	\item[(3)] Choose the radii $\sqrt{c_i}$; $i=1,\ldots,M$ appropriately. A general strategy for these selections will be discussed in the context
		of the applications.
	\item[(4)] Compute the Monte Carlo estimates $\widehat{\tilde\pi(\bA_i)}$; $i=1,\ldots,M$, in parallel processors. 
                Thus, $M$ parallel processors will obtain all the estimates simultaneously.
	\item[(5)] Instruct each processor to send its respective estimate to all the other processors.
	\item[(6)] Let $K$ be the required $iid$ sample size from the target distribution $\pi$. Split the job of obtaining $K$ $iid$ realizations into 
		parallel\\ processors, each processor scheduled to simulate a single realization at a time. That is, with $K$ parallel processors, $K$ realizations
		will be simulated simultaneously. In each processor, do the following:
		\begin{enumerate}	
			\item[(i)] Select $\pi_i$ with probability proportional to $\widehat{\tilde\pi(\bA_i)}$. 
			\item[(ii)]If $i=M$ for any processor, then return to Step (2), increase $M$ to $2M$, and repeat the subsequent steps
				(in Step (4) only $\widehat{\tilde\pi(\bA_i)}$; $i=M+1,\ldots,2M$, need to be computed).		
				Else
				\begin{enumerate}
					\item[(a)] Draw $T_i\sim \mbox{Geometric}(\hat p_i)$ with respect to (\ref{eq:geo}).
                        \item[(b)] Draw $\btheta^{(-T_i)}\sim Q_i(\cdot)$.
                        \item[(c)] Using $\btheta^{(-T_i)}$ as the initial value, carry the chain $\btheta^{(t+1)}\sim \tilde R_i(\btheta^{(t)},\cdot)$ 
				forward for $t=-T_i,-T_i+1,\ldots,-1$. 
%using the deterministic functional
%relationship $\btheta^{(t+1)}=\psi(\btheta^{(t)},\bU^{(t+1)})$ and the available sequence
%$\{\bU^{(t)};t=0,1,2,\ldots\}$.
                \item[(d)] From the current processor, send $\btheta^{(0)}$ to processor $0$ as a perfect realization from $\pi$.
				\end{enumerate}
		\end{enumerate}
	\item[(7)] Processor $0$ stores the $K$ $iid$ realizations $\left\{\btheta^{(0)}_1.\ldots,\btheta^{(0)}_K\right\}$ thus generated from the target distribution $\pi$.
\end{itemize}
\rmfamily
\botline
\end{algo}

Figures~\ref{fig:schema1}--\ref{fig:schema3}
provide a schematic illustration of the proposed methodology for
generating $iid$ samples from an arbitrary target distribution $\pi(\btheta)$.

Figure~\ref{fig:schema1} presents the conceptual
overview of the iid sampling algorithm. The target distribution is
decomposed into a sequence of concentric ellipsoidal or annular
regions $\{\bA_i\}$, each associated with probability $\pi(\bA_i)$.
A component distribution $\pi_i(\btheta)$ is selected with
probability proportional to $\pi(\bA_i)$, and a perfect sample is
drawn from it. Repeating this procedure yields $iid$ realizations
from the full target $\pi(\btheta)$.

{Figure~\ref{fig:schema2} illustrates the perfect
sampling mechanism within a chosen component $\bA_i$. The
Metropolis--Hastings transition kernel $P_i$ is expressed through the
minorization inequality $P_i\ge\hat p_iQ_i$, leading to
$P_i = \hat{p}_i Q_i + (1-\hat{p}_i)R_i$, where $Q_i$ denotes the
uniform distribution over $\bA_i$ and $R_i$ represents the residual.
This allows a split-chain representation in which a random
coalescence time $T_i$ is drawn from a geometric distribution, and
the chain is propagated forward to obtain a perfect draw
$\theta^{(0)}\sim\pi_i$.

Figure~\ref{fig:schema3} depicts the
parallel implementation architecture of the algorithm. Each worker
processor estimates its corresponding component weight
$\widehat{\pi}(\bA_i)$ and performs perfect sampling independently.
The master processor aggregates all draws
$\{\btheta_1,\dots,\btheta_K\}$ to form a collection of $iid$ samples
from the overall target distribution $\pi(\btheta)$.

\begin{figure}
	\centering
	\includegraphics[width=18cm,height=19cm]{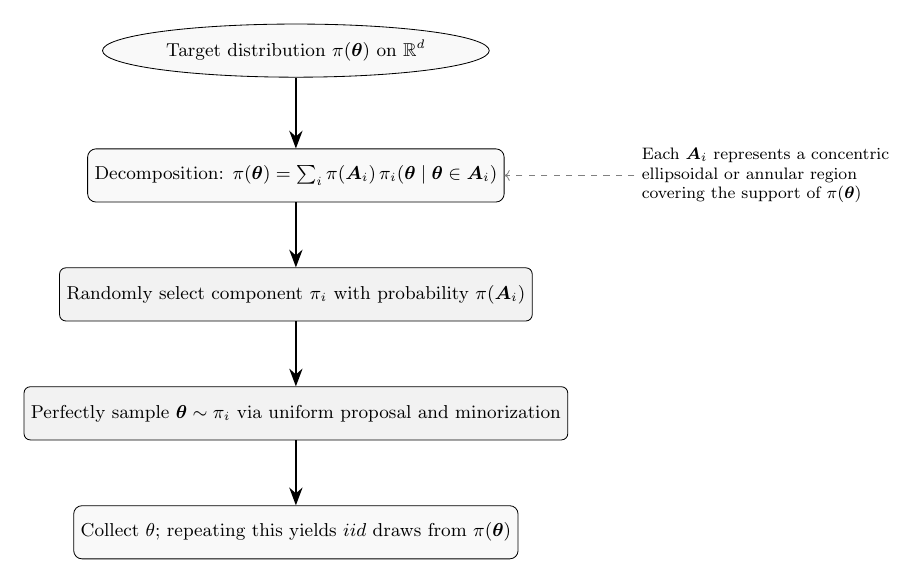}
	\caption{Schematic diagram: the conceptual overview.}
	\label{fig:schema1}
\end{figure}

\begin{figure}
	\centering
	\includegraphics[width=18cm,height=19cm]{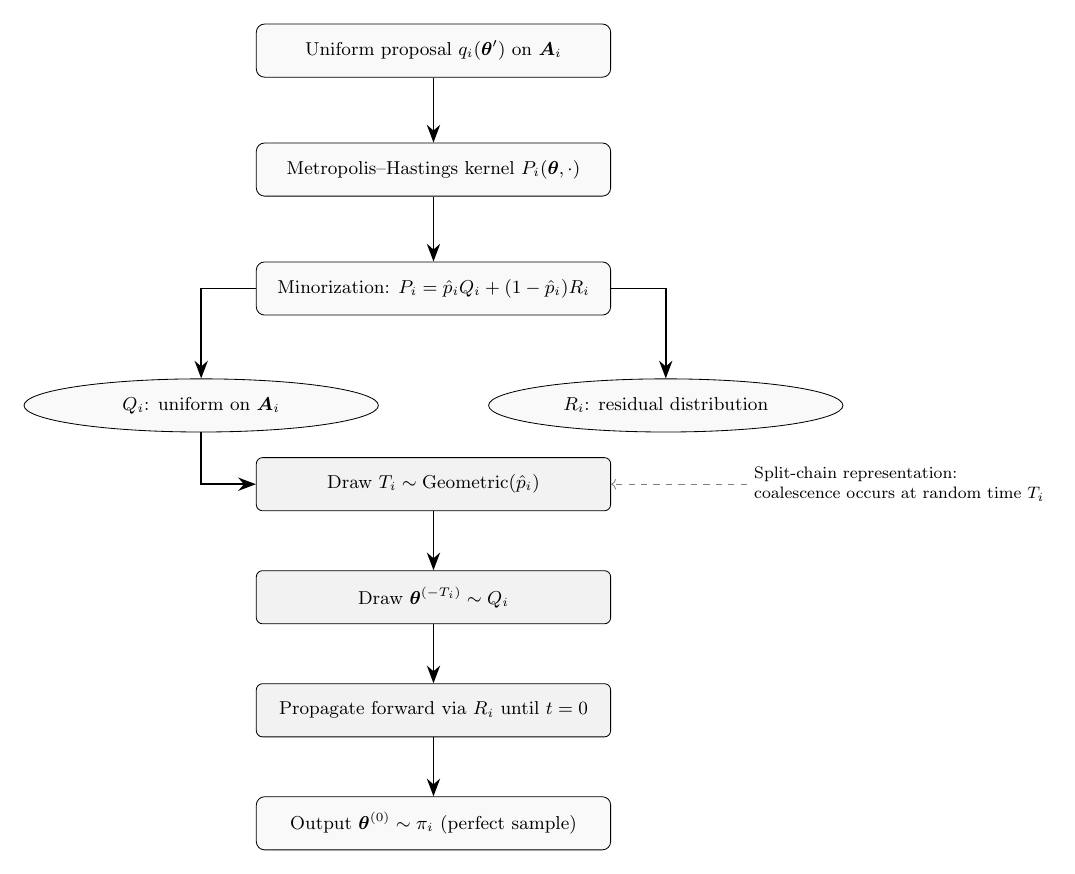}
	\caption{Schematic diagram: perfect sampling.}
	\label{fig:schema2}
\end{figure}

\begin{figure}
	\centering
	\includegraphics[width=18cm,height=19cm]{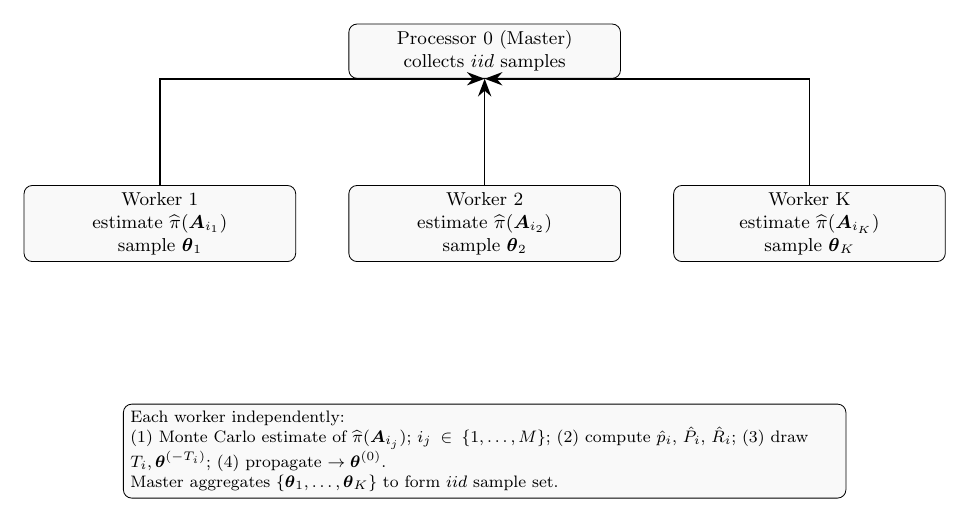}
	\caption{Schematic diagram: parallel implementation.}
	\label{fig:schema3}
\end{figure}

\section{Experiments with standard distributions}
\label{sec:simstudy}
To illustrate our $iid$ sampling methodology, we now apply the same to generate $iid$ samples from standard distributions, namely, 
normal, Student's $t$ (with $5$ degrees of freedom), Cauchy, and mixture of normals. We consider both univariate and multivariate versions of the distributions,
with dimensions $d=1,5,10,50,100$. For the location vector, we set $\bnu=(\nu_1,\ldots,\nu_d)^T$, with $\nu_i=i$, for $i=1,\ldots,d$. We denote the $d\times d$ 
scale matrix by $\bS$, whose $(i,j)$-th element is specified by $S_{ij}=10\times\exp\left\{-(i-j)^2/2\right\}$. 
We also consider a mixture of two normals for $d=50$ with the means being $\bnu_1=\bnu$ and $\bnu_2=2\bnu$, and the covariance matrix $\bS$ the same 
as the above for both the normal components. The mixing proportions
of the normals associated with $\bnu_1$ and $\bnu_2$ are $2/3$ and $1/3$, respectively. 
For illustrative purposes, we assume that $\bmu=\bnu$ and $\bSigma=\bS$ associated with the sets $\bA_i$. In the normal mixture setup,
there are two relevant sets of the form $\bA_i$ for each $i$, associated with the two mixture components. We denote these sets by $\bA_{1i}$ and $\bA_{2i}$, 
associated with $\bmu=\bnu_1$ and $\bmu=\bnu_2$, respectively, while $\bSigma=\bS$ in both the cases.
In $\hat p_i$, we set $\eta_i=10^{-5}$ in all the simulation experiments.

Details of our $iid$ simulation procedure and the results for the different standard distributions are presented below.
All our codes are written in C using the Message Passing Interface (MPI) protocol for parallel processing. We implemented our codes on a $80$-core VMWare 
provided by Indian Statistical Institute. The machine has $2$ TB memory and each core has about $2.8$ GHz CPU speed.

\subsection{Simulation from normals}
\label{subsec:normal}
For $d=1,5,10,50,100$, we set $\sqrt{c_1}=4$ and for $i=2,\ldots,M$, set $\sqrt{c_i}=\sqrt{c_1}+(i-1)/2$. With these choices of the radii, it turned out that $M=71$
was sufficient for all the dimensions considered. 
The choice of such forms of the radii has been guided by the insight that the target distribution is concentrated on $\bA_1$, and the sets $\bA_i$, for increasing $i$,
encapsulate the tail regions. %Since the normal distribution is light-tailed, 
Thus, setting $\sqrt{c_1}$ somewhat large and 
keeping the difference $\sqrt{c_i}-\sqrt{c_{i-1}}$ adequately small for $i>1$, seems to be a sensible strategy for giving more weight to the modal region
and less weight to the tail regions. Of course, it has to be borne in mind that a larger value of $\sqrt{c_1}$ than what is adequate can make $\hat p_1$
extremely small, since for larger radii $s_1$ would be smaller and $S_1$ larger (see (\ref{eq:minor1})). The explicit choices in our cases are based on experimentations after
fixing the forms of the radii for dimension $d$ as follows: $\sqrt{c_1}=r_d$ and for $i=2,\ldots,M$, $\sqrt{c_i}=\sqrt{c_1}+a_d\times (i-1)$. The values of $M$, $r_d$
and $a_d$ that yielded sufficiently large values of $\hat p_i$; $i=1,\ldots,M$, for Monte Carlo size $N_i=10,000$, are considered for our applications.
With a given choice of $M$, if $\bA_{M}$ is represented in the set of $iid$ realizations, then we increase $M$ to $2M$ and repeat the experiment, as suggested 
in Section \ref{subsec:ptmcmc} and made explicit in Algorithm \ref{algo:perfect}.
This basic method of selecting the radii and $M$ will remain the same in all our applications. %both for  and real data based.

With the Monte Carlo size $N_i=10,000$ used for the construction of $\hat\pi(\btheta)$, 
we then simulated $10,000$ $iid$ realizations 
from $\hat\pi(\btheta)$, which took $57$ minutes, $1$ hour $45$ minutes, $25$ minutes, $6$ minutes, and $1$ hour $39$ minutes on our VMWare, for $d=1,5,10,50,100$, respectively.
In all the cases, the true marginal densities and the correlation structures are very accurately captured by the $iid$ samples generated by our methodology.
For the sake of brevity, we display in Figure \ref{fig:normal_simstudy100} the results for only four co-ordinates in the context of $d=100$.
The exact and estimated correlation structures for the first $20$ co-ordinates are displayed in Figure \ref{fig:normal_corr_simstudy100}.
\begin{figure}
	\centering
	\subfigure [True and $iid$-based density for $\theta_1$.]{ \label{fig:normal1}
	\includegraphics[width=7.5cm,height=7.5cm]{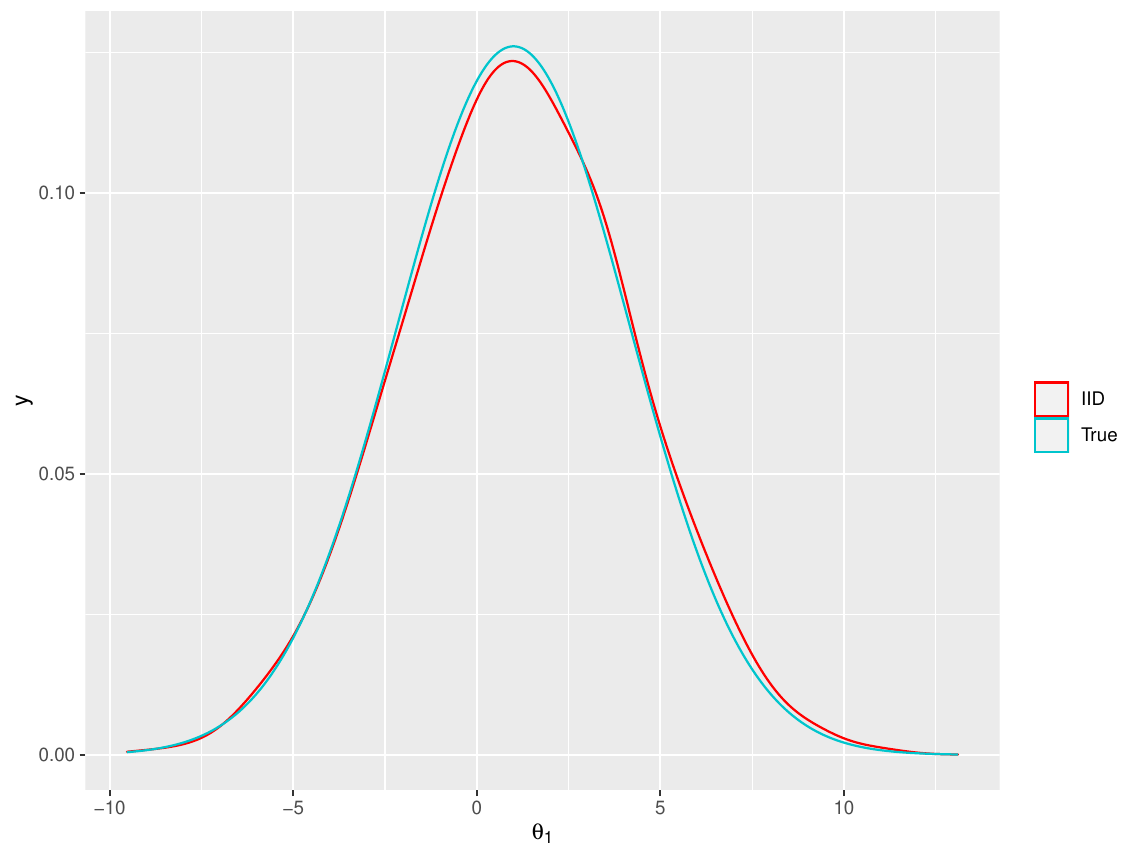}}
	\hspace{2mm}
	\subfigure [True and $iid$-based density for $\theta_{25}$.]{ \label{fig:normal25}
	\includegraphics[width=7.5cm,height=7.5cm]{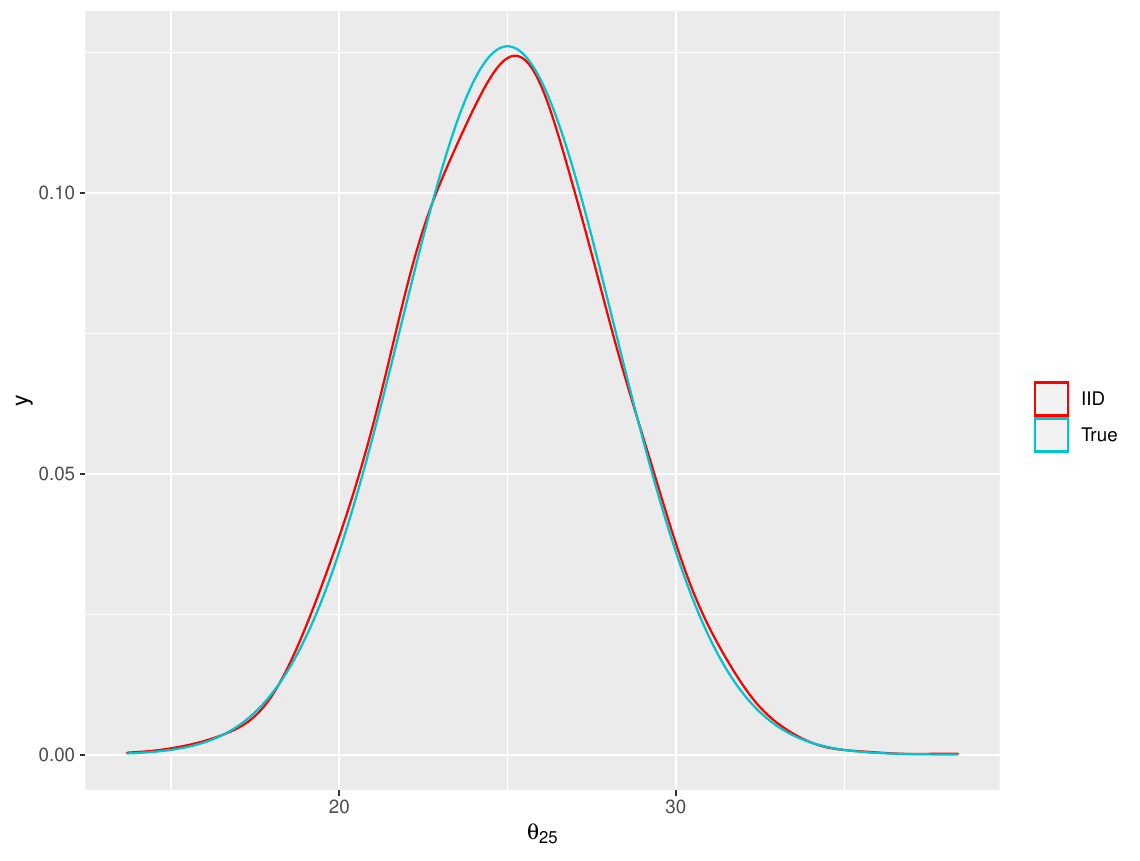}}\\
	\vspace{2mm}
	\subfigure [True and $iid$-based density for $\theta_{75}$.]{ \label{fig:normal75}
	\includegraphics[width=7.5cm,height=7.5cm]{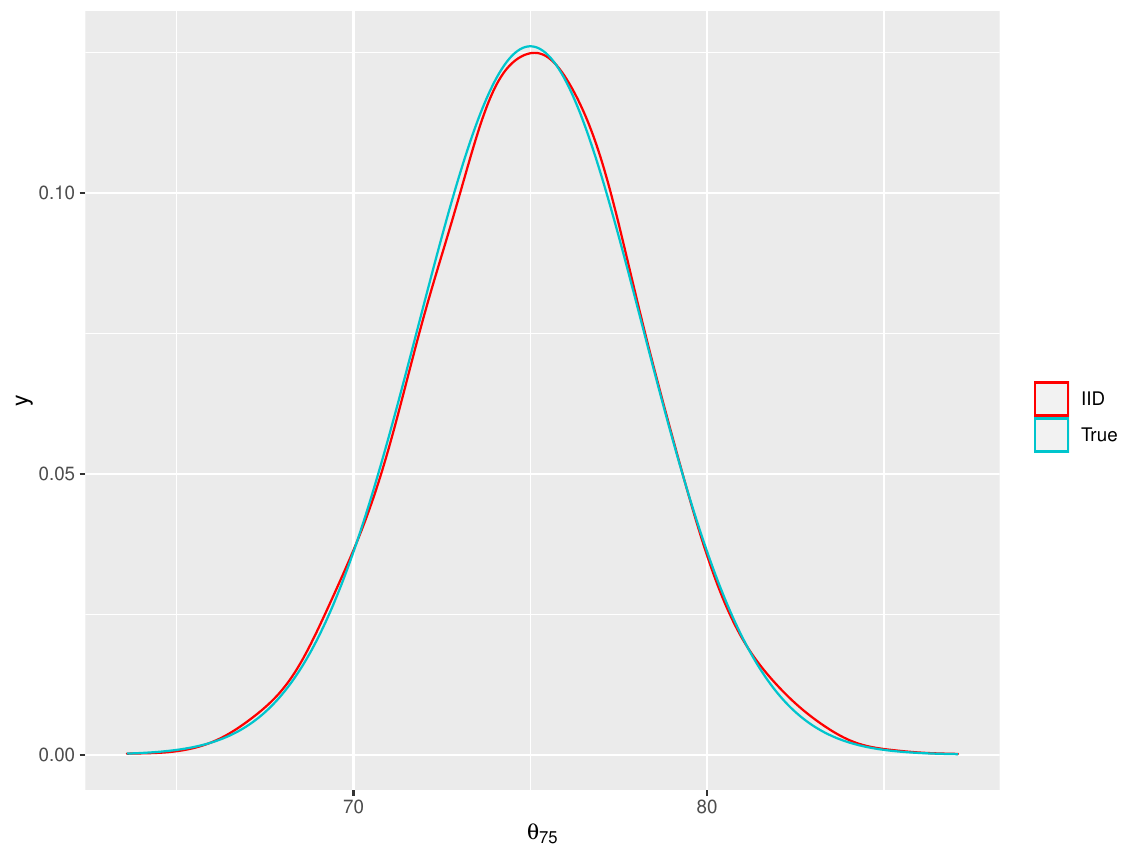}}
	\vspace{2mm}
	\subfigure [True and $iid$-based density for $\theta_{100}$.]{ \label{fig:normal100}
	\includegraphics[width=7.5cm,height=7.5cm]{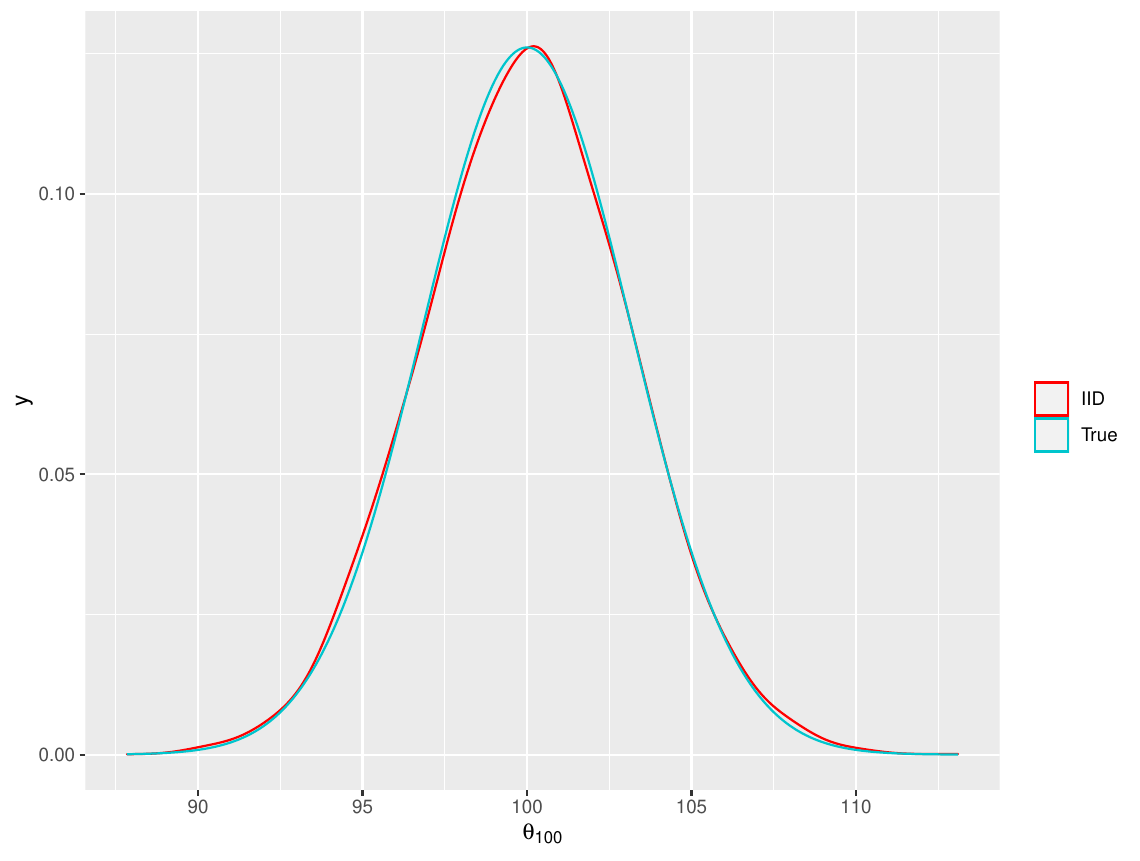}}
	\caption{Simulation from $100$-dimensional normal distribution. The red and blue colours denote the $iid$ sample based density and the true density, respectively.}
	\label{fig:normal_simstudy100}
\end{figure}

\begin{figure}
	\centering
	\subfigure [Exact correlation structure.]{ \label{fig:normal_corr_exact}
	\includegraphics[width=7.5cm,height=7.5cm]{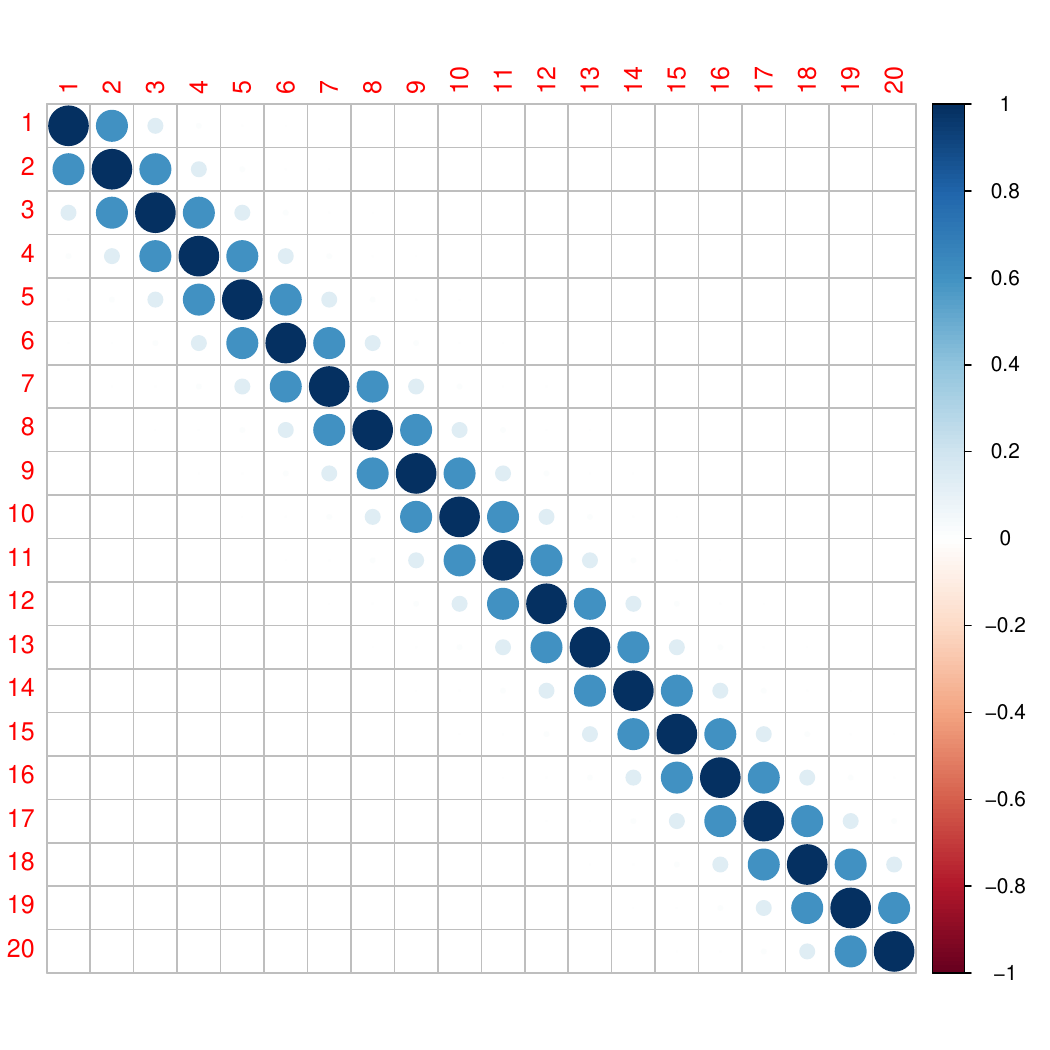}}
	\hspace{2mm}
	\subfigure [Correlation structure based on $iid$ samples.]{ \label{fig:normal_corr100}
	\includegraphics[width=7.5cm,height=7.5cm]{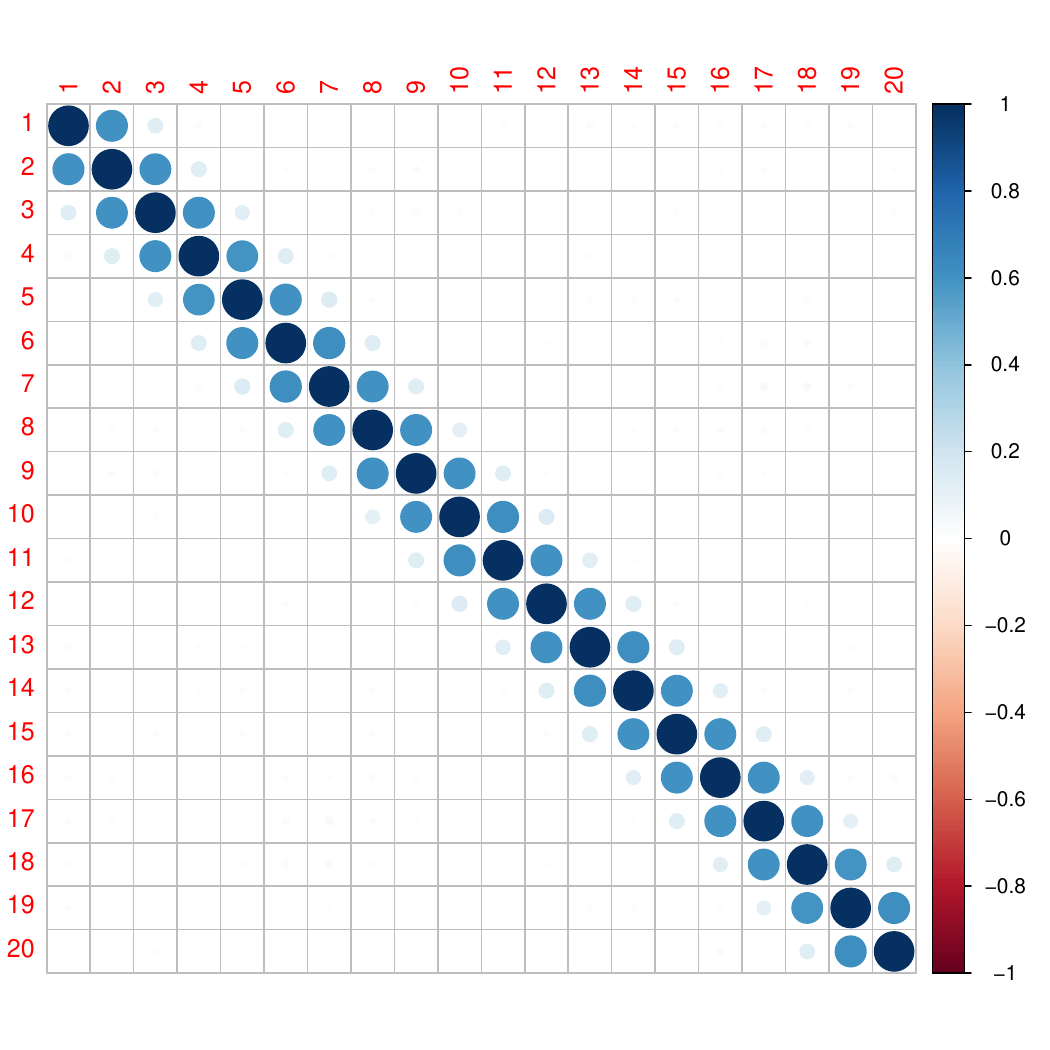}}
	\caption{Simulation from $100$-dimensional normal distribution. True and $iid$-based correlation structures for the first $20$ co-ordinates.}
	\label{fig:normal_corr_simstudy100}
\end{figure}

\subsection{Simulation from Student's $t$ with $5$ degrees of freedom}
\label{subsec:t}

In the Student's $t$ setup, $M=1000$ turned out to be sufficient for all the dimensions that we considered; this significantly larger value of $M$ compared to
that for the normal setup is due to the much thicker tails of the $t$ distribution with $5$ degrees of freedom. Depending upon the dimension,
we chose the radii differently for adequate performance of our methodology. The general form of our choice is the same as in the normal setup, that is,
$\sqrt{c_1}=r_d$ and $\sqrt{c_i}=\sqrt{c_1}+a_d\times (i-1)$, for $i=2,\ldots,M$. Following our experimentations as discussed in the normal simulation context, 
we set $r_1=5$ and $r_d=4$, for $d=5,10,50,100$. As regards $a_d$, we set
$a_1=3.801$, $a_5=2.1654$, $a_{10}=2.5$, $a_{50}=0.52$ and $a_{100}=0.52$.
%Here, for $d=1$, we set $\sqrt{c_1}=5$ and for $i=1,\ldots,M$, with $M=1000$, set $\sqrt{c_i}=\sqrt{c_1}+3.801\times i$. 
%For $d=5$, we set $\sqrt{c_1}=4$ and for $i=1,\ldots,M$, with $M=1000$, set $\sqrt{c_i}=\sqrt{c_1}+2.1654\times i$.
%For $d=10$, we set $\sqrt{c_1}=4$ and for $i=1,\ldots,M$, with $M=1000$, set $\sqrt{c_i}=\sqrt{c_1}+2.5\times i$.
%For $d=50$, we set $\sqrt{c_1}=4$ and for $i=1,\ldots,M$, with $M=1000$, set $\sqrt{c_i}=\sqrt{c_1}+0.52\times i$.
%For $d=100$, we set $\sqrt{c_1}=4$ and for $i=1,\ldots,M$, with $M=1000$, set $\sqrt{c_i}=\sqrt{c_1}+0.52\times i$.
As before, we used Monte Carlo size $N_i=10,000$ for the construction of $\hat\pi(\btheta)$, and then simulated $10,000$ $iid$ realizations 
from $\hat\pi(\btheta)$. 
The times taken for $d=1,5,10,50,100$ are $4$ minutes, $29$ minutes, $1$ hour $3$ minutes, less than a minute, and less than a minute, respectively.
The reason for the significantly less time here compared to the normal setup is again attributable to the thick tails of the $t$ distribution; the flatter tails
ensure that the infimum $s_i$ and supremum $S_i$ are not much different, entailing that $\hat p_i$ is reasonably large (see (\ref{eq:minor1})), so that simulation from
the geometric distribution (\ref{eq:geo}) yielded significantly smaller values of the coalescence time $T_i$ compared to the normal setup. Thus, in spite of the larger
value of $M$, these small values of $T_i$ led to much quicker overall computation.

Again, the $iid$ simulations turned out to be highly reliable in all the cases.
Figure \ref{fig:t_simstudy100} shows the results for four co-ordinates in the context of $d=100$, and
the exact and estimated correlation structures for the first $20$ co-ordinates are displayed in Figure \ref{fig:t_corr_simstudy100}.

\begin{figure}
	\centering
	\subfigure [True and $iid$-based density for $\theta_1$.]{ \label{fig:t1}
	\includegraphics[width=7.5cm,height=7.5cm]{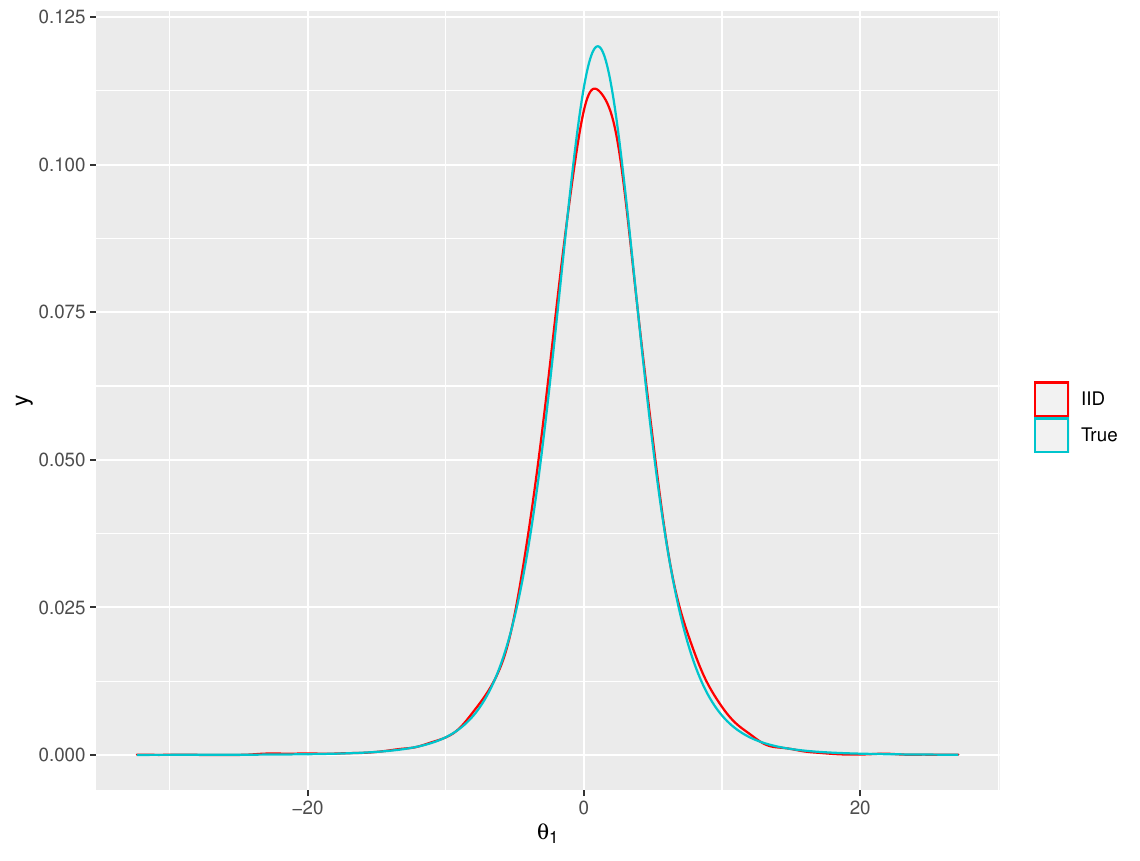}}
	\hspace{2mm}
	\subfigure [True and $iid$-based density for $\theta_{25}$.]{ \label{fig:t25}
	\includegraphics[width=7.5cm,height=7.5cm]{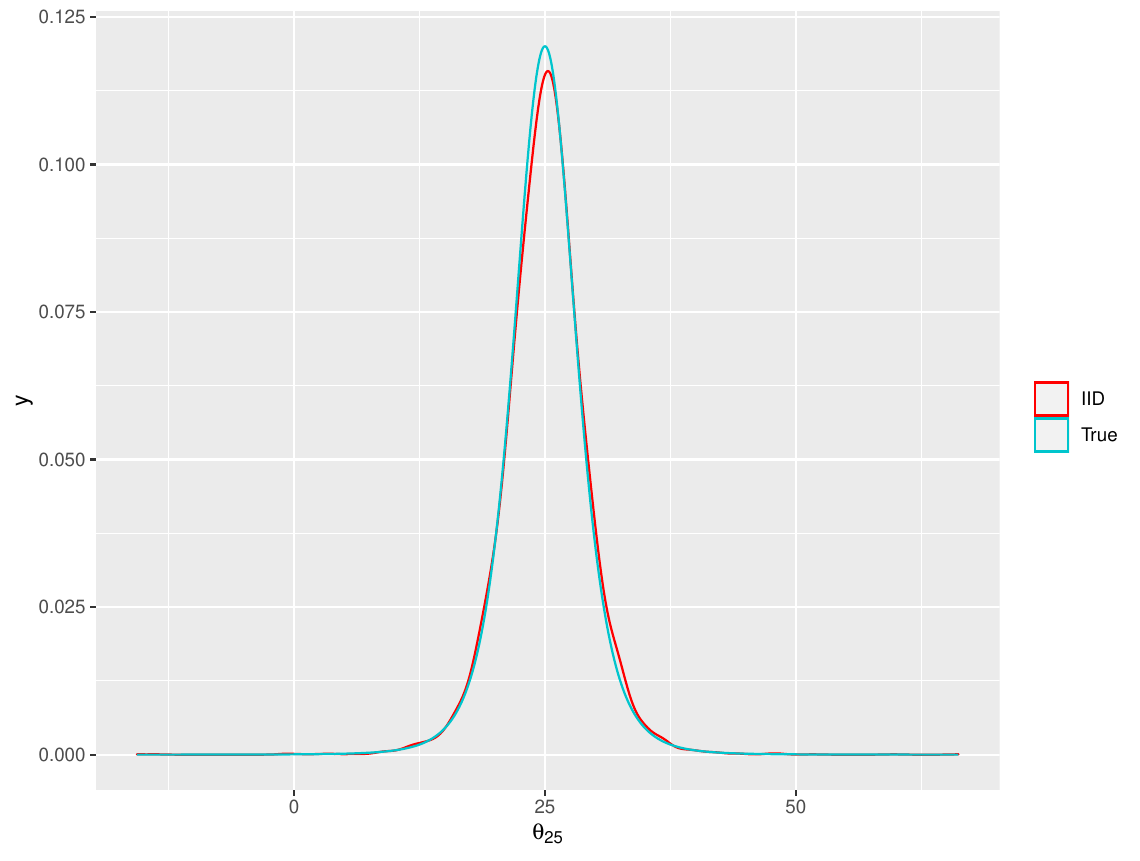}}\\
	\vspace{2mm}
	\subfigure [True and $iid$-based density for $\theta_{75}$.]{ \label{fig:t75}
	\includegraphics[width=7.5cm,height=7.5cm]{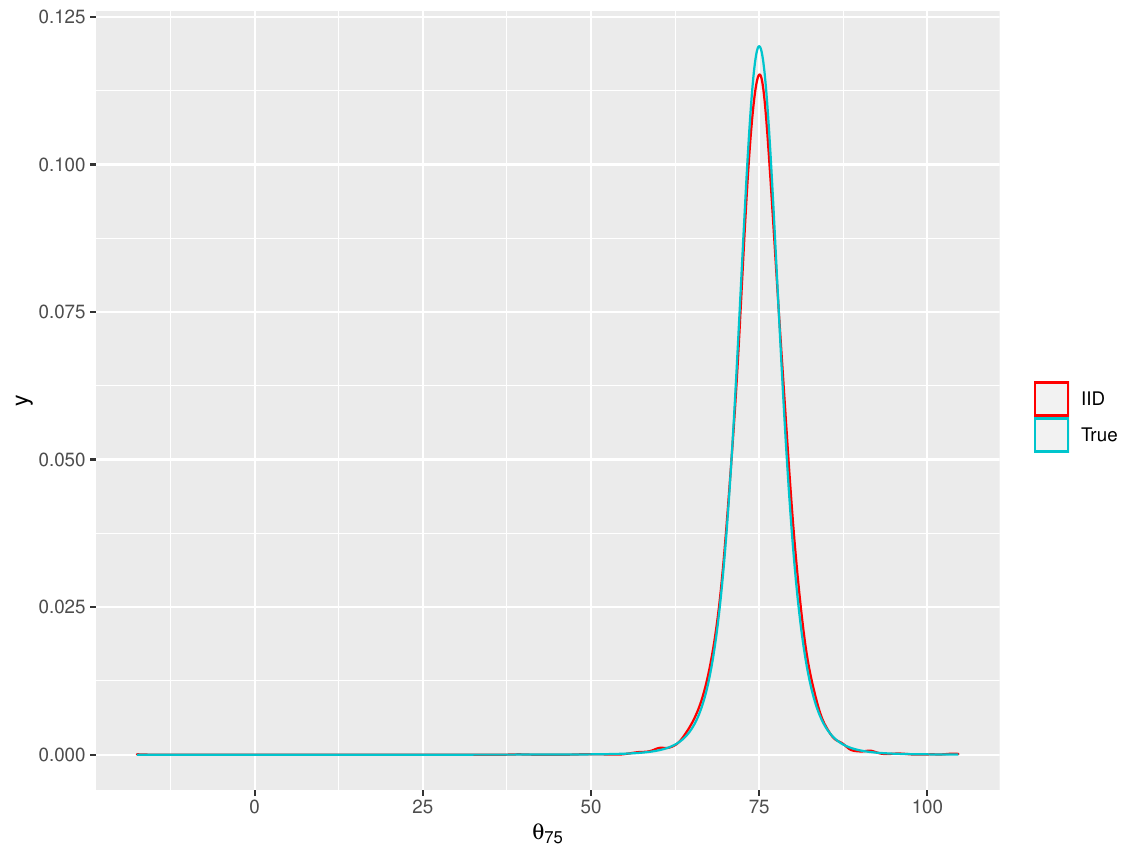}}
	\vspace{2mm}
	\subfigure [True and $iid$-based density for $\theta_{100}$.]{ \label{fig:t100}
	\includegraphics[width=7.5cm,height=7.5cm]{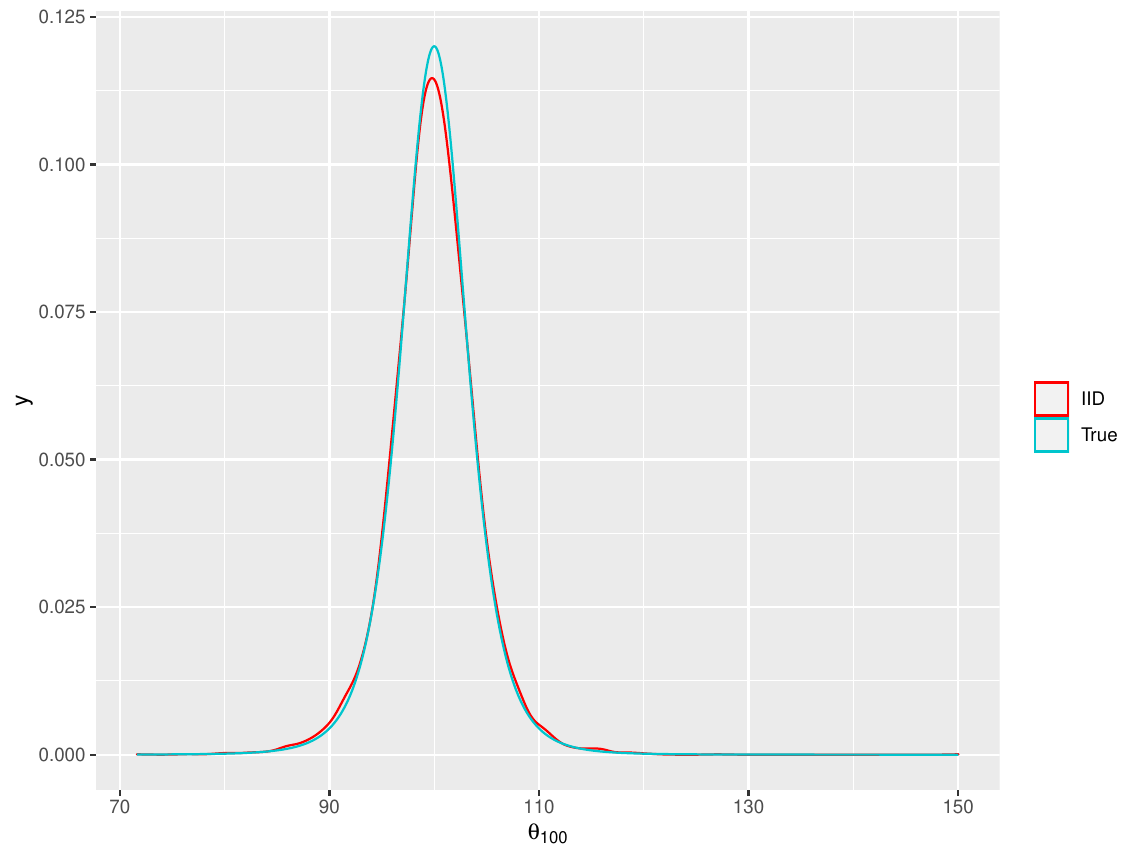}}
	\caption{Simulation from $100$-dimensional Student's $t$ distribution with $5$ degrees of freedom. 
	The red and blue colours denote the $iid$ sample based density and the true density, respectively.}
	\label{fig:t_simstudy100}
\end{figure}

\begin{figure}
	\centering
	\subfigure [Exact correlation structure.]{ \label{fig:t_corr_exact}
	\includegraphics[width=7.5cm,height=7.5cm]{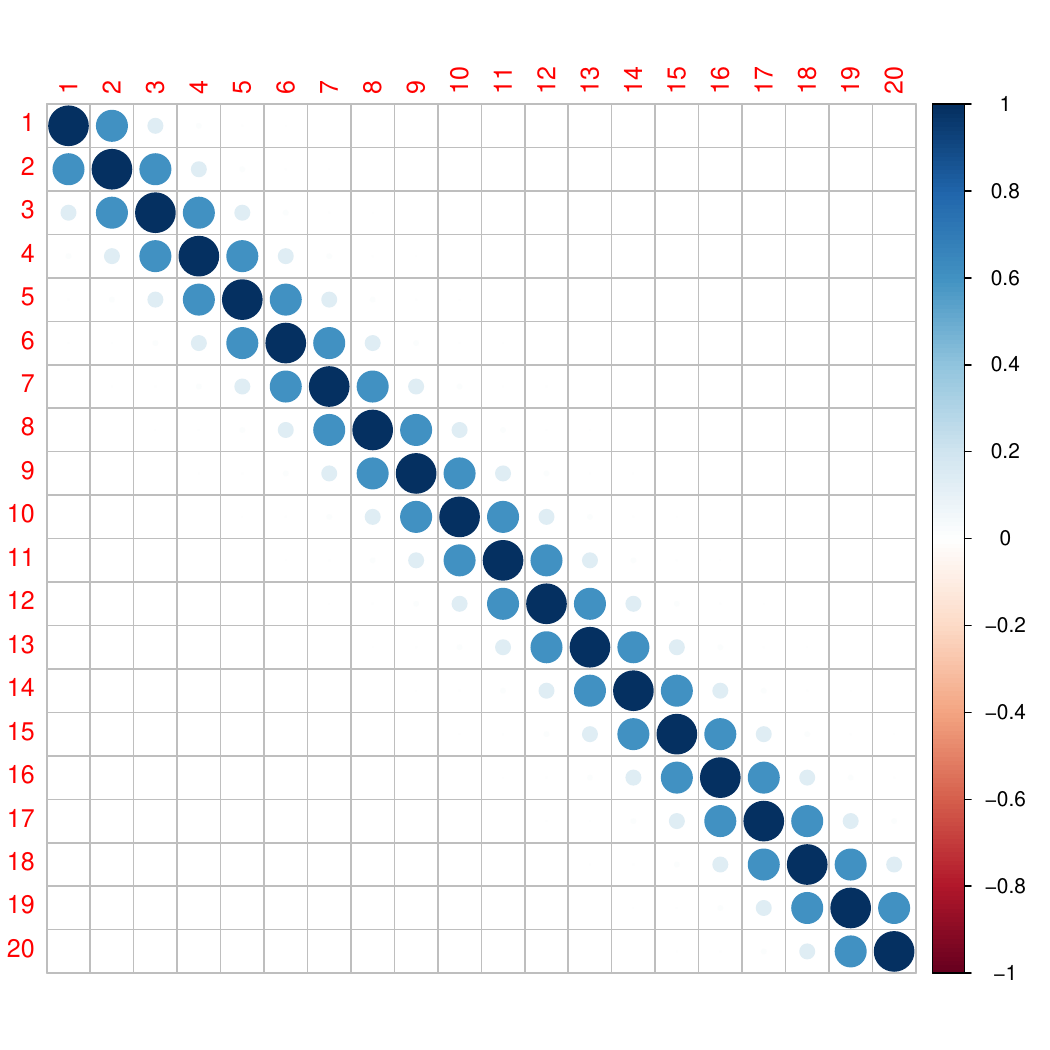}}
	\hspace{2mm}
	\subfigure [Correlation structure based on $iid$ samples.]{ \label{fig:t_corr}
	\includegraphics[width=7.5cm,height=7.5cm]{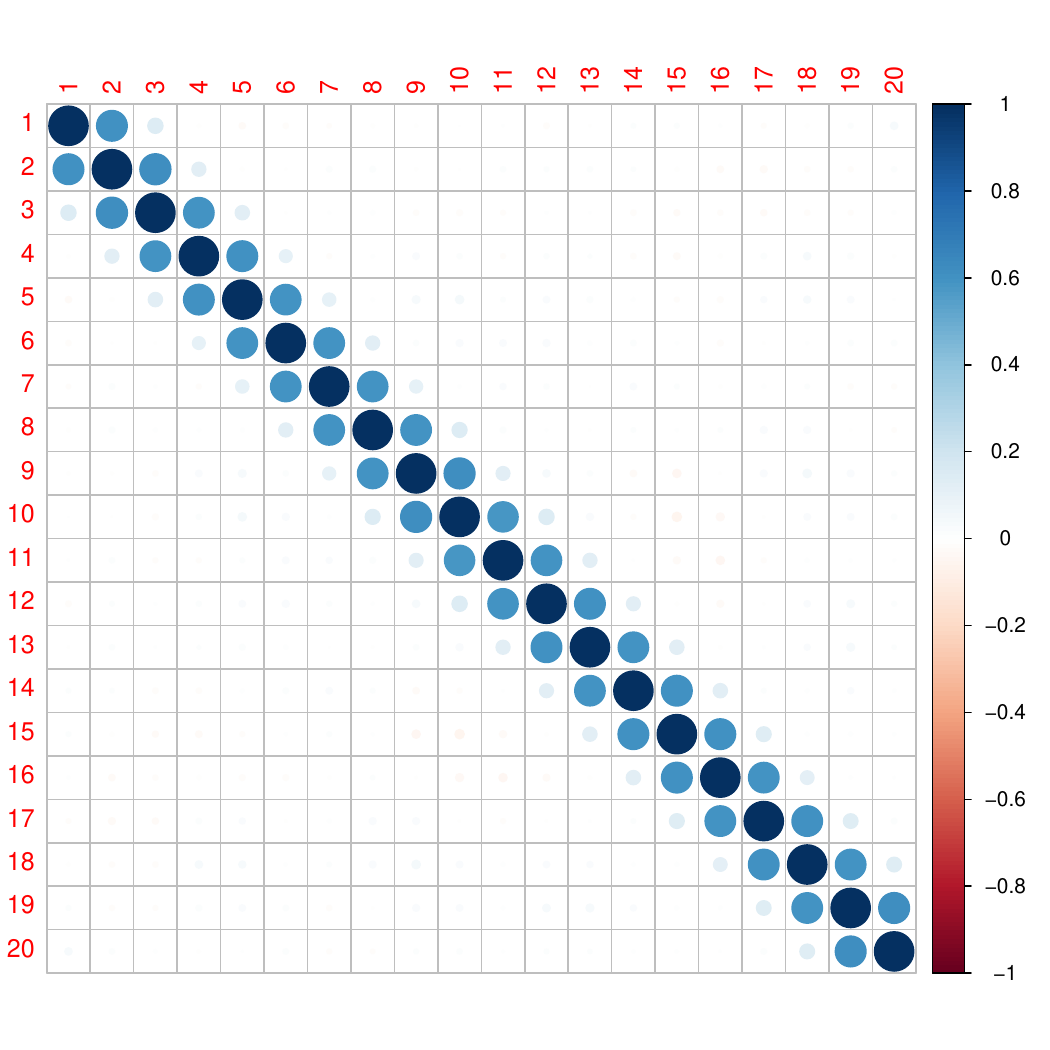}}
	\caption{Simulation from $100$-dimensional Student's $t$ distribution with $5$ degrees of freedom. 
	True and $iid$-based correlation structures for the first $20$ co-ordinates.}
	\label{fig:t_corr_simstudy100}
\end{figure}

\subsection{Simulation from Cauchy}
\label{subsec:cauchy}

In this case, we set
$\sqrt{c_1}=r_d$ and $\sqrt{c_i}=\sqrt{c_1}+a_d\times (i-1)$, for $i=2,\ldots,M_d$. Thus, even the value of $M$ now depends upon $d$. 
Following our experimentations we obtained the following: $(r_1=5,a_1=3.801,M_1=2000)$, $(r_5=0.5,a_5=0.5,M_5=3000)$, $(r_{10}=0.5,a_{10}=0.5,M_{10}=3000)$,  
$(r_{50}=4,a_{50}=0.52,M_{50}=2000)$, $(r_{100}=4,a_{100}=0.52,M_{100}=2576)$.
Note that the values of $M_d$ are at least twice that for the Student's $t$ distribution, which is clearly due to the much thicker tail of Cauchy compared to $t$.

As before, using Monte Carlo size $N_i=10,000$ for the construction of $\hat\pi(\btheta)$, we then simulated $10,000$ $iid$ realizations 
from $\hat\pi(\btheta)$. In this case, the times taken for $d=1,5,10,50,100$ are $2$ minutes, $2$ minutes, $2$ minutes, less than a minute, and $2$ minutes, respectively.
Notice the significantly less time taken in this setup compared to both normal and Cauchy. Given the thickest tail of Cauchy, 
the reason for this is the same as provided in Section \ref{subsec:t} that explains lesser time for $t$ compared to normal.  

As expected, the $iid$ simulations accurately represented the underlying Cauchy distributions. Depictions of the true and $iid$ sample based marginal densities
of four co-ordinates are presented by Figure \ref{fig:cauchy_simstudy100} in the context of $d=100$. 
Since the covariance structure does not exist for Cauchy, we do not present the correlation structure comparison unlike the previous cases. 

\begin{figure}
	\centering
	\subfigure [True and $iid$-based density for $\theta_1$.]{ \label{fig:c1}
	\includegraphics[width=7.5cm,height=7.5cm]{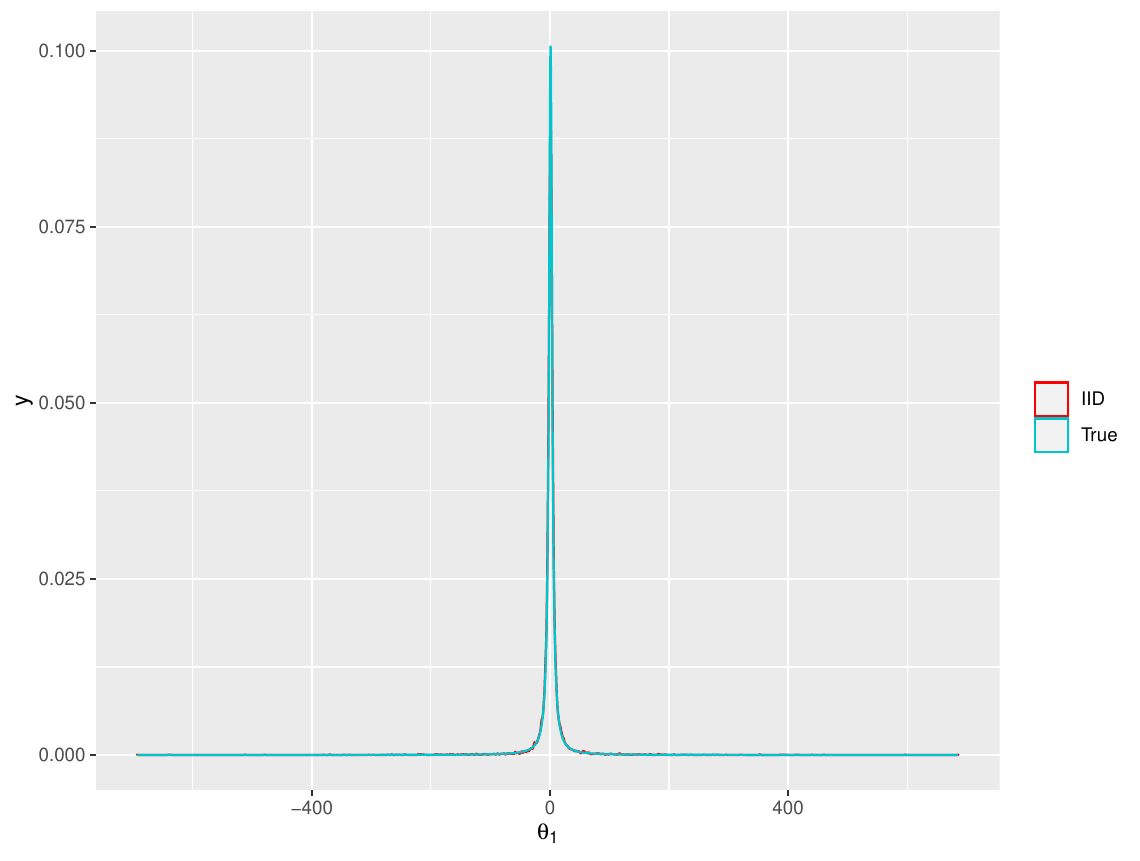}}
	\hspace{2mm}
	\subfigure [True and $iid$-based density for $\theta_{25}$.]{ \label{fig:c25}
	\includegraphics[width=7.5cm,height=7.5cm]{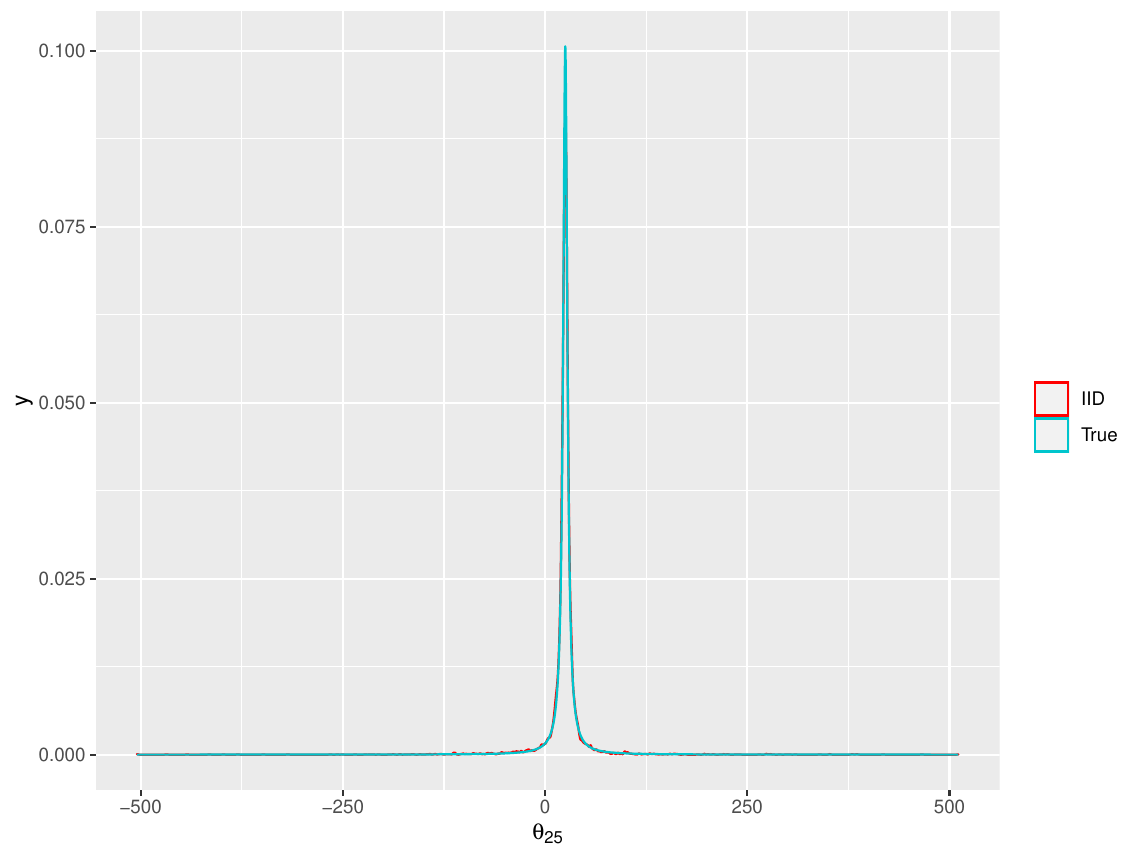}}\\
	\vspace{2mm}
	\subfigure [True and $iid$-based density for $\theta_{75}$.]{ \label{fig:c75}
	\includegraphics[width=7.5cm,height=7.50cm]{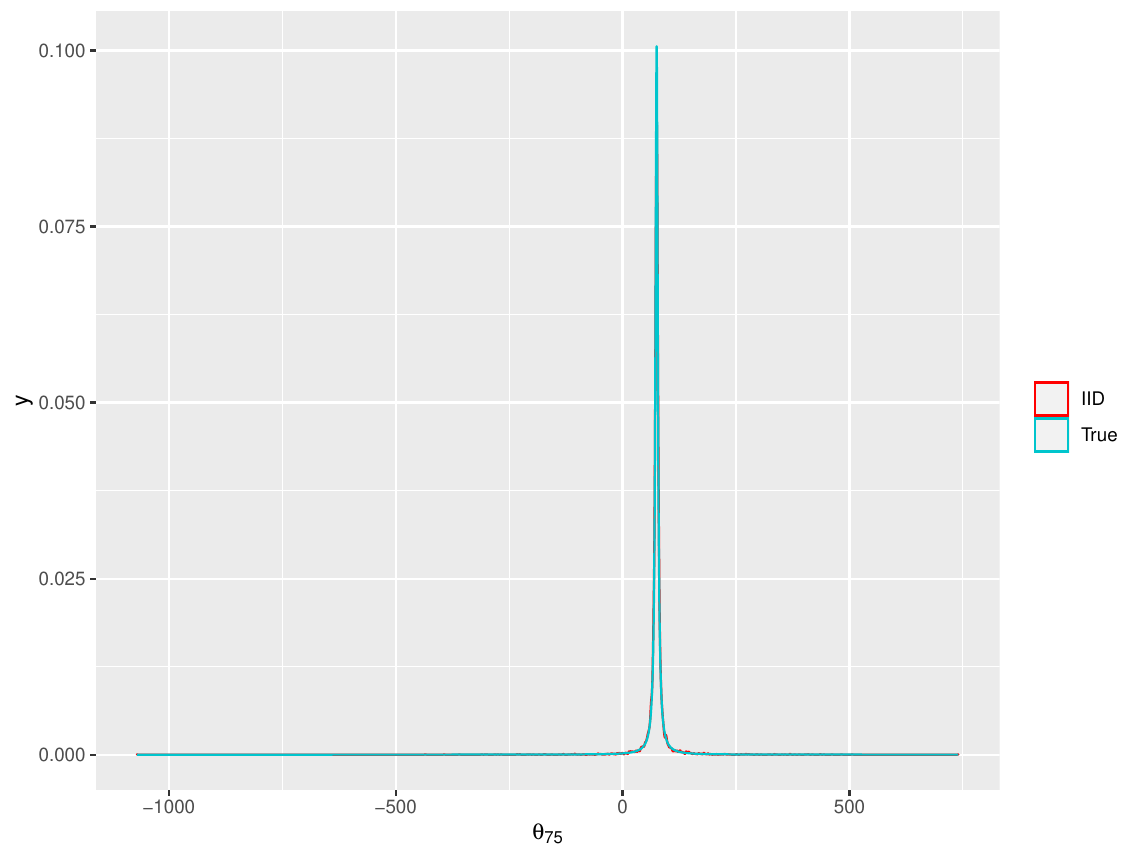}}
	\vspace{2mm}
	\subfigure [True and $iid$-based density for $\theta_{100}$.]{ \label{fig:c100}
	\includegraphics[width=7.5cm,height=7.5cm]{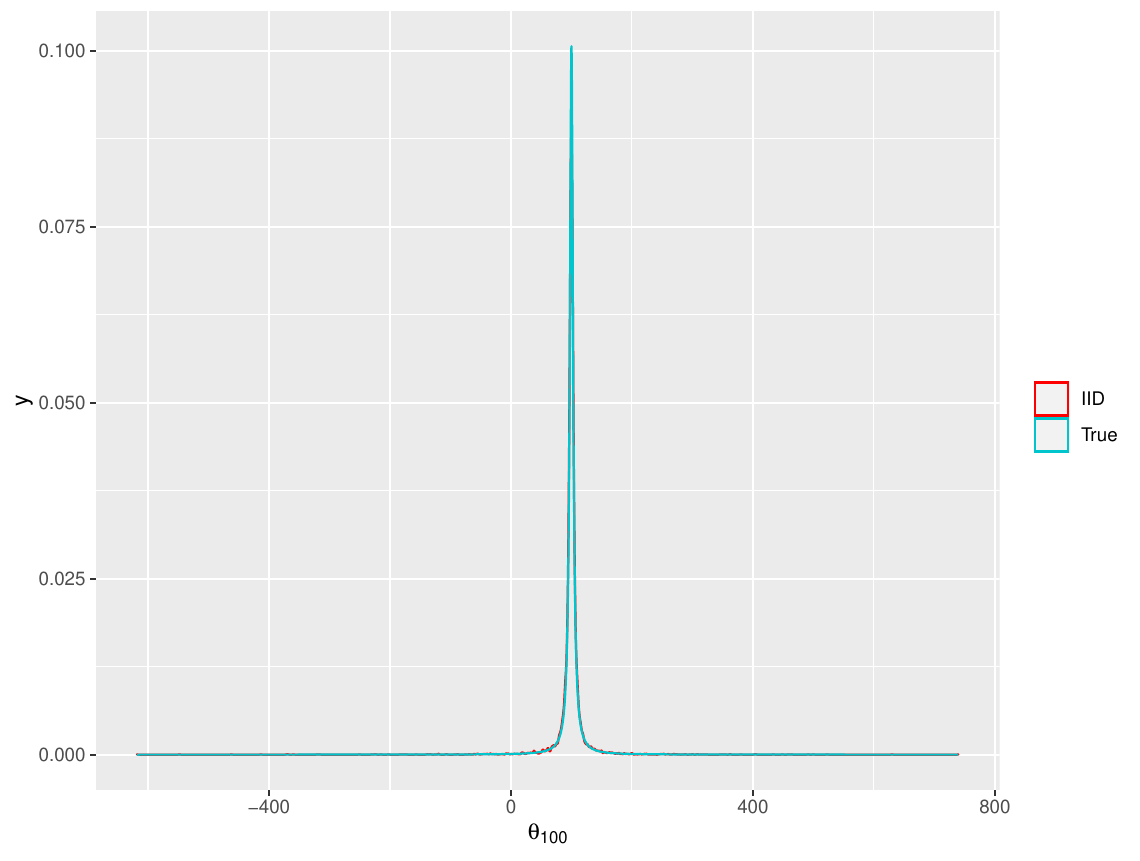}}
	\caption{Simulation from $100$-dimensional Cauchy distribution. 
	The red and blue colours denote the $iid$ sample based density and the true density, respectively.}
	\label{fig:cauchy_simstudy100}
\end{figure}

\subsection{Simulation from normal mixture}
\label{subsec:normix}
In this case, for both $\bA_{1i}$ and $\bA_{2i}$, we chose $r_{50}=4$ and $a_{50}=0.5$. As in the normal setup detailed in Section \ref{subsec:normal},
$M=71$ turned out to be sufficient. We set $N_i=10,000$ as the Monte Carlo size, as in the previous cases. 
For drawing each $iid$ realization, assuming known mixing probabilities $2/3$ and $1/3$, we first selected either of the two mixing densities, indexed by $j=1,2$. 
Then, with the corresponding sequence of sets $\bA_{ji}$, we proceeded with perfect sampling from the mixing density indexed by $j$.
The time taken for the entire implementation was $7$ minutes for generating $10,000$ $iid$ realizations.

As before, the $iid$ realizations very accurately represented the true mixture distribution, Figure \ref{fig:normix_simstudy50} bearing some testimony.
Figure \ref{fig:normix_corr_simstudy50} compares the true correlation structure and that estimated by our $iid$ method. As expected, the estimated structure
yielded by our $iid$ method has been highly accurate. Here it is worth remarking that the true correlation structure is also an estimated structure, based
on $10,000$ direct $iid$ realizations from the normal mixture.

Details of the method for generating perfect $iid$ realizations from any general multimodal target distribution $\pi$, is provided in \ctn{Bhattacharya21a}. 

%using TMCMC it is possible to identify the modes and the modal regions. Assuming $m$ modes with corresponding
%(presumably non-compact) modal regions $\bB_1,\ldots,\bB_m$, we can first represent $\pi$ as the following mixture:
%\begin{equation*}
%	\pi(\btheta)=\sum_{j=1}^m\pi(\bB_j)\pi_j(\btheta),
%\end{equation*}
%where
%\begin{equation*}
%	\pi_j(\btheta)=\frac{\pi(\btheta)}{\pi(\bB_j)}I_{\bB_j}(\btheta).
%\end{equation*}
%To simulate from $\pi$, we shall first select $\pi_j$ with probability proportional to $\widehat{\pi(\bB_j)}$, the estimate
%of $\pi(\bB_j)$ obtained from either TMCMC or Monte Carlo (up to a normalizing constant), and then apply Algorithm \ref{algo:perfect} to $\pi_j$,
%after decomposing it into a mixture of distributions on compact sets $\bB_j\cap\bA_{ji}$ (ignoring possible null sets with respect to the Lebesgue measure 
%in the intersection), in the same way as (\ref{eq:p1}).

\begin{figure}
	\centering
	\subfigure [True and $iid$-based density for $\theta_1$.]{ \label{fig:n1}
	\includegraphics[width=7.5cm,height=7.5cm]{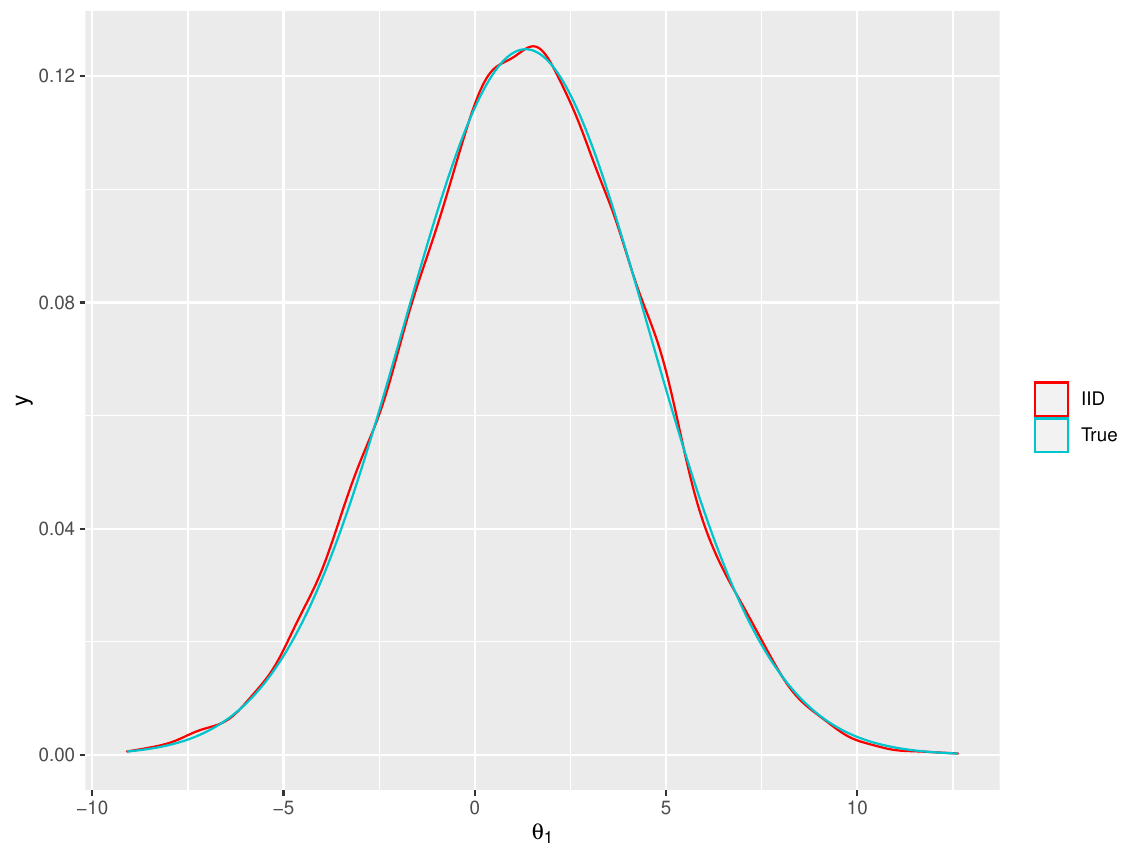}}
	\hspace{2mm}
	\subfigure [True and $iid$-based density for $\theta_{10}$.]{ \label{fig:n10}
	\includegraphics[width=7.5cm,height=7.5cm]{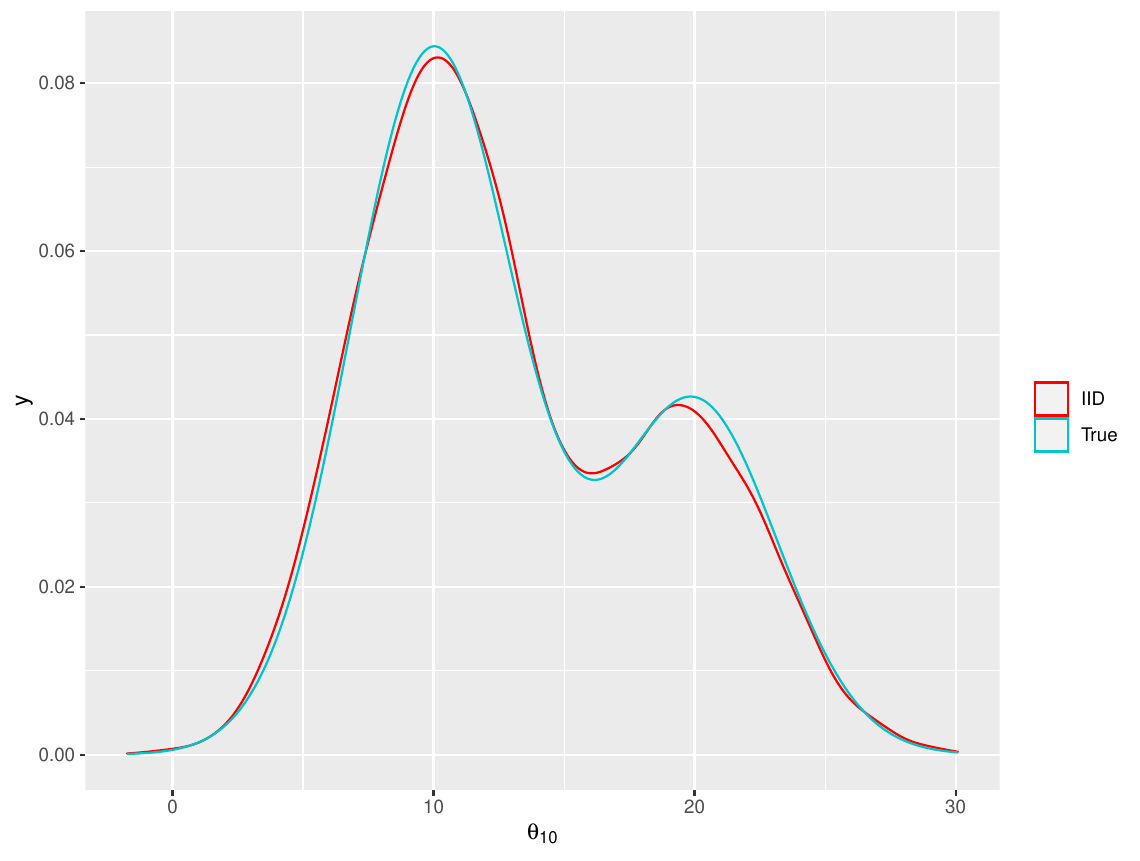}}\\
	\vspace{2mm}
	\subfigure [True and $iid$-based density for $\theta_{25}$.]{ \label{fig:n25}
	\includegraphics[width=7.5cm,height=7.5cm]{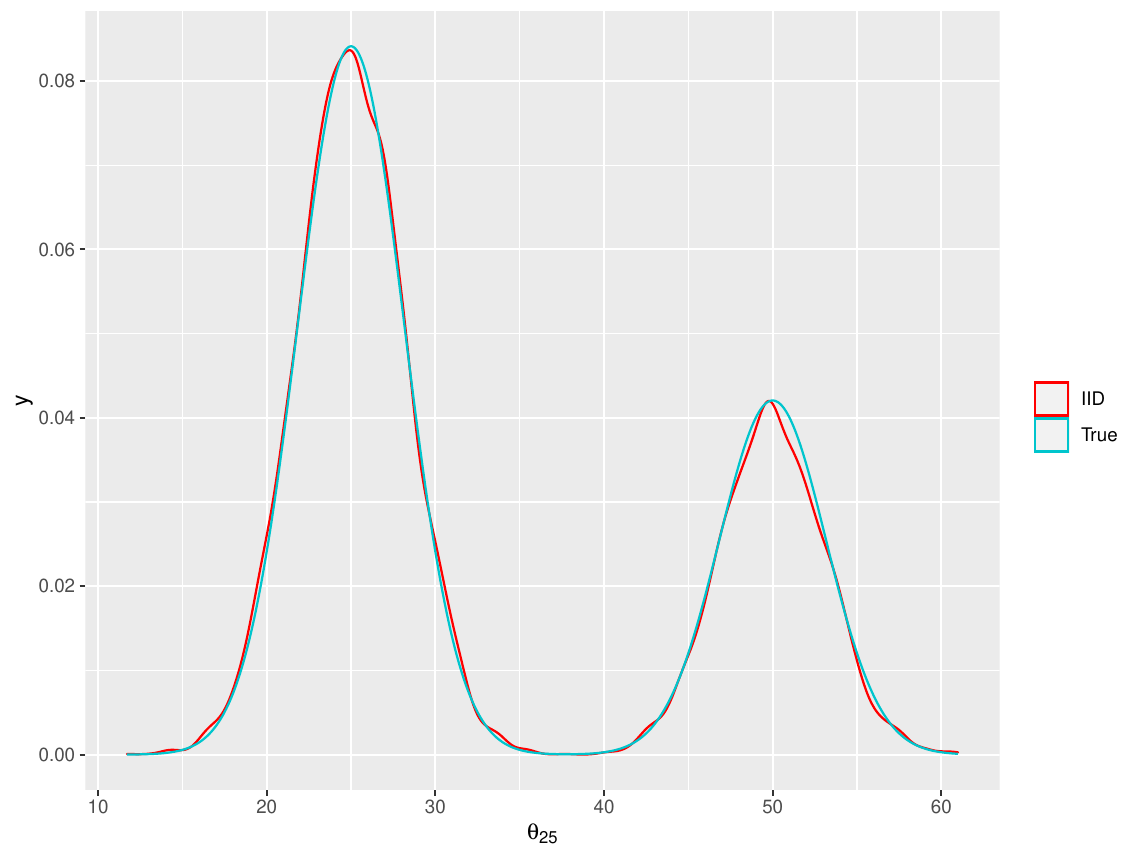}}
	\vspace{2mm}
	\subfigure [True and $iid$-based density for $\theta_{50}$.]{ \label{fig:n50}
	\includegraphics[width=7.5cm,height=7.5cm]{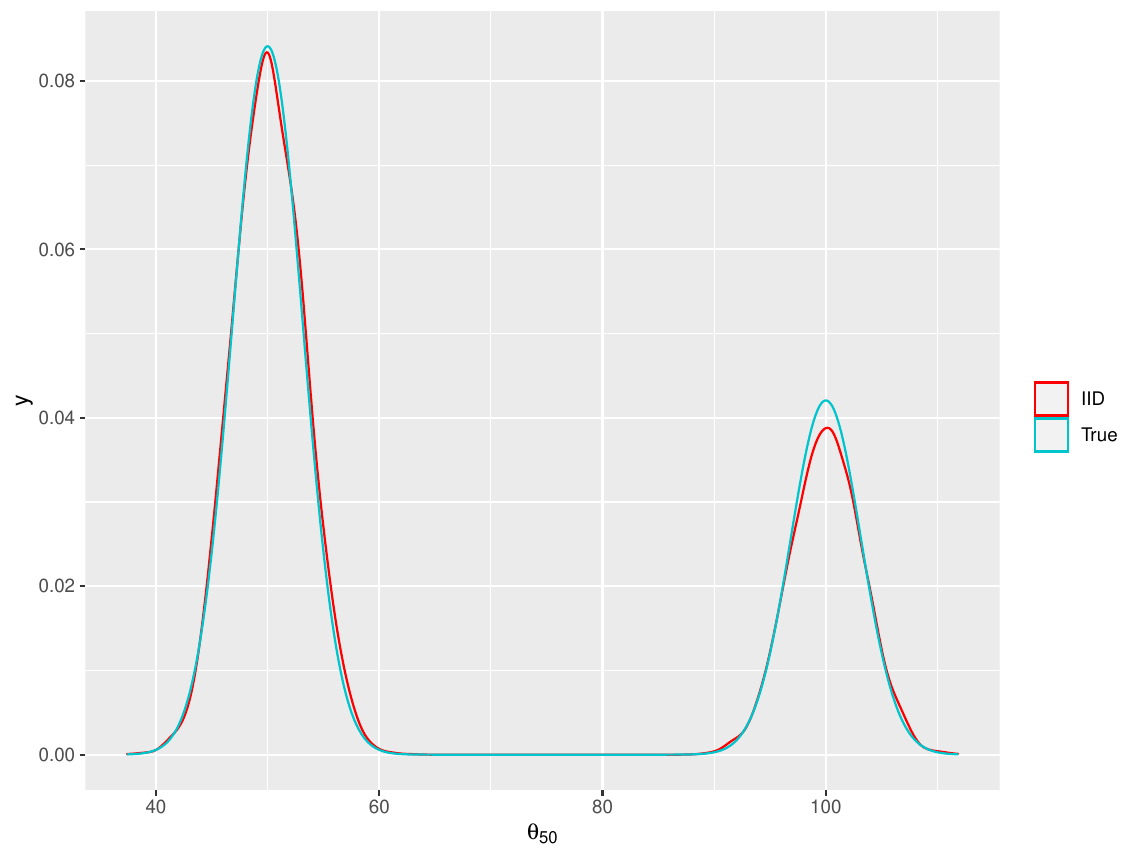}}
	\caption{Simulation from $50$-dimensional mixture normal distribution. 
	The red and blue colours denote the $iid$ sample based density and the true density, respectively.}
	\label{fig:normix_simstudy50}
\end{figure}

\begin{figure}
	\centering
	\subfigure [Exact correlation structure.]{ \label{fig:normix_corr_exact}
	\includegraphics[width=7.5cm,height=7.5cm]{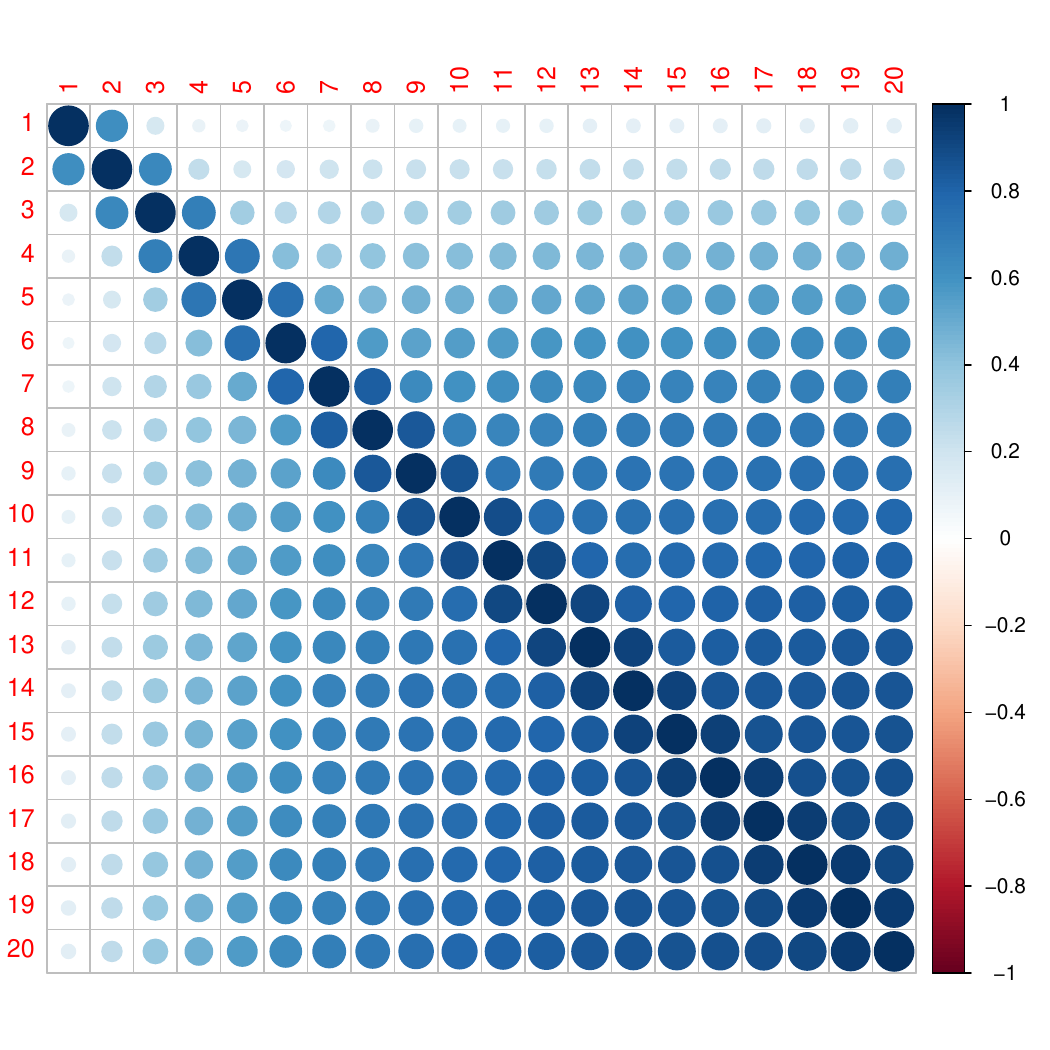}}
	\hspace{2mm}
	\subfigure [Correlation structure based on $iid$ samples.]{ \label{fig:normix_corr}
	\includegraphics[width=7.5cm,height=7.5cm]{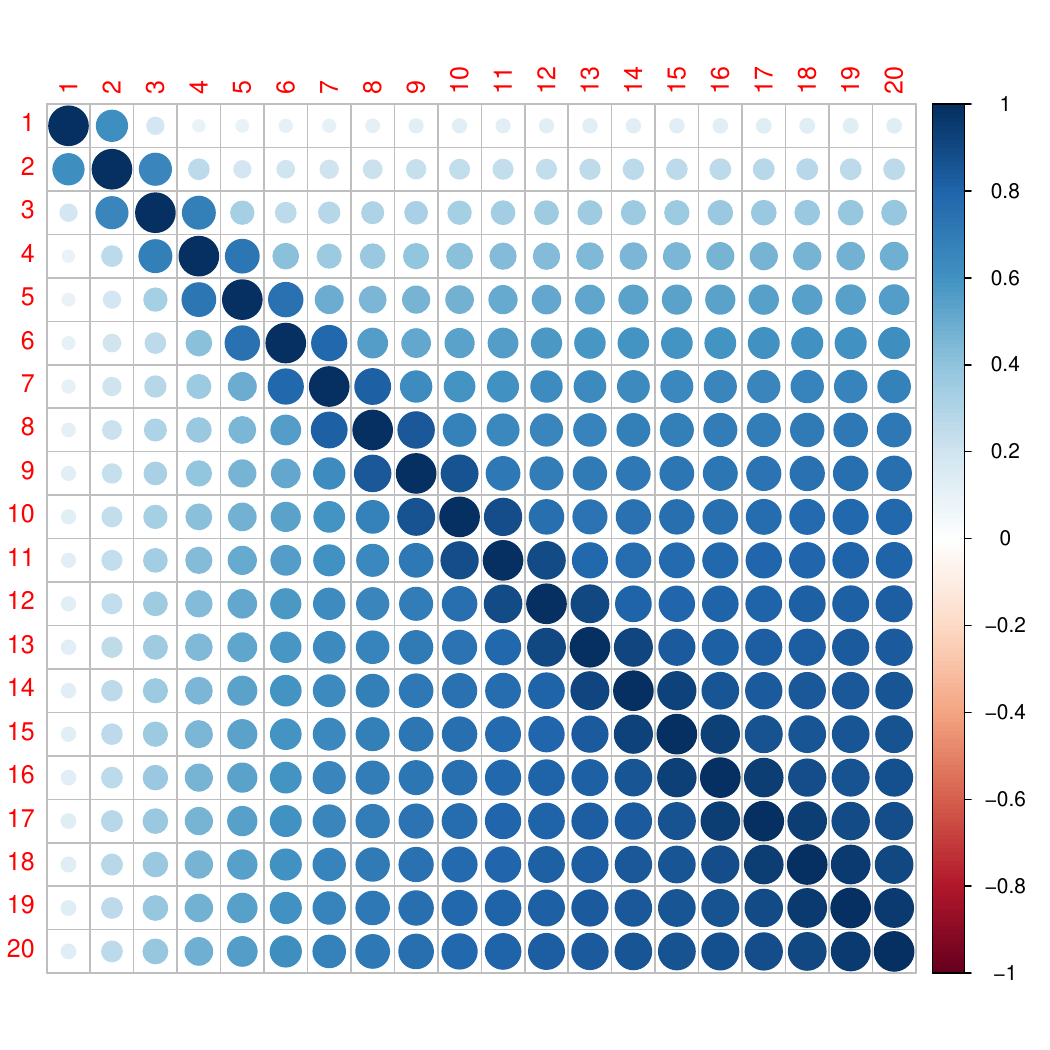}}
	\caption{Simulation from $50$-dimensional mixture normal distribution. 
	True and $iid$-based correlation structures for the first $20$ co-ordinates.}
	\label{fig:normix_corr_simstudy50}
\end{figure}

\section{Experiments with posterior distributions given real data}
\label{sec:realdata}

We now consider the application of our $iid$ method to posterior distributions given real datasets. 
Our first application in this regard will be the well-known Challenger space shuttle problem 
(see, for example, \ctn{Dalal89}, \ctn{Martz92}, \ctn{Robert04} and \ctn{Dutta14}).
The data can be found in \ctn{Robert04} and in the supplement of \ctn{Dutta14}.
Here we consider drawing $iid$ realizations from the two-parameter posterior distribution corresponding to a logit likelihood
and compare the $iid$ based and TMCMC based posteriors. 

For our second application, we shall consider the Salmonella example from Chapter 6.5.2 of \ctn{Lunn12}. The data, three-parameter Poisson log-linear model, the priors 
and the BUGS/JAGS codes are also available at \url{http://runjags.sourceforge.net/quickjags.html}. 
In this example as well, we shall compare the $iid$ based and TMCMC based posteriors.

Our third application is on the really challenging, spatial data and model setup related to radionuclide count data on Rongelap Island
(\ctn{Diggle97}, \ctn{Diggle98}). The model for the radionuclide counts 
is a Poisson log-linear model encapsulating a latent Gaussian process to account for spatial dependence. The model consists of $160$ parameters.
The difficulty of MCMC convergence for the posterior analysis in this problem is very well-known and several research works have been devoted to this 
(see, for example, \ctn{Rue01}, \ctn{Chris04}, \ctn{Chris06}).
Very encouragingly, even in this challenging problem, we are able to very successfully obtain $iid$ samples from the posterior, after using a specific
diffeomorphic transformation to the posterior to render it sufficiently thick-tailed to yield reasonably large values of $\hat p_i$. As in the previous examples, here
also we shall compare the $iid$ and TMCMC based posteriors.

\subsection{Application to the Challenger dataset}
\label{subsec:challenger}
In 1986, the space shuttle Challenger exploded during take off, killing the seven astronauts aboard. 
The explosion was the result of an O-ring failure, a splitting of a ring of rubber that seals the parts of 
the ship together. The accident was believed to be caused by the unusually cold weather ($31\degree$F or $0\degree$C) 
at the time of launch, as there is reason to believe that the O-ring failure probabilities increase as 
temperature decreases. 

%Let \[ \eta_i = \b_1 + \b_2x_i \] where $x_i = t_i/\max ~t_i$, $t_i$'s being the temperature at flight 
%time (degrees F), $i=1,\ldots,n$,where $n=23$. 
In this regard, for $i=1,\ldots,n$, where $n=23$, let $y_i$ denote the indicator variable denoting failure 
of the O-ring. Also let $x_i = t_i/\max ~t_i$, where $t_i$ is the temperature (degrees F) at flight time $i$.
We assume that for $i=1,\ldots,n$, $y_i\sim Bernoulli(p(x_i))$ independently, where $p(x_i)=\exp(\alpha+\beta x_i)\big/(1+\exp(\alpha+\beta x_i))$.
As regards the prior for $(\alpha,\beta)$, we set $\pi(\alpha,\beta)\equiv 1$ for $\alpha,\beta\in\mathbb R$. 

For TMCMC, we consider the additive transformation detailed in Section 4 of \ctn{Dutta14}, with scaling constants in the additive transformation being
$7.944$ and $9.762$ respectively for $\alpha$ and $\beta$. %the other details remaining exactly the same. 
After accepting the proposed additive TMCMC move
or the previous state in accordance with the TMCMC acceptance probability,
we conduct another mixing-enhancement step in a similar vein as in \ctn{Liu00} which, given the current accepted state, proposes another move
that gives either forward transformation to both the parameters or backward transformation to both the parameters, with equal probability.
Either this final move, or the last accepted additive TMCMC move is accepted in accordance with the TMCMC acceptance probability. 
With this strategy, we run our chain for $2\times 10^5$ iterations, discarding the first $10^5$ as burn-in. 
We estimate $\bmu$ and $\bSigma$ required for our
sets $\bA_i$ using the stored $10^5$ iterations after burn-in.

In order to select the radii, we set as before, with $d=2$,
$\sqrt{c_1}=r_d$ and $\sqrt{c_i}=\sqrt{c_1}+a_d\times (i-1)$, for $i=2,\ldots,M$.
Our experiments indicated that $M=85$, $r_d=2$ and $a_d=0.02$ are appropriate choices. 
Now, since 
$\sqrt{c_2}-\sqrt{c_1}=2\times 0.02$ and $\sqrt{c_i}-\sqrt{c_{i-1}}=0.02$ for $i\geq 3$, rejection sampling in order to simulate uniformly
from $\bA_i$ for $i\geq 2$ is costly computationally, since the probability of any point $\tilde\btheta$ drawn uniformly from 
the ellipsoid $\{\btheta:(\btheta-\bmu)^T\bSigma^{-1}(\btheta-\bmu)\leq c_i\}$ to satisfy $(\tilde\btheta-\bmu)^T\bSigma^{-1}(\btheta-\tilde\bmu)> c_{i-1}$,
is small. To handle this, we adopt the following strategy. First observe that in this situation, $\bA_i$ here is well approximated by the region close to the surface of the 
above ellipsoid. For sufficiently large value of $\tilde d$, $\tilde\bY=\sqrt{c_i}U^{1/\tilde d}\bX/\|\bX\|$ 
with $\bX=(X_1,\ldots,X_d)^T$ where $X_k\stackrel{iid}{\sim}N(0,1)$, for $k=1,\ldots,d$, $U\sim U(0,1)$,
represents the uniform distribution essentially about the surface
of the ball centered around $\bzero$ and with radius $\sqrt{c_i}$, so that $\tilde\btheta=\bmu+\bB\tilde\bY$ essentially represents the desired uniform distribution
on $\bA_i$. Thus, this method completely avoids rejection sampling. Setting $\tilde d=10^5$ turned out to yield quite accurate results
with this strategy. Indeed, comparison with actual rejection sampling showed that the final results are almost indistinguishable.

Instead of setting $N_i=10,000$ as the Monte Carlo size as in our experiments with the standard distributions, here we set $N_i=5000$. One obvious reason for this is  
to reduce the computing time, since there are $M=10^5$ set $\bA_i$ where the Monte Carlo method needs to be applied to. Another reason is that, due to the narrow
regions covered individually by the current sets $\bA_i$, not much Monte Carlo realizations are necessary. Again, we compared the results with actual rejection
sampling by setting $N_i=10,000$, and again the final results turned out to be almost the same.

There is another subtle issue that deserves mention. The above strategy for avoiding rejection sampling does not guarantee that none of the points sampled from
$\bA_i$ will fall in $\bA_{i-1}$, for $i\geq 2$. Indeed, a few points must percolate into $\bA_{i-1}$, but are still counted for computation of 
the infimum and supremum of $\tilde\pi(\btheta)$ in $\bA_i$. These points thus render the effect of lowering the value of the ratio of the 
actual infimum and supremum. To counter this effect, we set $\eta_i=0$ in the formula for $\hat p_i$. 
Note however, that the Monte Carlo estimate of $\tilde\pi(\bA_i)$ is unaffected by this percolation since the effect of the few points, 
when divided by the Monte Carlo size, is washed away.

As before, we sampled $10,000$ $iid$ realizations from this Challenger posterior.
The implementation time of our $iid$ method has been $22$ minutes for this problem.

Figure \ref{fig:challenger} compares the $iid$ based and TMCMC based marginal posteriors of $\alpha$ and $\beta$. Here we thinned the 
$10^5$ TMCMC realizations by using one in every $10$ for comparing with our $10,000$ $iid$ realizations.
Observe that the $iid$ based densities are in very close agreement with those based on TMCMC. 

The TMCMC and $iid$ based posterior correlations between $\alpha$ and $\beta$ are $-0.99771$ $-0.99761$, respectively, exhibiting close agreement.

%Figure \ref{fig:challenger_corr} shows that the correlation structures obtained by TMCMC and $iid$ sampling are not very different.
\begin{figure}
	\centering
	\subfigure [TMCMC and $iid$-based density for $\alpha$.]{ \label{fig:ch1}
	\includegraphics[width=7.5cm,height=7.5cm]{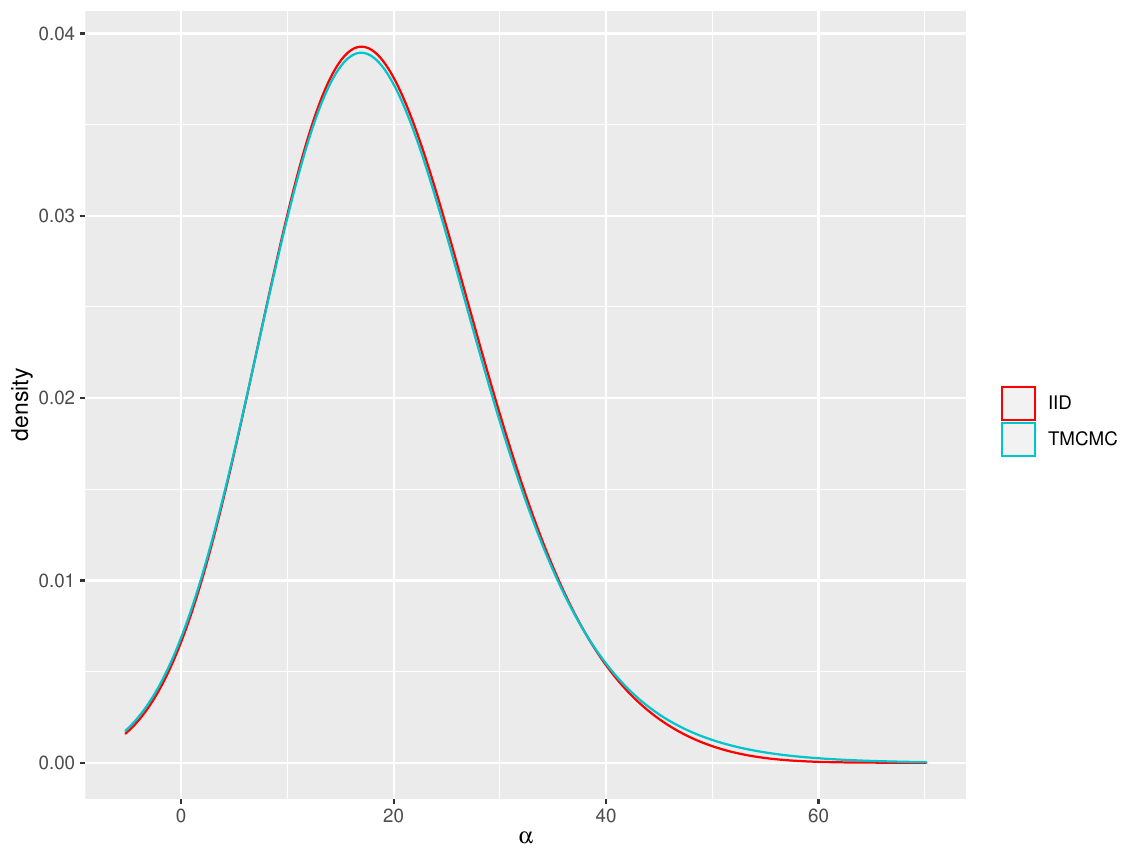}}
	\hspace{2mm}
	\subfigure [TMCMC and $iid$-based density for $\beta$.]{ \label{fig:ch2}
	\includegraphics[width=7.5cm,height=7.5cm]{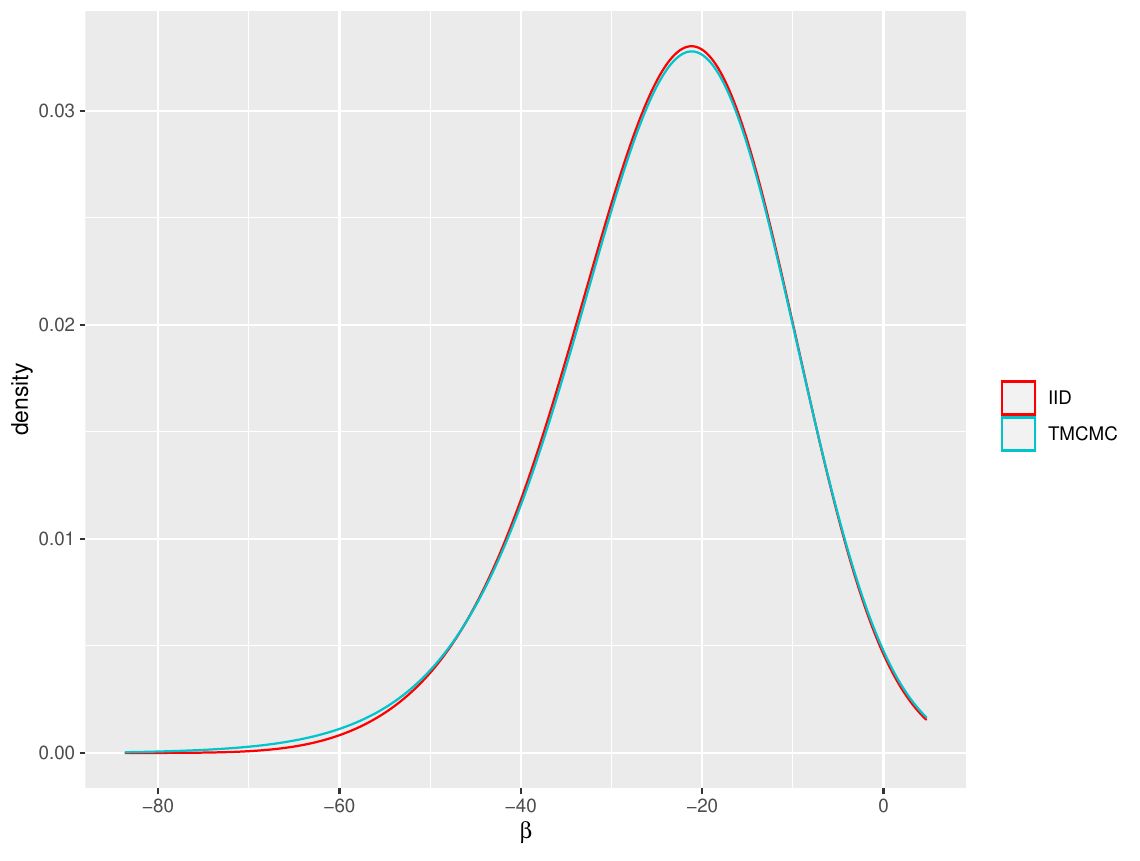}}
	\caption{Challenger posterior. 
	The red and blue colours denote the $iid$ sample based density and the TMCMC based density, respectively.}
	\label{fig:challenger}
\end{figure}

%\begin{figure}
%	\centering
%	\subfigure [TMCMC based correlation structure.]{ \label{fig:ch_corr_exact}
%	\includegraphics[width=7.5cm,height=7.5cm]{figures_challenger/corr_tmcmc.pdf}}
%	\hspace{2mm}
%	\subfigure [Correlation structure based on $iid$ samples.]{ \label{fig:ch_corr}
%	\includegraphics[width=7.5cm,height=7.5cm]{figures_challenger/corr_iid.pdf}}
%	\caption{Challenger posterior. 
%	TMCMC and $iid$-based correlation structures.}
%	\label{fig:challenger_corr}
%\end{figure}

\subsection{Application to the Salmonella dataset}
\label{subsec:salmonella}

Here the data concerned is a mutagenicity assay data on Salmonella. In the relevant experiment, 
three plates have each been processed at various doses of quinoline and subsequently
the number of revertant colonies of TA98 Salmonella are measured; see \ctn{Breslow84}, \ctn{Lunn12}.

For $i=1,\ldots,6$ and $j=1,2,3$, letting $y_{ij}$ denote the number of colonies observed on plate $j$ at dose $x_i$, the model considered (see \ctn{Lunn12}) is
$y_{ij}\sim Poisson(\mu(x_i))$, where $\log(\mu_i)=\alpha+\beta+\gamma(x_i+10)+\gamma x_i$. 
The priors for $\alpha,\beta,\gamma$ are independent zero-mean normal distributions with standard deviation $100$.

We first consider an additive TMCMC application to this problem, with the scaling constants in the additive
transformation for $\alpha,\beta,\gamma$ being $0.21832$, $0.056563$ and $0.00024192$, respectively. The forward and backward transformations are considered with
equal probability. The rest of the details remain the same as in the Challenger problem. Quite encouraging TMCMC convergence properties are indicated by our
informal diagnostics. As before, $\bmu$ and $\bSigma$ are estimated from the stored
$10^5$ TMCMC realizations, after ignoring the first $10^5$ realizations as burn-in.

For application of our $iid$ method, we set, for $d=3$, $r_d=3$, $a_d=0.02$, and $M=200$, as suggested by our experiments. 
Again, we avoided rejection sampling in the same way as in the Challenger case, and considered $N_i=5000$, drawing $10,000$ realizations from the three-parameter posterior. 
The time for implementation here is only a minute.

Figure \ref{fig:salmonella} %and \ref{fig:salmonella_corr} 
compares TMCMC and $iid$ sampling with respect to density estimation of the parameters %and correlation structure estimation,
with $10^5$ TMCMC realizations thinned to $10,000$ realizations by considering one in every $10$. Once again, the $iid$ results are in close agreement
with the TMCMC results.

The estimated correlation structures, expectedly, are not much different with
respect to the competing methods. The TMCMC estimates of the posterior correlations between $(\alpha,\beta)$, $(\alpha,\gamma)$ and $(\beta,\gamma)$
are $-0.9627896$, $0.7810736$ and $-0.8901597$, respectively, while the respective $iid$ estimates are $-0.9625148$, $0.7774780$ and $-0.8887078$.
\begin{figure}
	\centering
	\subfigure [TMCMC and $iid$-based density for $\alpha$.]{ \label{fig:po1}
	\includegraphics[width=7.5cm,height=7.5cm]{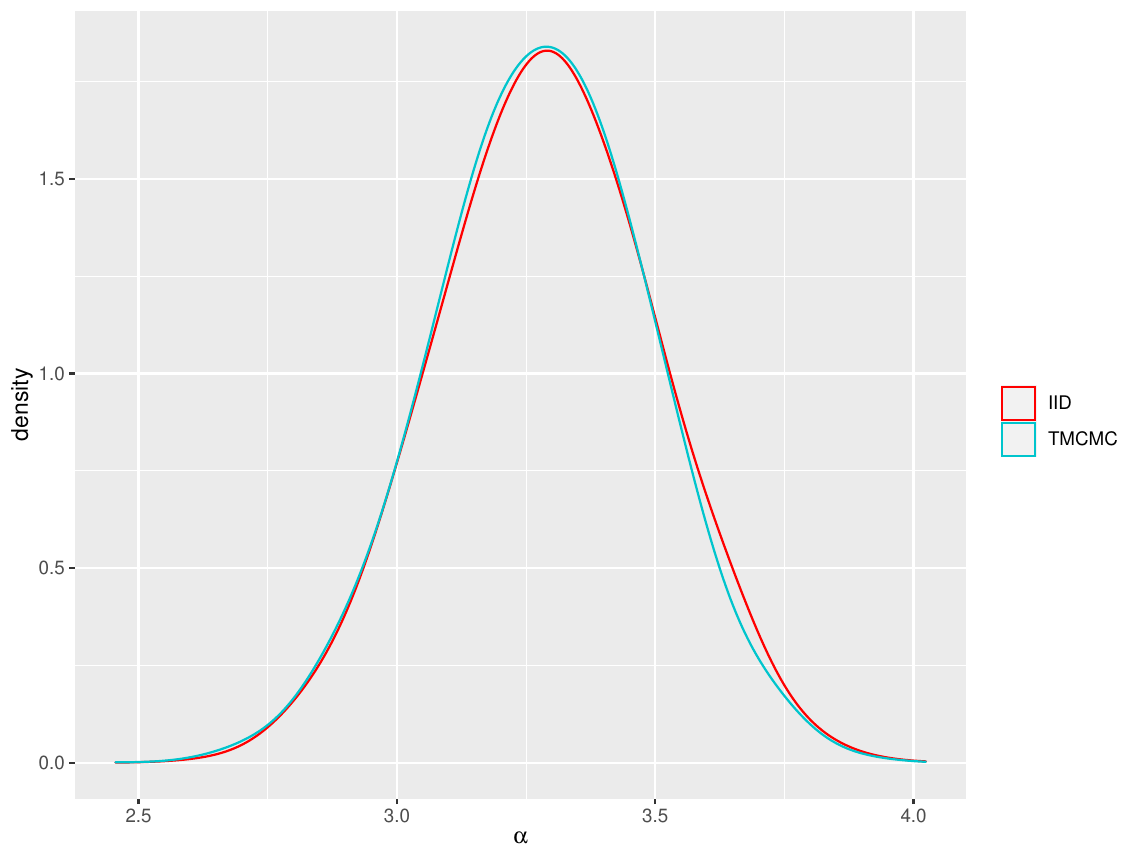}}
	\hspace{2mm}
	\subfigure [TMCMC and $iid$-based density for $\beta$.]{ \label{fig:po2}
	\includegraphics[width=7.5cm,height=7.5cm]{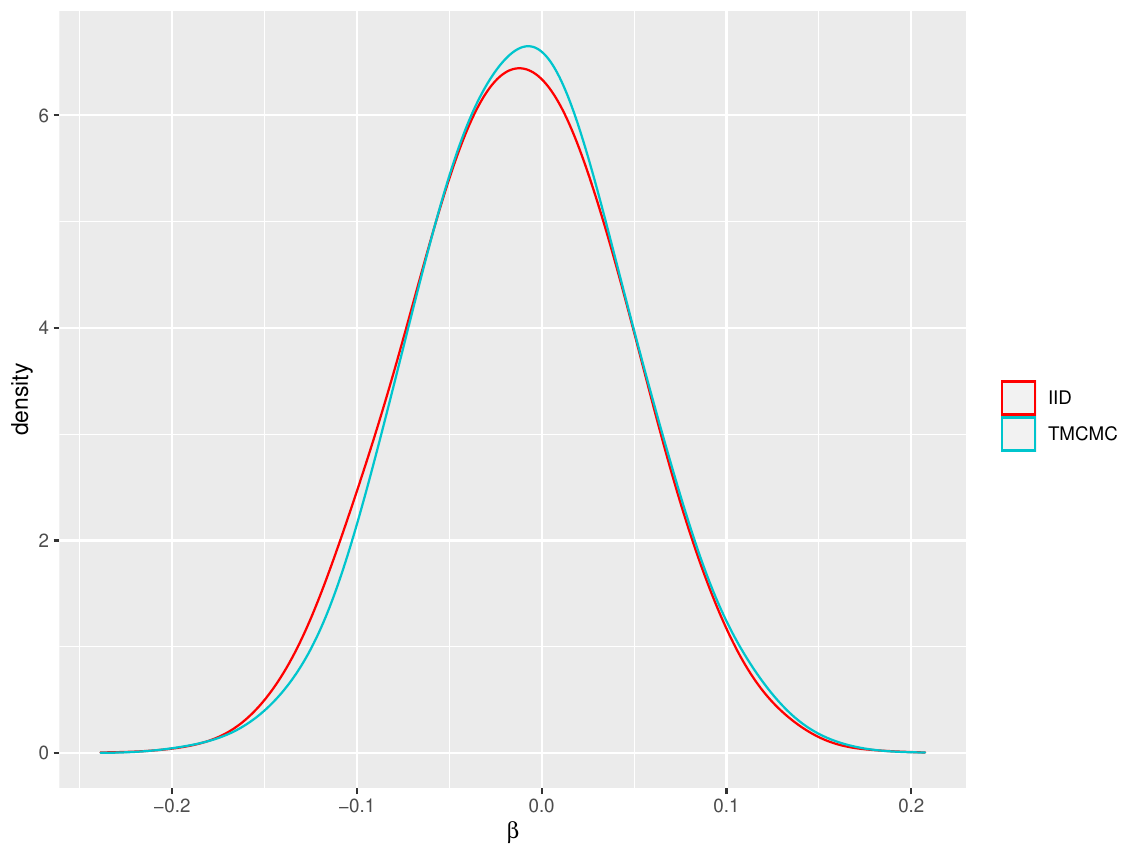}}\\
	\vspace{2mm}
	\subfigure [TMCMC and $iid$-based density for $\gamma$.]{ \label{fig:po3}
	\includegraphics[width=7.5cm,height=7.5cm]{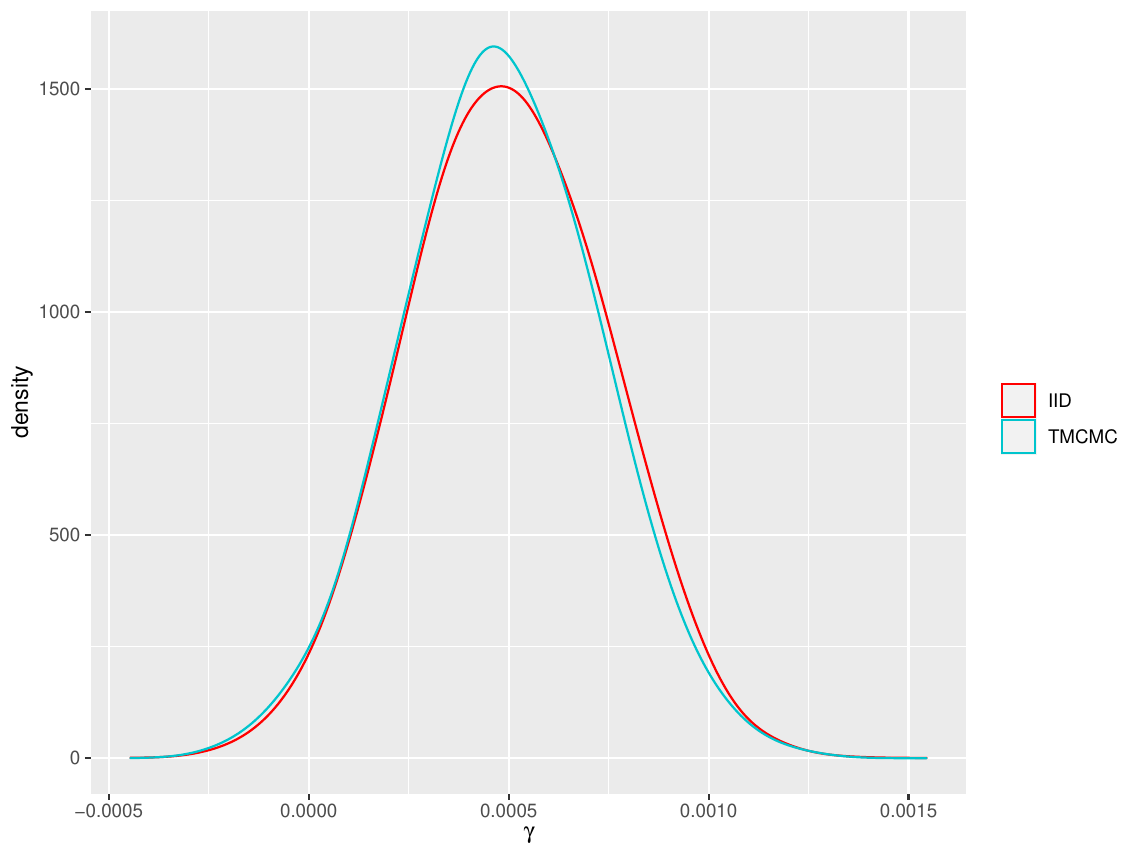}}
	\caption{Salmonella posterior. 
	The red and blue colours denote the $iid$ sample based density and the TMCMC based density, respectively.}
	\label{fig:salmonella}
\end{figure}

%\begin{figure}
%	\centering
%	\subfigure [TMCMC based correlation structure.]{ \label{fig:po_corr_exact}
%	\includegraphics[width=7.5cm,height=7.5cm]{figures_poisson/corr_tmcmc.pdf}}
%	\hspace{2mm}
%	\subfigure [Correlation structure based on $iid$ samples.]{ \label{fig:po_corr}
%	\includegraphics[width=7.5cm,height=7.5cm]{figures_poisson/corr_iid.pdf}}
%	\caption{Salmonella posterior. 
%	TMCMC and $iid$-based correlation structures.}
%	\label{fig:salmonella_corr}
%\end{figure}

\subsection{Application to the Rongelap Island dataset}
\label{subsec:rongelap}

We now consider application of our $iid$ sampling idea to the challenging spatial statistics problem involving radionuclide count data on Rongelap Island.
Following \ctn{Diggle98}, we model the count data, for $i=1,\ldots,157$, as
$$Y_i\sim Poisson(\mu_i),$$ where $$\mu_i=t_i\exp\{\beta+S(\bx_i)\},$$ $t_i$ being the duration of observation
at location $\bx_i$, $\beta$ is an unknown parameter and $S(\cdot)$ is a zero-mean Gaussian process
with isotropic covariance function of the form 
$$Cov\left(S(\tilde\bx_1),S(\tilde\bx_2)\right)
=\sigma^2\exp\{-\alpha\parallel\tilde\bx_1-\tilde\bx_2\parallel\}$$
for any two locations $\tilde\bx_1,\tilde\bx_2$. In the above, 
%$\parallel\cdot\parallel$ denotes the Euclidean distance between two locations, and 
$(\sigma^2, \alpha)$ are unknown parameters.
We assume uniform priors on the entire parameter space corresponding to
$(\beta, \log(\sigma^2), \log(\alpha))$. Since there are $157$ latent Gaussian variables $S(\bx_i)$ and three other unknowns $(\alpha,\beta,\sigma^2)$,
there are in all $160$ unknowns in this problem.

To first consider a TMCMC implementation, we adopted the same additive TMCMC sampler of \ctn{Dey17} specific to this spatial count data problem
(see Section 8 of their article), but also enhanced its convergence properties by adding the step with the flavour of \ctn{Liu00} in the same way
as in the previous applications regarding Challenger and Salmonella. After discarding the first $10^6$ realizations as burn-in, we stored one in $100$ in the next
$10^6$ iterations, to obtain $10,000$ realizations, which we used for estimation of $\bmu$ and $\bSigma$ needed for $\bA_i$ of our $iid$ sampler. 
The TMCMC exercise took about $1$ hour $40$ minutes.

Now, with $d=160$, application of the $iid$ method requires appropriate specification of $r_d$ and $a_d$ such that $\hat p_i$ are sufficiently large.
However, in this challenging application, all our experiments failed to yield significant $\hat p_i$. To handle this challenge, we decided to flatten
the posterior distribution in a way that its infimum and the supremum are reasonably close (so that $\hat p_i$ are adequately large) on all the $\bA_i$.
This requires an appropriate flattening bijective transformation. As it turns out, the MCMC literature already contains instances of such transformations
with stronger properties such that the transformation and its inverse are continuously differentiable. Such transformations are called diffeomorphisms.
In fact, diffeomorphisms with stronger properties have been considered in \ctn{Johnson12a}; see also \ctn{Dey16} who implement such diffeomorphisms in the TMCMC context.
However, in the above works, the transformations are meant to reduce thick-tailed target distributions to thin-tailed ones. 
Here we need just the opposite; the Rongelap Island posterior
needs to be converted to a thick-tailed distribution for our $iid$ sampling purpose. As can be anticipated, the inverse transformations for thick to thin tail
conversions will be appropriate here. Details follow.

\subsubsection{A suitable diffeomorphic transformation and the related method}
\label{subsubsec:diffeo}

Observe that if $\pi_{\btheta}$, the multivariate target density of some random vector $\btheta$ is of interest, then  
\begin{align}
\pi_{\bgamma}(\bgamma)=\pi_{\btheta}\left(h(\bgamma)\right)\left|\mbox{det}~\nabla h(\bgamma)\right|
\label{eq:transformed_target}
\end{align}
is the density of $\bgamma=h^{-1}(\btheta)$,
where $h$ is a diffeomorphism.
In the above, $\nabla h(\bgamma)$ denotes the gradient of $h$ at $\bgamma$ and $\mbox{det}~\nabla h(\bgamma)$ stands for the determinant of the gradient of $h$ at $\bgamma$.

\ctn{Johnson12a} obtain conditions on $h$ which make $\pi_{\bgamma}$ super-exponentially light.
Specifically, they define the following isotropic function $h:\mathbb R^d\mapsto\mathbb R^d$:
\begin{equation}
	h(\bgamma)=\left\{\begin{array}{cc}f(\|\bgamma\|)\frac{\bgamma}{\|\bgamma\|}, & \bgamma\neq \bzero\\
		0, & \bgamma=\bzero,
\end{array}\right.
\label{eq:isotropy}
\end{equation}
for some function $f: (0,\infty)\mapsto (0,\infty)$.
\ctn{Johnson12a} confine attention to isotropic diffeomorphisms, that is, functions of the form $h$ where 
both $h$ and $h^{-1}$ are continuously differentiable, with the further property that 
$\mbox{det}~\nabla h$ and  $\mbox{det}~\nabla h^{-1}$ are also continuously 
differentiable. In particular, if $\pi_{\bbeta}$ is only sub-exponentially light, then the following form of $f:[0,\infty)\mapsto [0,\infty)$ given by
\begin{equation}
f(x)=\left\{\begin{array}{cc}e^{bx}-\frac{e}{3}, & x>\frac{1}{b}\\
x^3\frac{b^3e}{6}+x\frac{be}{2}, & x\leq \frac{1}{b},
\end{array}\right.
\label{eq:diffeo2}
\end{equation}
where $b>0$, ensures that the transformed density $\pi_{\bgamma}$ of the form (\ref{eq:transformed_target}),
is super-exponentially light.

In our current context, the $d=160$ dimensional target distribution $\pi_{\btheta}$ needs to be converted to some thick-tailed distribution $\pi_{\bgamma}$. 
Hence, we apply the transformation $\bgamma=h(\btheta)$, the inverse of the transformation considered in \ctn{Johnson12a}.  
Consequently, the density of $\bgamma$ here becomes
\begin{align}
	\pi_{\bgamma}(\bgamma)=\pi_{\btheta}\left(h^{-1}(\bgamma)\right)\left|\mbox{det}~\nabla h(\bgamma)\right|^{-1},
\label{eq:transformed_target2}
\end{align}
where $h$ is the same as (\ref{eq:isotropy}) and $f$ is given by (\ref{eq:diffeo2}).
We also give the same transformation to the uniform proposal density (\ref{eq:proposal}), so that the new proposal density now becomes
\begin{equation}
	q_i(\bgamma')=\frac{1}{\mathcal L(\bA_i)}I_{\bA_i}(h^{-1}(\bgamma'))\left|\mbox{det}~\nabla h(\bgamma')\right|^{-1}.
	\label{eq:proposal2}
\end{equation}
With (\ref{eq:proposal2}) as the proposal density for (\ref{eq:transformed_target2}), the proposal will not cancel in the acceptance ratio
of the Metropolis-Hastings acceptance probability. For any set $\bA$, let $h(\bA)=\left\{h(\btheta):\btheta\in\bA\right\}$.
Also, now let $s_i=\underset{\bgamma\in h(\bA_i)}{\inf}~\frac{\tilde\pi_{\bgamma}(\bgamma)}{q_i(\bgamma)}$ and 
$S_i=\underset{\bgamma\in h(\bA_i)}{\sup}~\frac{\tilde\pi_{\bgamma}(\bgamma)}{q_i(\bgamma)}$, where $\tilde\pi_{\bgamma}(\bgamma)$ is the same as (\ref{eq:transformed_target2})
but without the normalizing constant.
Then, with (\ref{eq:proposal2}) as the proposal density we have 
\begin{align}
	P_i(\bgamma,h(\mathbb B\cap\bA_i))&\geq\int_{h(\mathbb B\cap\bA_i)}
	\min\left\{1,\frac{\tilde\pi_{\bgamma}(\bgamma')/q_i(\bgamma')}{\tilde\pi_{\bgamma}(\bgamma)/q_i(\bgamma)}\right\}q_i(\bgamma')d\bgamma'\notag\\
	%&\geq\left(\frac{s_i}{S_i}\right)\times\frac{\mathcal L(\mathbb B\cap\bA_i)}{\mathcal L(\bA_i)}\notag\\
	&\geq p_i~Q_i(h(\mathbb B \cap\bA_i)),\notag
\end{align}
where $p_i=s_i/S_i$ and $Q_i$ is the probability measure corresponding to (\ref{eq:proposal2}).
The rest of the details remain the same as before with necessary modifications pertaining to the new proposal density (\ref{eq:proposal2}) and the new 
Metropolis-Hastings acceptance ratio
with respect to (\ref{eq:proposal2}) incorporated in the subsequent steps. Once $\bgamma$ is generated from (\ref{eq:transformed_target2}) we
transform it back to $\btheta$ using $\btheta=h^{-1}(\bgamma)$.

\subsubsection{Results with the diffeomorphic transformation}
\label{subsubsec:diffeo_results}
With the transformed posterior density corresponding to the form (\ref{eq:transformed_target}), our experiments suggested that 
$M=10,000$, $b=0.01$, $r_d=7.8$ and $a_d=0.0005$ are appropriate, and rendered $\hat p_i$ significantly large for the transformed target.
With $N_i=5000$ and the technique of avoiding rejection sampling remaining the same as in the previous two applications, generation of $10,000$ $iid$ realizations
took $30$ minutes in our parallel processors.

Figures \ref{fig:dtm1} and \ref{fig:dtm2} show that the marginal density estimates with the TMCMC based and the $iid$ based samples are in agreement, as in the
previous examples.
That the correlation structures obtained
by both the methods are also in close agreement, is borne out by Figure \ref{fig:dtm3}.

\begin{figure}
	\centering
	\subfigure [TMCMC and $iid$-based density for $\alpha$.]{ \label{fig:dn1}
	\includegraphics[width=7.5cm,height=7.5cm]{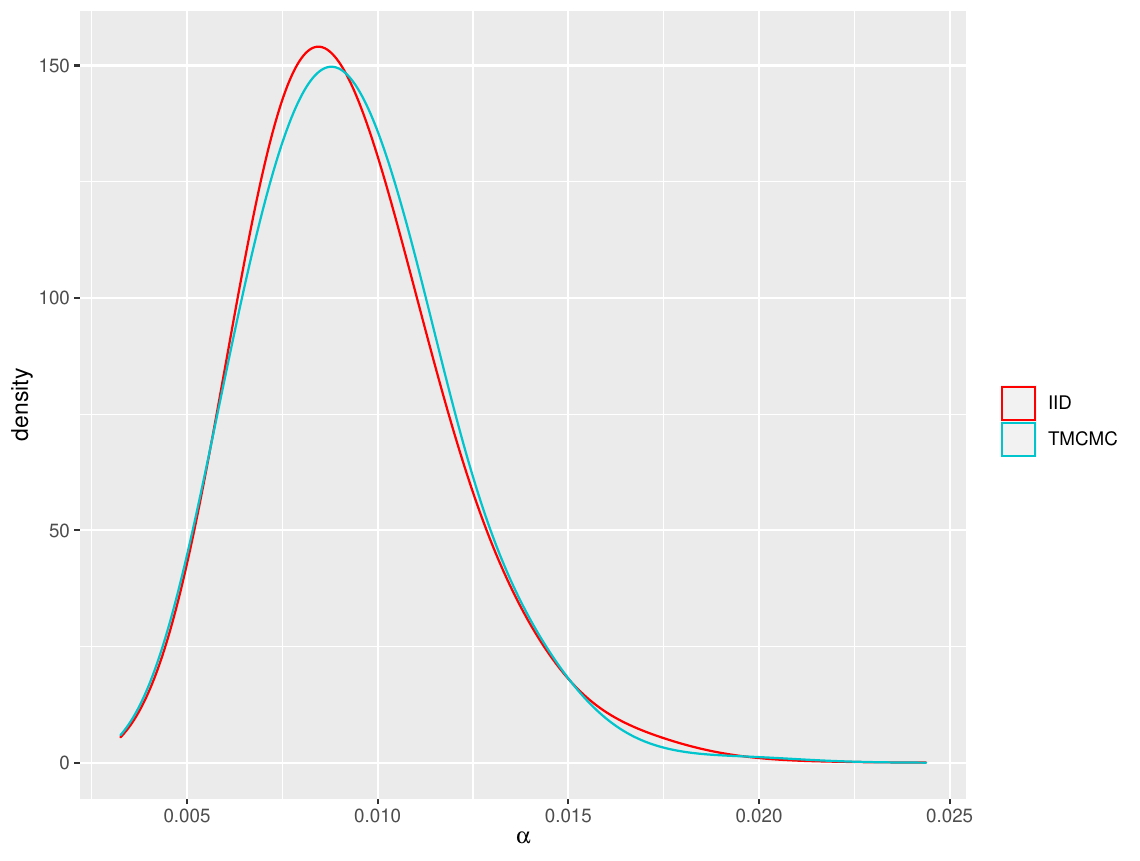}}
	\hspace{2mm}
	\subfigure [TMCMC and $iid$-based density for $\beta$.]{ \label{fig:dn10}
	\includegraphics[width=7.5cm,height=7.5cm]{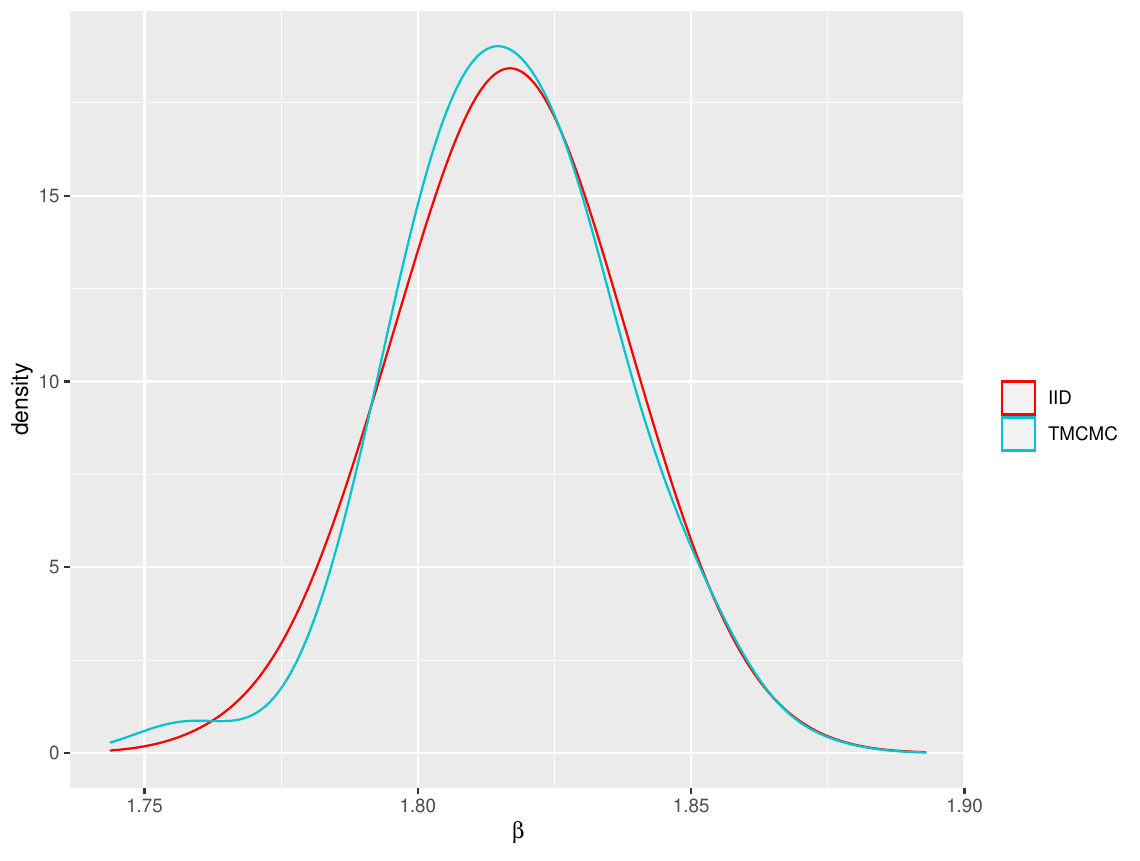}}\\
	\vspace{2mm}
	\subfigure [TMCMC and $iid$-based density for $\sigma^2$.]{ \label{fig:dn25}
	\includegraphics[width=7.5cm,height=7.5cm]{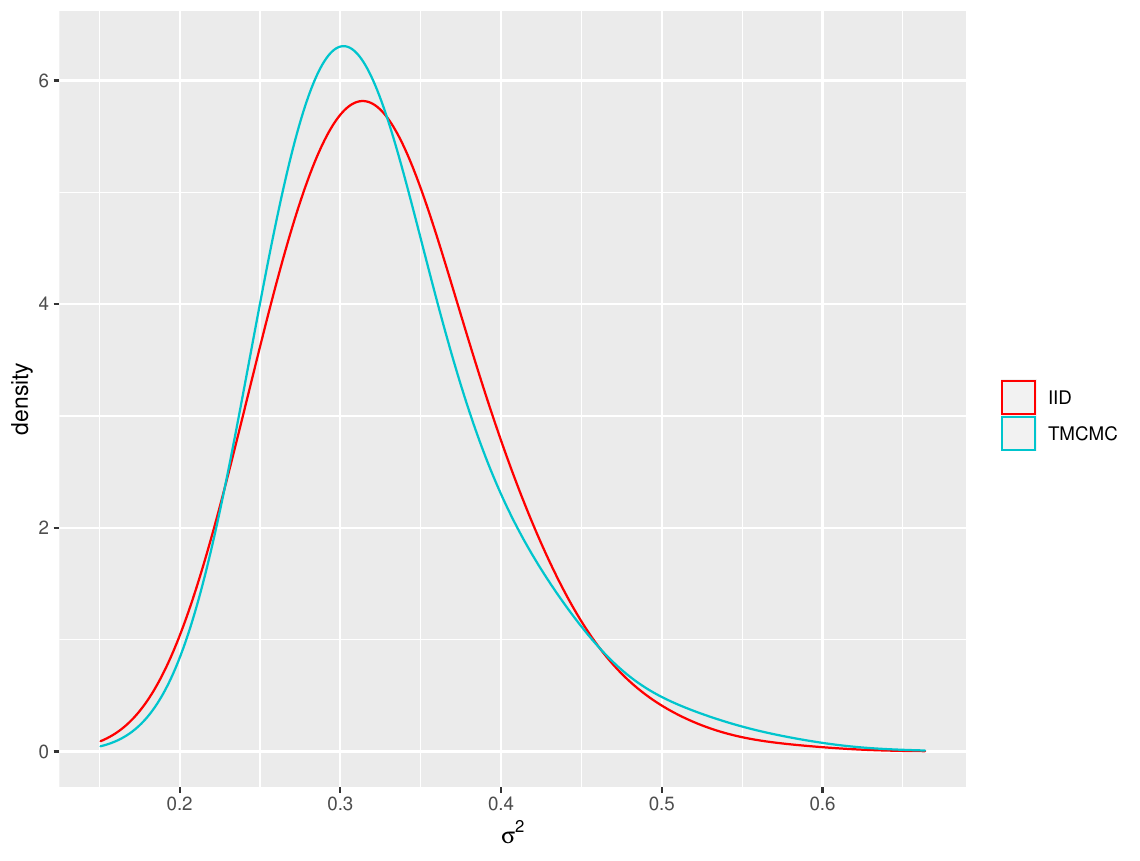}}
	\vspace{2mm}
	\subfigure [TMCMC and $iid$-based density for $S_1$.]{ \label{fig:dn50}
	\includegraphics[width=7.5cm,height=7.5cm]{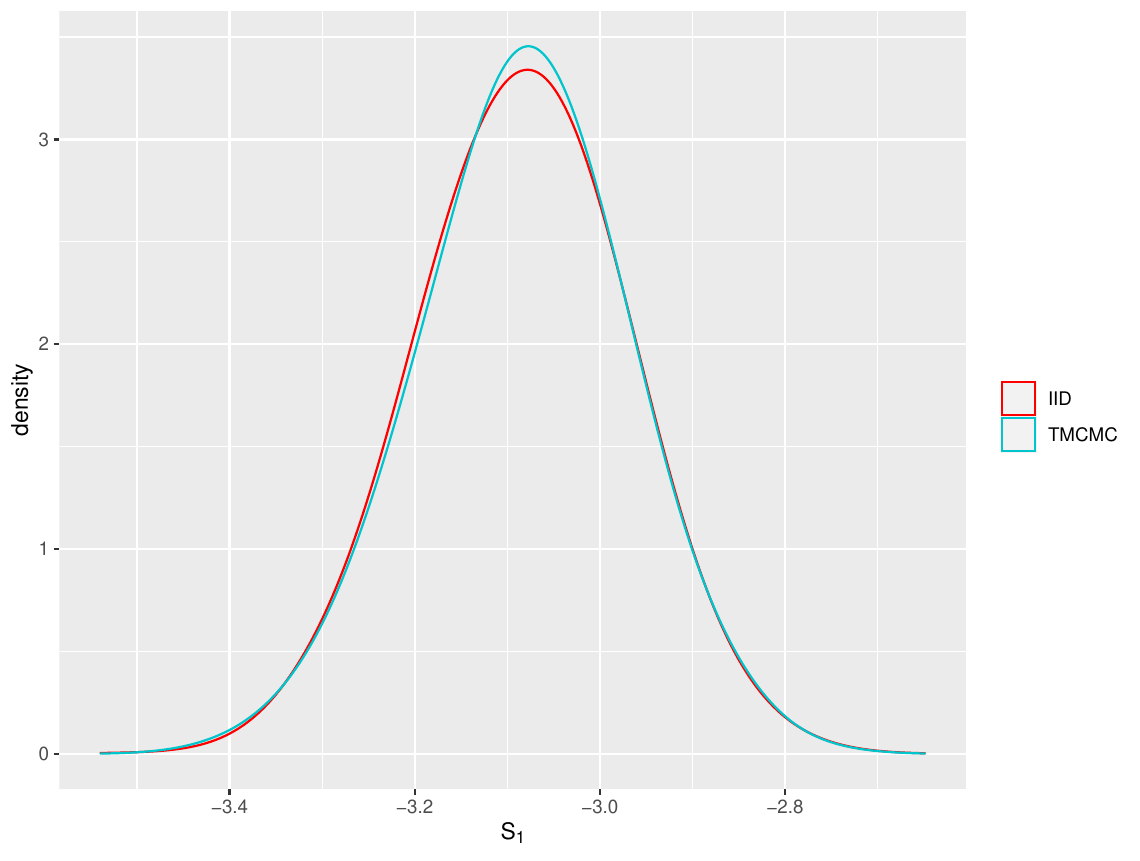}}
	\caption{Rongelap Island posterior. 
	The red and blue colours denote the $iid$ sample based density and the TMCMC based density, respectively.}
	\label{fig:dtm1}
\end{figure}

\begin{figure}
	\centering
	\subfigure [TMCMC and $iid$-based density for $S_{10}$.]{ \label{fig:n1d}
	\includegraphics[width=7.5cm,height=7.5cm]{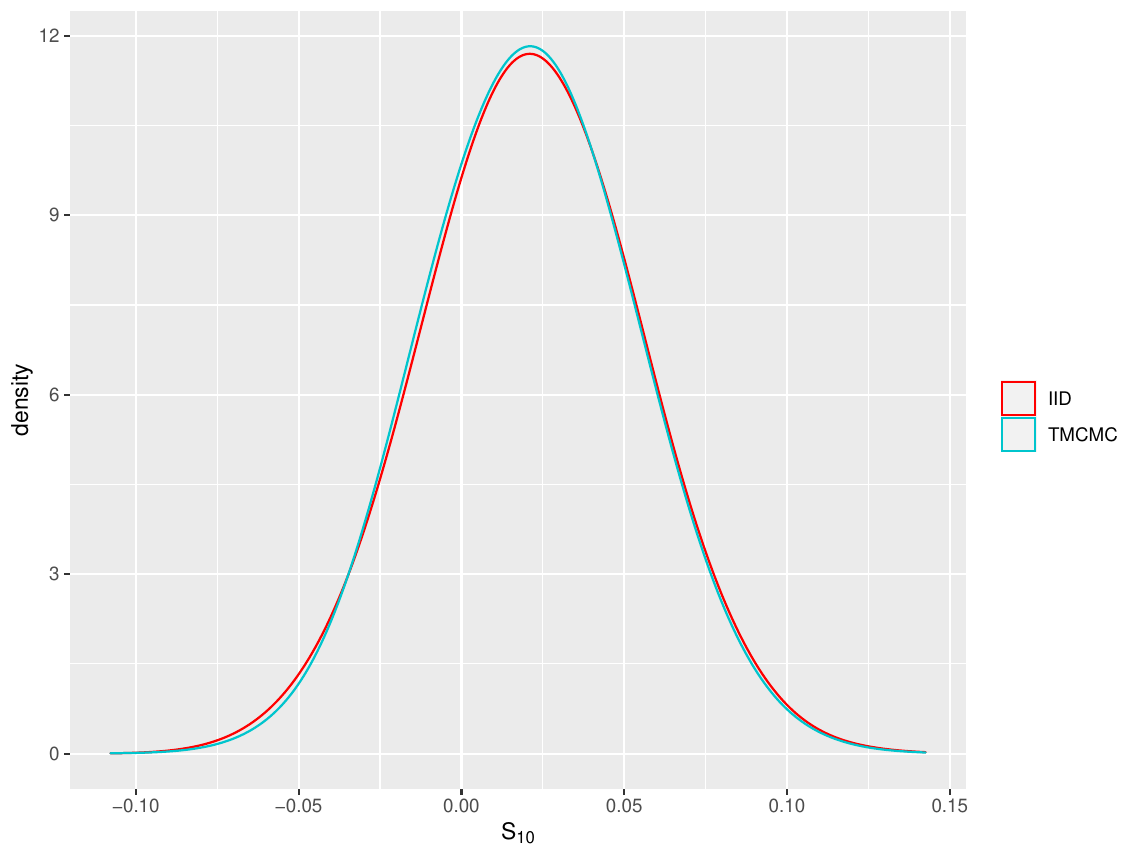}}
	\hspace{2mm}
	\subfigure [TMCMC and $iid$-based density for $S_{50}$.]{ \label{fig:n10d}
	\includegraphics[width=7.5cm,height=7.5cm]{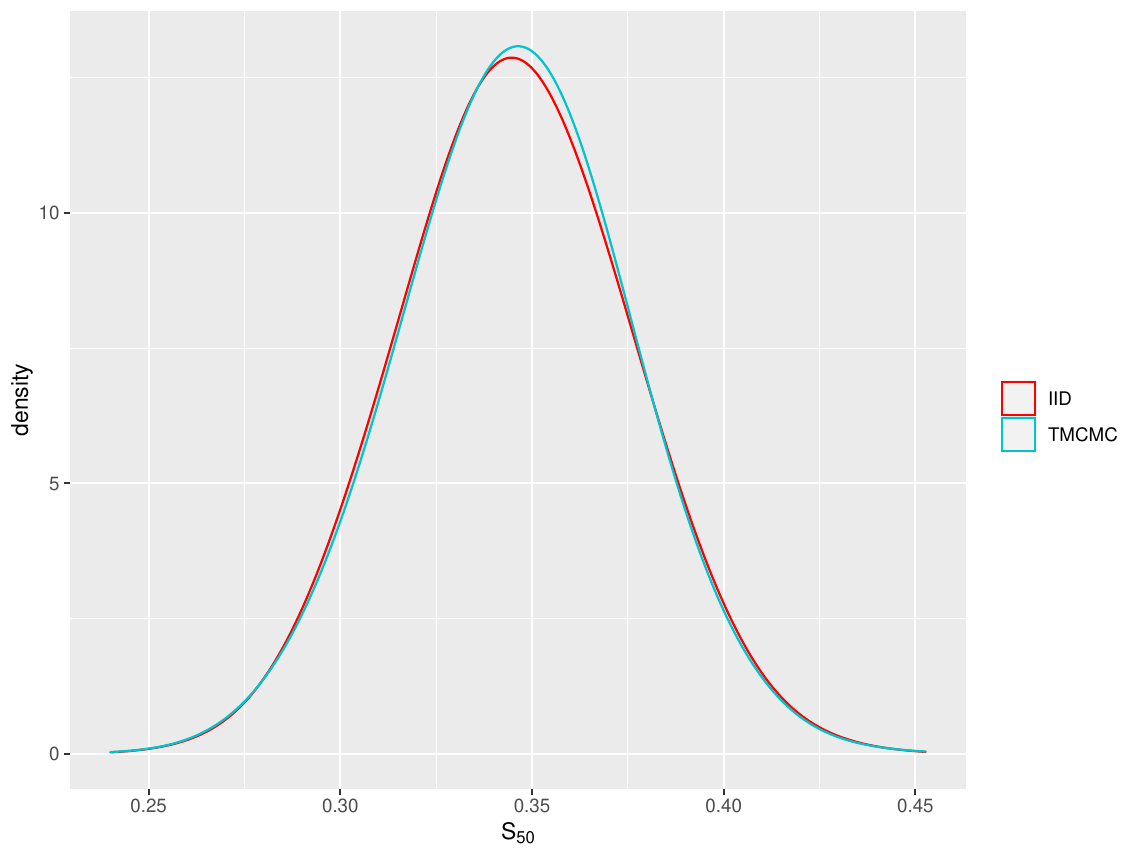}}\\
	\vspace{2mm}
	\subfigure [TMCMC and $iid$-based density for $S_{100}$.]{ \label{fig:n25d}
	\includegraphics[width=7.5cm,height=7.5cm]{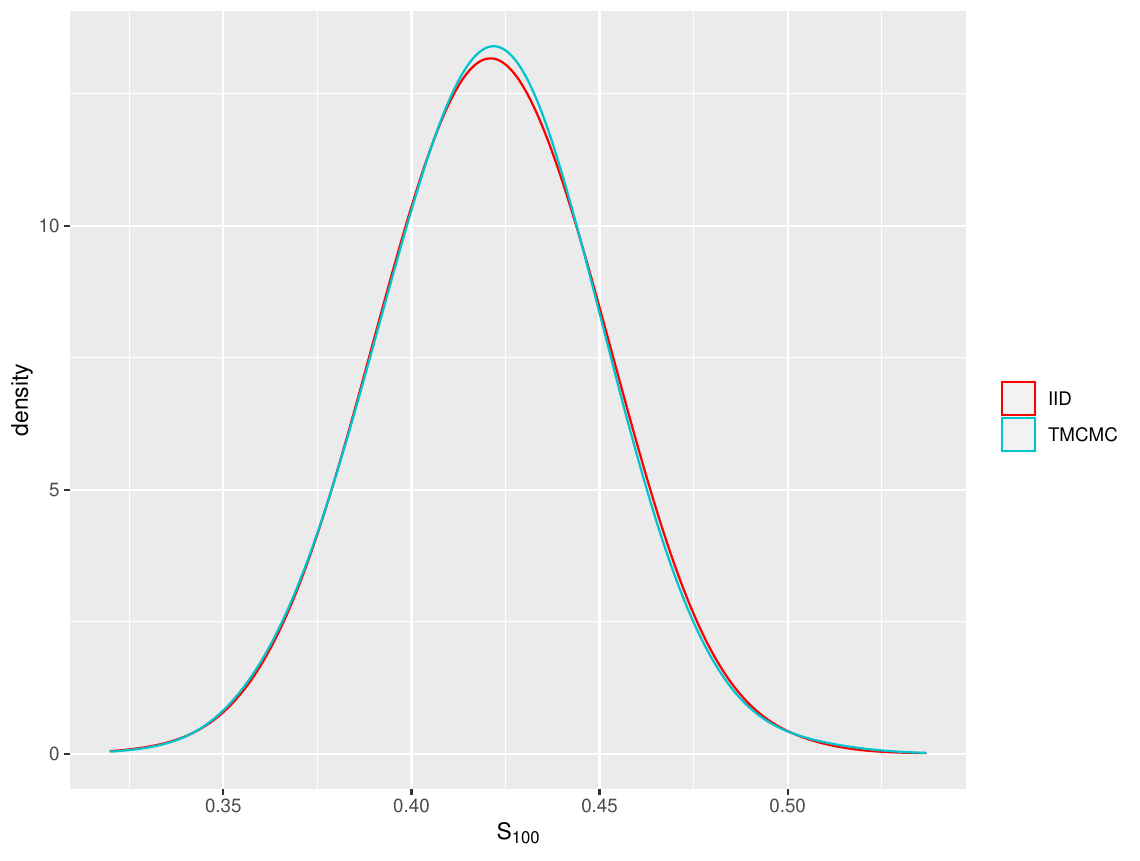}}
	\vspace{2mm}
	\subfigure [TMCMC and $iid$-based density for $S_{157}$.]{ \label{fig:n50d}
	\includegraphics[width=7.5cm,height=7.5cm]{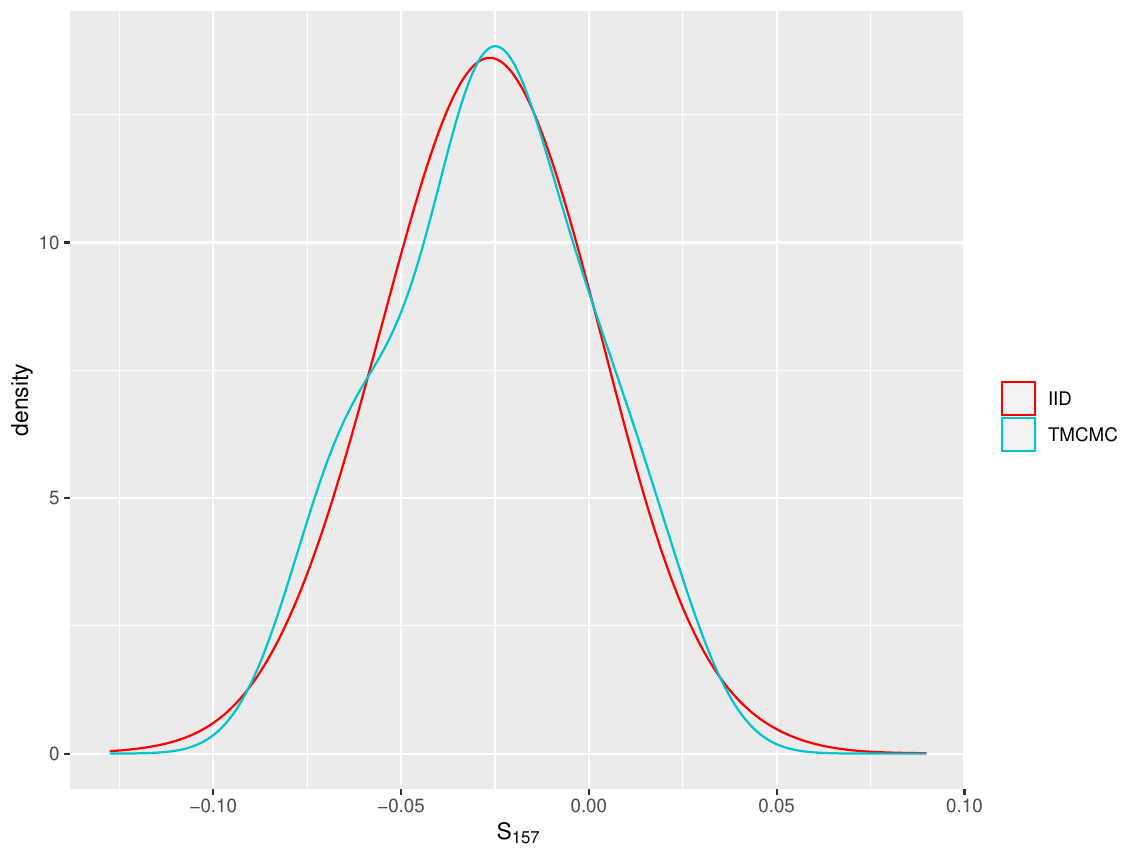}}
	\caption{Rongelap Island posterior. 
	The red and blue colours denote the $iid$ sample based density and the TMCMC based density, respectively.}
	\label{fig:dtm2}
\end{figure}

\begin{figure}
	\centering
	\subfigure [TMCMC based correlation structure.]{ \label{fig:dtm_corr_tmcmc}
	\includegraphics[width=7.5cm,height=7.5cm]{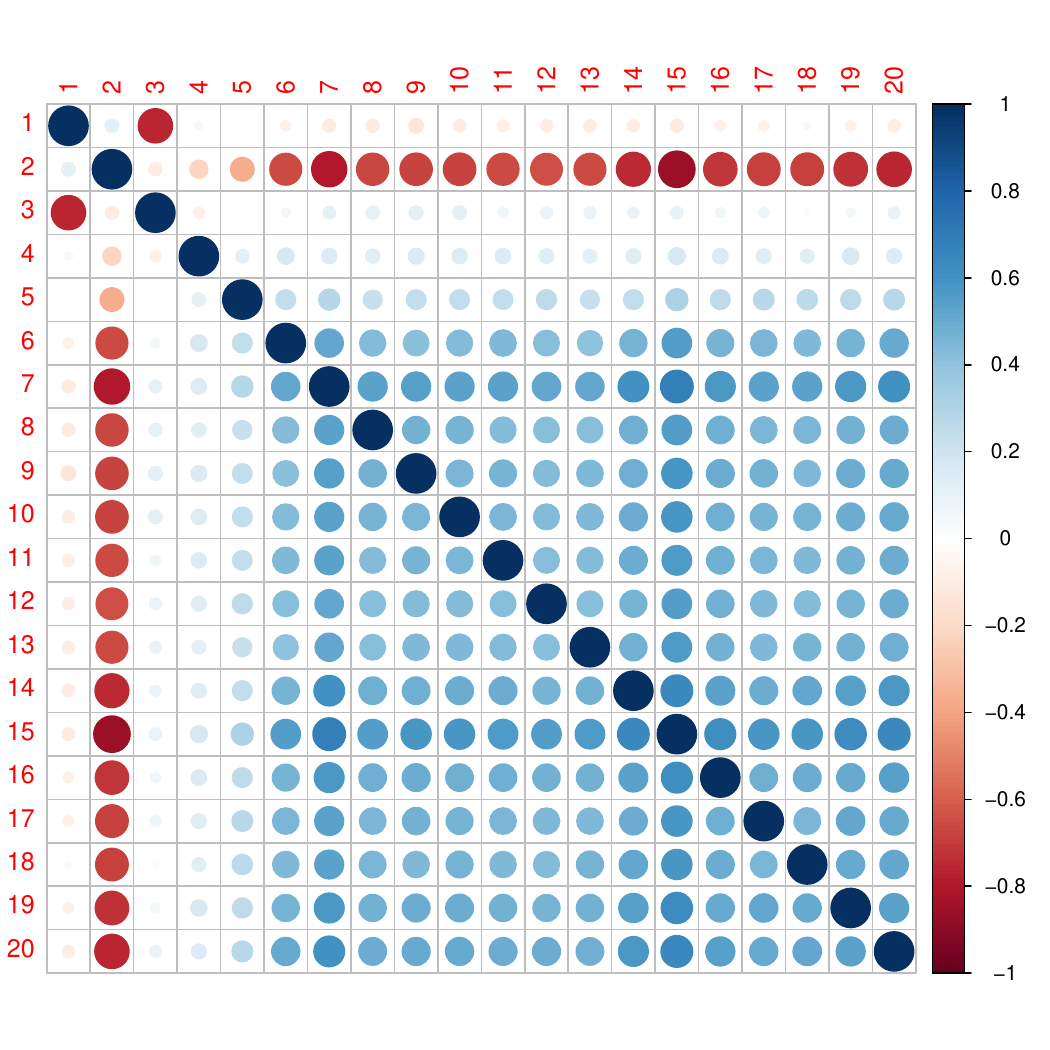}}
	\hspace{2mm}
	\subfigure [Correlation structure based on $iid$ samples.]{ \label{fig:dtm_corr}
	\includegraphics[width=7.5cm,height=7.5cm]{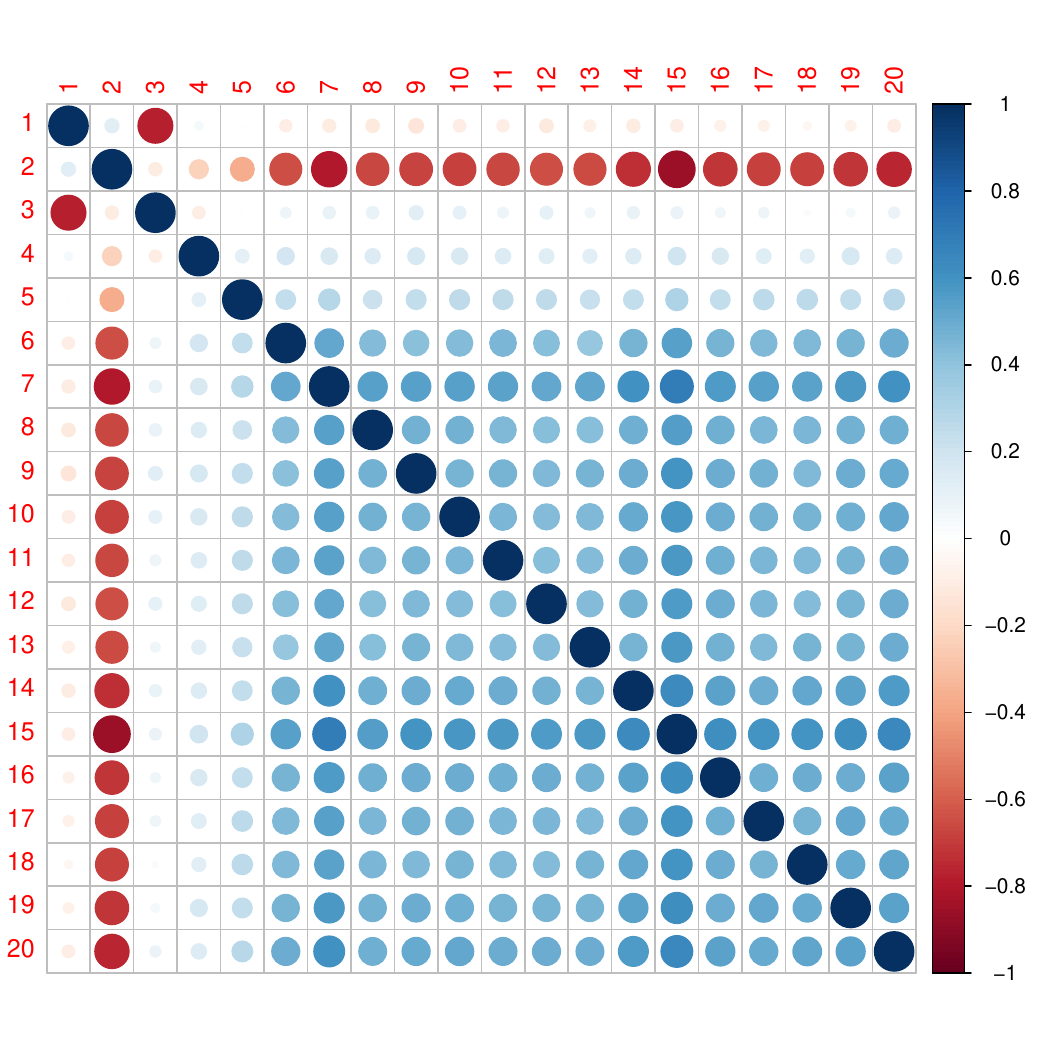}}
	\caption{Rongelap Island posterior. 
	TMCMC and $iid$-based correlation structures for the first $20$ co-ordinates.}
	\label{fig:dtm3}
\end{figure}

\section{Summary and conclusion}
\label{sec:conclusion}
In this article, we have proposed a general methodology for producing $iid$ realizations from any distribution on $\mathbb{R}^d$, for arbitrary $d \ge 1$. 
Our key idea is to represent the target distribution as an infinite mixture of component distributions defined on a central ellipsoid and on successive ellipsoidal 
annuli. The compactness of these sets allowed us to establish a minorization inequality for the Metropolis--Hastings algorithm with uniform proposals, which 
in turn facilitated a tractable perfect sampling scheme.

By integrating these results, we developed a generic parallel algorithm for exact $iid$ sampling from general target distributions. Our approach is simple, widely 
applicable, and avoids the computational burden of coupling from the past. It also scales naturally with modern parallel computing environments.

Through experiments, we demonstrated that the method performs effectively on a variety of challenging problems, including high-dimensional settings and complex 
posterior distributions. In particular, we showed that 10,000 $iid$ realizations can be generated quickly and reliably, even in dimensions as high as 160.

The impact of this foundational work has been substantial, spawning a sequence of subsequent developments that extend the $iid$ sampling methodology to increasingly 
challenging contexts. \ctn{Bhattacharya21a} extended the method to general multimodal and variable-dimensional distributions, illustrating its applicability 
on mixtures of 50-dimensional normals and variable-dimensional normal mixtures for the acidity dataset. 
The methodology was further extended to doubly intractable posterior distributions, 
where normalizing constants depend on parameters and are analytically intractable. By combining Monte Carlo and importance sampling approximations with Gaussian process 
interpolations, the $iid$ sampling framework achieved highly accurate and efficient sampling for complex posteriors, including two-dimensional Ising and Strauss models, 
two-dimensional normal-gamma posteriors, and a 100-dimensional autologistic model (\ctn{Bhattacharya21b}).
The work was further extended to the framework of Bayesian nonparametrics, 
enabling exact $iid$ sampling from Dirichlet process mixture posteriors--a class of distributions notoriously difficult to sample satisfactorily using MCMC methods 
due to their infinite-dimensional and discrete nature. The $iid$ sampling method was validated using the novel and efficient Dirichlet process
mixture setup of \ctn{Bhattacharya08} on benchmark datasets such as enzyme, acidity, and galaxy data, demonstrating both 
computational efficiency and accuracy, with $10,000$ $iid$ realizations generated in minutes using parallel computing (\ctn{Bhattacharya22}). 
Most recently, \ctn{Durba25} extended the $iid$ sampling procedure to a novel Bayesian nonparametrics theory
based on random normalizing flows. Such flows, governed by random diffeomorphisms, are  induced by compositions of monotone Gaussian processes, leading to
highly complex, non-standard distributions on non-Euclidean spaces. Implementation of the $iid$ sampling theory in such a setup, along with its comparison with MCMC,
emphatically brought out the convergence challenges of the latter in high dimensions and the necessity of $iid$ sampling.

Together, these developments underscore the versatility and power of the $iid$ sampling methodology. They illustrate a coherent trajectory of research that 
extends the original framework from high-dimensional Euclidean distributions to doubly intractable distributions and nonparametric Bayesian posteriors. 
This body of work provides a practical and theoretically rigorous solution to exact $iid$ sampling, offering promising directions for both theoretical 
research and applied Bayesian computation, particularly in contexts where assessing convergence of MCMC methods is difficult.

\section*{Acknowledgments}
We are sincerely grateful to the Editors for very valuable feedback. We also thank ChatGPT for some help with proofreading.

%\newpage

\renewcommand\baselinestretch{1.3}
\normalsize
\bibliography{irmcmc}

@MISC{Durba25,
  author = {D. Bhattacharya and S. Roy and S. Bhattacharya},
  title = {{B}ayesian {N}onparametrics {W}ith {R}andom {N}ormalizing {F}lows},
  year = {2025},
  note = {Available at \url{https://www.researchgate.net/publication/395337211_Bayesian_Nonparametrics_With_Random_Normalizing_Flows}}
}

@ARTICLE{Murdoch98,
               TITLE = "{E}xact sampling for a continuous state",
	       AUTHOR = "D. Murdoch and P. J. Green", 
	       JOURNAL = "Scandinavian Journal of Statistics", 
               VOLUME = "25",
	       NUMBER  = "",
	       PAGES  = "483--502",
               YEAR    =  1998,
 }

@ARTICLE{Dalal89,
               TITLE = "{R}isk {A}nalysis of the {S}pace {S}huttle: pre-{C}hallenger {P}rediction of
	          {F}ailure",
	       AUTHOR = "S. R. Dalal and E. B. Fowlkes and B. Hoadley", 
	       JOURNAL = "Journal of the American Statistical Association", 
               VOLUME = "84",
	       NUMBER  = "",
	       PAGES  = "945--957",
               YEAR    =  1989,
 }

@ARTICLE{Martz92,
               TITLE = "{T}he {R}isk of {C}atastrophic {F}ailure of the {S}olid {R}ocket
	             {B}oosters on the {S}pace {S}huttle",
	       AUTHOR = "H. F. Martz and W. J. Zimmer", 
	       JOURNAL = "The American Statistician", 
               VOLUME = "46",
	       NUMBER  = "",
	       PAGES  = "42--47",
               YEAR    =  1992,
 }

@MISC{Bhattacharya21a,
  author = {S. Bhattacharya},
  title = {{IID} {S}ampling from {I}ntractable {M}ultimodal and {V}ariable-{D}imensional {D}istributions},
  year = {2021},
  note = {arXiv:2109.12633v2}
}

@MISC{Bhattacharya21b,
  author = {S. Bhattacharya},
  title = {{IID} {S}ampling from {D}oubly {I}ntractable {D}istributions},
  year = {2021},
  note = {arXiv:2112.07939v1}
}

@MISC{Bhattacharya22,
  author = {S. Bhattacharya},
  title = {{IID} {S}ampling from {P}osterior {D}irichlet {P}rocess {M}ixtures},
  year = {2022},
  note = {arXiv:2206.09233v1}
}

@article{Liu00,
  title={{G}eneralized {G}ibbs {S}ampler and {M}ultigrid {M}onte {C}arlo for {B}ayesian {C}omputation},
  author={J. S. Liu and S. Sabatti},
  journal={Biometrika},
  volume={87},
  number={},
  pages={353--369},
  year={2000},
  publisher={},
  note={}
}

@BOOK{Giraud15,
  title={{I}ntroduction to {H}igh-{D}imensional {S}tatistics},
  publisher={CRC Press},
  year={2015},
  author={C. Giraud},
  address={Boca Raton, FL},
  note={}
}

@ARTICLE{Johnson12a,
               TITLE = "{V}ariable {T}ransformation to {O}btain {G}eometric {E}rgodicity in the
	       {R}andom-{W}alk {M}etropolis {A}lgorithm",
	       AUTHOR = "L. T. Johnson and Geyer", 
	       JOURNAL = "The Annals of Statistics", 
               VOLUME = "40",
	       NUMBER  = "",
	       PAGES  = "3050--3076",
               YEAR    =  2012,
 }

@ARTICLE{Dey16,
  author = {K. K. Dey and S. Bhattacharya},
  title = {{O}n {G}eometric {E}rgodicity of {A}dditive and {M}ultiplicative {T}ransformation {B}ased {M}arkov {C}hain {M}onte {C}arlo in {H}igh {D}imensions},
  journal = {Brazilian Journal of Probability and Statistics},
  year = {2016},
  volume = {30},
  pages = {570--613},
  note = {Also available at ``http://arxiv.org/pdf/1312.0915.pdf''}
}

@ARTICLE{Dey17,
  author = {K. K. Dey and S. Bhattacharya},
  title = {{A} {B}rief {T}utorial on {T}ransformation {B}ased {M}arkov {C}hain {M}onte {C}arlo and {O}ptimal {S}caling of the {A}dditive {T}ransformation},
  journal = {Brazilian Journal of Probability and Statistics},
  year = {2017},
  volume = {31},
  pages = {569--617},
  note = {Also available at ``http://arxiv.org/abs/1307.1446''}
}

@ARTICLE{Bhattacharya08,
  author = {S. Bhattacharya},
  title = {{G}ibbs {S}ampling {B}ased {B}ayesian {A}nalysis of {M}ixtures with {U}nknown {N}umber of {C}omponents},
  journal = {Sankhya. Series B},
  year = {2008},
  volume = {70},
  pages = {133--155}
}

@ARTICLE{Propp96,
  author = {J. G. Propp and D. B. Wilson},
  title = {{E}xact {S}ampling with {C}oupled {M}arkov {C}hains and {A}pplications to {S}tatistical {M}echanics},
  journal = {Random Structures and Algorithms},
  year = {1996},
  volume = {9},
  pages = {223--252}
}

@ARTICLE{Sabya12,
  author = {S. Mukhopadhyay and S. Bhattacharya},
  title = {{P}erfect {S}imulation for {M}ixtures with {K}nown and {U}nknown {N}umber of {C}omponents},
  journal = {Bayesian Analysis},
  year = {2012},
  volume = {7},
  pages = {675--714}
}

@ARTICLE{Dutta14,
  author = {S. Dutta and S. Bhattacharya},
  title = {{M}arkov {C}hain {M}onte {C}arlo {B}ased on {D}eterministic {T}ransformations},
  journal = {Statistical Methodology},
  year = {2014},
  volume = {16},
  pages = {100--116},
  note = {Also available at  http://arxiv.org/abs/1106.5850. Supplement available at http://arxiv.org/abs/1306.6684}
}

@ARTICLE{Chris04,
               TITLE = "{M}onte {C}arlo {M}aximum {L}ikelihood in {M}odel-{B}ased {G}eostatistics",
	       AUTHOR = "O. F. Christensen", 
	       JOURNAL = "Journal of Computational and Graphical Statistics",
               VOLUME = "13",
	       NUMBER  = "",
	       PAGES  = "702--718",
               YEAR    =  2004,
 }

@ARTICLE{Chris06,
               TITLE = "{R}obust {M}arkov {C}hain {M}onte {C}arlo {M}ethods for {S}patial
	           {G}eneralized {L}inear {M}ixed {M}odels",
	       AUTHOR = "O. F. Christensen", 
	       JOURNAL = "Journal of Computational and Graphical Statistics", 
               VOLUME = "15",
	       NUMBER  = "",
	       PAGES  = "1--17",
               YEAR    =  2006,
 }

@ARTICLE{Diggle98,
               TITLE = "{M}odel-{B}ased {G}eostatistics (with discussion)",
	       AUTHOR = "P. J. Diggle and J. A. Tawn and R. A. Moyeed", 
	       JOURNAL = "Applied Statistics", 
               VOLUME = "47",
	       NUMBER  = "",
	       PAGES  = "299--350",
               YEAR    =  1998,
 }

@INPROCEEDINGS{Diggle97,
         AUTHOR  = "P. J. Diggle and J. A. Tawn and R. A. Moyeed",
	 TITLE   = "{G}eostatistical {A}nalysis of {R}esidual {C}ontamination from {N}uclear {W}eapons {T}esting",
	 BOOKTITLE = "Statistics for Environment 3: Pollution Assessment and Control",
	 EDITOR    = "V. Barnet and K. F. Turkman",
	 PUBLISHER = "Chichester: Wiley",
	 PAGES   = "89--107",
	 YEAR    =  1997,
 }

@ARTICLE{Breslow84,
  author = {N. Breslow},
  title = {{E}xtra-{P}oisson {V}ariation in {L}og-{L}inear {M}odels},
  journal = {Applied Statistics},
  year = {1984},
  volume = {33},
  pages = {38--44}
}

@BOOK{Lunn12,
  title = {{T}he {BUGS} {B}ook: {A} {P}ractical {I}ntroduction to {B}ayesian {A}nalysis},
  publisher = {CRC Press},
  year = {2012},
  author = {D. Lunn and C. Jackson and N. Best and A. Thomas and D. Spiegelhalter},
  address = {Boca Raton, Florida}
}

@BOOK{Robert04,
  title = {{M}onte {C}arlo {S}tatistical {M}ethods},
  publisher = {Springer-Verlag},
  year = {2004},
  author = {C. P. Robert and G. Casella},
  address = {New York}
}

@ARTICLE{Rue01,
  author = {H. Rue},
  title = {{F}ast sampling of {G}aussian {M}arkov random fields},
  journal = {Journal of the Royal Statistical Society. Series B},
  year = {2001},
  volume = {63},
  pages = {325--338}
}
%\bibliography{irmcmc,references,references1}

%\newpage
%\input{figures}

\end{document}